\documentclass[sigconf]{acmart} 

\setcopyright{none}
\acmDOI{}
\acmISBN{}
\acmConference{Conference Name}{Month Year}{City, Country}
\acmYear{}
\copyrightyear{}
\usepackage{array}
\usepackage{textcomp}
\usepackage{verbatim}
\usepackage{graphicx}
\usepackage{longtable}
\usepackage{rotating}
\usepackage{booktabs}
\usepackage{tabularx}
\usepackage{makecell}
\usepackage{balance}
\usepackage{multirow}
\usepackage{listings}
\usepackage[table]{xcolor}
\usepackage{pifont}
\usepackage{textcase}
\usepackage[ruled,vlined,linesnumbered]{algorithm2e}

\begin{document}

\title{eDySec: A Deep Learning-based Explainable Dynamic Analysis Framework for Detecting Malicious Packages in PyPI Ecosystem}

\author{Sk Tanzir Mehedi}
\authornote{Corresponding authors (tanzir.mehedi@hdr.qut.edu.au)}
\affiliation{
  \institution{Queensland University of Technology}
  \city{Brisbane}
  \state{QLD}
  \country{Australia}
}

\author{Raja Jurdak}
\affiliation{
  \institution{Queensland University of Technology}
  \city{Brisbane}
  \state{QLD}
  \country{Australia}
}

\author{Chadni Islam}
\affiliation{
  \institution{Edith Cowan University}
  \city{Joondalup}
  \state{WA}
  \country{Australia}
}

\author{Abu Bakar Siddique Mahi}
\affiliation{
 \institution{Research Graduate School, BUBT}
 \city{Dhaka}
 \country{Bangladesh}
 }

\author{Gowri Ramachandran}
\affiliation{
  \institution{Queensland University of Technology}
  \city{Brisbane}
  \state{QLD}
  \country{Australia}
}

\renewcommand{\shortauthors}{Tanzir et al.}

\begin{abstract}
  The security of open-source software repositories is increasingly threatened by next-gen software supply chain attacks. These attacks include multiphase malware execution, remote access activation, and dynamic payload generation. Traditional Machine Learning (ML) detectors struggle to detect these attacks due to the high-dimensional and sparse nature of dynamic behavioral data, including system calls, network traffic, directory access patterns, and dependency logs. As a result, these data characteristics degrade the performance, stability, and explainability of ML models. These challenges have made Deep Learning (DL) a promising alternative, given its success across various domains and its potential for modeling complex patterns. This paper presents eDySec, a DL-based efficient, stable, and explainable framework for dynamic behavioral analysis to detect malicious packages. Using the QUT-DV25 dataset, which captures both install-time and post-installation behaviors of packages, we evaluate DL models and investigate feature sets to identify the most discriminative attributes for enabling efficient malicious package detection. Additionally, model stability analysis and explainable AI techniques are incorporated into the detection pipeline to enable stable, and transparent interpretations of model decisions. Experimental results demonstrate that eDySec significantly outperforms the state-of-the-art frameworks. Specifically, it halves feature dimensionality while lowering false positives by 82\% and false negatives by 79\%. It also improves accuracy by 3\%, achieves near-perfect stability, and maintains an inference latency of 170ms per package. Further analysis reveals that feature and model selection play a critical role, as certain combinations degrade performance. Ultimately, this study advances the understanding of the strengths and limitations of dynamic analysis for the development of eDySec against next-gen software supply chain attacks.
\end{abstract}


\maketitle

\newcommand{\best}[1]{\cellcolor{green!10}\textbf{#1}}
\newcommand{\tc}[1]{\cellcolor{black!10}\textbf{#1}}
\arrayrulecolor{black}

\section{Introduction}
\label{introduction}

Today’s software supply chains resemble intricate webs, where each component can serve as a potential entry point for attackers~\cite{milligan2023cybersecurityventures}. Among these components, open-source packages are especially critical, serving as foundational building blocks for the entire ecosystem~\cite{milligan2023cybersecurityventures, cisa_opensource_security}. A vulnerability in even a single malicious package can cascade throughout the ecosystem, compromising countless applications~\cite{gesellchen2021securing}. The Sonatype Open Source Malware Index (2025) reports that over 845,000 malicious packages have been identified across ecosystems, including 16,279 malicious packages targeting major repositories such as Python Package Index (PyPI) and Node Package Manager (NPM)~\cite{sonatype2025malware845k}. Among these ecosystems, PyPI has emerged as a primary target due to its scale and accessibility~\cite{ladisa2023crosslang, mehedi2025dysec}.

Since 2021, malicious package activity has increasingly affected PyPI, which now receives more than 500 inbound malware reports each month~\cite{fiedler2024aiocpa}. A notable example is LiteLLM, a PyPI package first released on 27 July 2023 (v0.1.0)~\cite{pypi_litellm}. Within 2.8 years, it released 1,284 updates, averaging nearly one to two releases per day~\cite{pypi_litellm}. However, on 24 March 2026, two compromised versions of the package, v1.82.7 and v1.82.8, were published on PyPI~\cite{sonatype2026litellm}. These releases contained a sophisticated multi-stage credential-stealing payload targeting cloud services, Continuous Integration and Continuous Delivery (CI/CD) pipelines, and Application Programming Interface (API) credentials~\cite{sonatype2026litellm}. As a result, any application depending on the package, directly or transitively, became vulnerable upon updating.

This incident reflects a broader trend in next-gen software supply chain attacks, which involve multiphase malware execution, remote access activation, dynamic payload generation, and geofencing-based attacks~\cite{sonatype2026litellm, coker2025lotl, mehedi2025dysec, safety2026grokwrapper, prajapati2025silentsync}. The severity of these attacks not only compromises privacy and security but also results in significant economic impact~\cite{fox2025supplychaincosts}. According to Cybersecurity Ventures, the annual global cost of software supply chain attacks was estimated at \$59 billion in 2025 and is projected to reach \$138 billion by 2031, an increase of 133.9\%~\cite{milligan2023cybersecurityventures, fox2025supplychaincosts}. Despite these alarming trends, effective detection remains a significant challenge ~\cite{milligan2023cybersecurityventures}. Recent studies demonstrate the usefulness of metadata~\cite{halder2024memptec}, static-code~\cite{ladisa2023crosslang, gandhi2024pyguardex, sun2024ea4mp, samaana2025mlpypi, zhang2025cerebro, gao2025malguard}, and hybrid analysis~\cite{gao2024pyradar, huang2024donapi, iqbal2026clampd, guo2026pyguard, iqbal2025pypiguard, yan2026pypimaldet} for malicious package detection. However, these methods remain constrained by their reliance on artifacts available prior to execution. Consequently, they cannot detect malicious logic that emerges during install-time and post-installation, which is a common characteristic of next-gen software supply chain attacks~\cite{mehedi2025dysec, sonatype2026litellm}. More recently, dynamic analysis~\cite{mehedi2025dysec, tan2026synthchain} has addressed these limitations by providing direct visibility into install-time and post-installation activity, thereby offering a more realistic basis for identifying evasive malicious packages.

Nevertheless, achieving the full potential of dynamic analysis remains a major challenge, as the feature space is high-dimensional, sparse, and highly variable~\cite{mehedi2025dysec, zheng2024oscar}. This challenge is further complicated by the adaptive and context-dependent nature of next-gen malicious behaviors, which may manifest through mechanisms such as multiphase malware execution, remote access activation, and dynamic payload generation. As a result, feature extraction becomes particularly difficult, since the most informative behavioral signals may only appear under specific execution conditions~\cite{mehedi2025dysec}. Although recent studies have explored dynamic behavioral datasets for malicious package detection~\cite{mehedi2025dysec, tan2026synthchain}, rigorous feature-level analysis remains limited. For instance, rule-based methods have incorporated dynamic behaviors and reported an F1-score of 91\%, but their reliance on predefined rules constrains adaptability and weakens generalization to evolving attack patterns~\cite{zheng2024oscar}. This direction was further extended using a traditional Machine Learning (ML)-based method~\cite{mehedi2025dysec}. However, despite achieving 95.99\% accuracy, this method still produces a relatively high number of false positives and false negatives~\cite{mehedi2025dysec}. This limitation indicates that conventional ML models remain inadequate for identifying the most discriminative dynamic features and for training efficient detection models to understand complex behavioral patterns, including system calls, network traffic, directory access patterns, and dependency logs.

A further challenge is to ensure the stability of malicious package detection models, which is essential in rigorous Deep Learning (DL) evaluation frameworks for assessing model stability and consistency~\cite{demsar2006stability, benavoli2017stability, dror2018stability, garcia2010stability}. However, it remains underexplored and inconsistently evaluated in prior studies~\cite{mehedi2025dysec, tan2026synthchain, gao2024pyradar, gandhi2024pyguardex, huang2024donapi, zhang2025cerebro}. A subset of studies~\cite{ladisa2023crosslang, halder2024memptec, sun2024ea4mp, samaana2025mlpypi, iqbal2026clampd} provides partial evidence through repeated experiments and feature sensitivity analyses. However, these evaluations are largely qualitative and limited to metadata, static, and hybrid-based methods, without reporting standardized measures required for rigorous stability assessment.

In addition to identifying discriminative features for efficient malicious package detection and ensuring model stability, explainability is also essential. It helps reveal why detection models classify specific behaviors as malicious, thereby improving transparency and interpretability. While DL models offer strong predictive capability~\cite{lecun2015dl, markus2019dl, goodfellow2016dl}, their black-box nature limits their applicability in security-critical settings, where transparent and interpretable decisions are essential. However, explainability analysis of malicious package detection models remains insufficiently investigated in the literature~\cite{ladisa2023crosslang, mehedi2025dysec, gao2024pyradar, sun2024ea4mp, yan2026pypimaldet}. A few studies~\cite{gao2025malguard, iqbal2026clampd} partially address this issue by incorporating LIME-based explanations or using multimodal feature representations; however, their analyses remain limited and do not extend to recent dynamic behavioral data. In sum, the evolving complexity of dynamic behavioral data, combined with the lack of rigorous feature analysis, stability assessment, and transparent interpretation of model decisions, creates a dangerous illusion of security and highlights the need for a more comprehensive analytical framework.

\begin{figure}[htpb]
    \centering
    \includegraphics[width=0.88\linewidth]{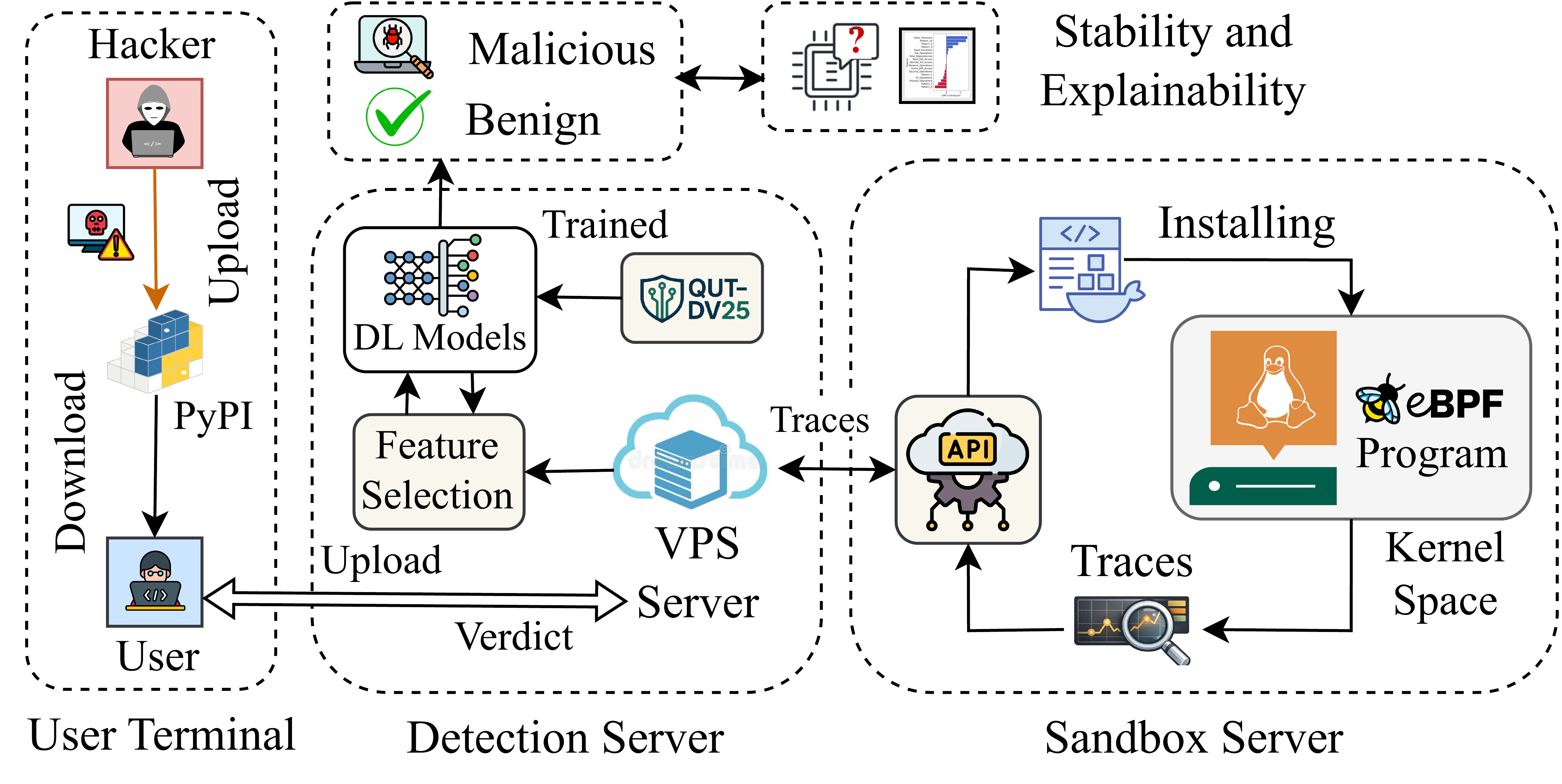}
    \caption{Overall system architecture of the proposed eDySec framework for detecting malicious packages.}
    \Description{}
    \label{fig:motivation}
\end{figure}

To address these challenges, this study presents eDySec, an efficient, stable, and explainable DL-based \textbf{e}xtended \textbf{Dy}namic \textbf{Sec}urity analysis framework for detecting malicious packages. Figure~\ref{fig:motivation} outlines the overall system architecture of the proposed framework. Specifically, it consists of three interacting components: the user terminal, the detection server, and the sandbox server. A user-uploaded package is first received by the virtual private server (VPS)-based detection server and then forwarded to the sandbox server, where an extended Berkeley Packet Filter (eBPF) kernel program collects install-time and post-installation behavioral traces and returns them to the detection server via an API. These traces are then processed using feature selection methods and DL models to determine whether the package is benign or malicious. To identify the most effective detection models, eDySec investigates classical, pre-attention, and attention-based DL architectures. It also incorporates statistical, FLAML, and probabilistic feature selection methods to identify discriminative behavioral features derived from raw traces, including system calls, network traffic, directory access patterns, and dependency logs. In addition, model stability analysis and explainable Artificial Intelligence (XAI) techniques are integrated into the detection pipeline to support stable and transparent interpretation of model decisions. To ensure reproducibility and facilitate further research, the source code and execution instructions are publicly available~\footnote{eDySec GitHub: \url{https://github.com/tanzirmehedi/eDySec}}. The key contributions of this study are as follows:

\begin{itemize}

    \item It proposes the  eDySec framework that includes DL models and feature selection mechanisms for detecting software supply chain attacks and assesses their effectiveness and limitations using a dynamic analysis dataset.

    \item The framework integrates model stability and explainable AI techniques into the detection pipeline, enabling near-perfect stable and transparent interpretation of model decisions by identifying the features that most influence each prediction.
    
    \item Experimental results demonstrate that eDySec significantly outperforms the state-of-the-art framework. Specifically, it reduces feature dimensionality by 52.78\% while lowering false positives by 82\% and false negatives by 79\%. It also improves accuracy by 3\% and maintains a low inference latency of 170ms per package.
    
\end{itemize}

This study is structured as follows: Section~\ref{label:related work} reviews existing methods for malicious package detection. Section~\ref{label:framework} presents the eDySec framework, and Section~\ref{label:design} describes dataset, experimental design, and implementation details. Section~\ref{label:evaluation} reports the results and their implications, while Section~\ref{label:limitation} discusses threats to validity and limitations. Finally, Section~\ref{label:conclusion} concludes the paper.

\section{Related Work}
\label{label:related work}

This section provides an overview of recent studies on software supply chain attack detection, along with the scope and limitations of existing datasets. It also assesses the impact of detection model stability and explainability on dynamic behavioral data, before introducing the proposed framework.

\subsection{ML and DL-based Detectors}

Recent studies on software supply chain attack detection have been dominated by ML and DL-based methods, with most methods focusing on PyPI and npm ecosystems~\cite{zheng2024oscar, zhang2025cerebro, iqbal2026clampd, guo2026pyguard}. Early studies primarily rely on metadata and static analysis. For instance, MeMPtec~\cite{halder2024memptec} demonstrates that metadata-based methods can achieve high accuracy using features with varying levels of manipulability, while static-based methods such as Ladisa et al.~\cite{ladisa2023crosslang}, PyGuardEX~\cite{gandhi2024pyguardex}, Samaana et al.~\cite{samaana2025mlpypi}, and MalGuard~\cite{gao2025malguard} apply statistical features, source code characteristics, structural representations, and graph-based features to detect malicious packages. Additionally, static methods such as Ea4mp~\cite{sun2024ea4mp} and Cerebro~\cite{zhang2025cerebro} employ transformer-based and sequence modeling techniques to capture code semantics. Although these methods achieve strong performance, their dependence on metadata and static code-level features prevents them from detecting next-gen supply chain attacks that remain dormant until install-time and post-installation phases~\cite{ohm2020backstabber, guo2023pypi, mehedi2025dysec}.

Recent studies have shifted toward hybrid analysis to address these limitations. Methods such as PyRadar~\cite{gao2024pyradar}, DONAPI~\cite{huang2024donapi}, PypiGuard~\cite{iqbal2025pypiguard}, and PyPIMalDet~\cite{yan2026pypimaldet} combine metadata and static features using ML-based and meta-learning frameworks. CLAMPD-Net~\cite{iqbal2026clampd} further extends this paradigm through multimodal DL fusion by integrating metadata, graph embeddings, and semantic representations, while PYGUARD~\cite{guo2026pyguard} utilize Retrieval-Augmented Generation (RAG)-based Large Language Models (LLMs) for semantic reasoning. However, despite aggregating diverse feature sources, these methods still rely primarily on metadata and static information, limiting their ability to capture critical behaviors that emerge during the install-time and post-installation phases, such as multiphase execution, remote access activation, conditional triggers, and dynamic payload generation~\cite{mehedi2025dysec, zheng2024oscar, mehedi2025qutdv25}. As a result, they remain vulnerable to sophisticated dormant attacks that activate only under specific execution conditions.

Dynamic analysis has therefore emerged as a more promising yet less explored direction, as it provides stronger behavioral visibility by capturing install-time and post-installation activities. For example, OSCAR~\cite{zheng2024oscar} incorporates dynamic behaviors through rule-based analysis and reports an F1-score of 91\%, but its reliance on predefined rules limits adaptability and generalization. DySec~\cite{mehedi2025dysec} further extends this direction by using install-time and post-installation kernel and user-level traces with traditional ML methods to detect malicious behavior. Although it achieves 95.99\% overall accuracy using a Random Forest (RF) model, it still yields a relatively high number of false positives and false negatives~\cite{mehedi2025dysec}. This limitation arises because traditional ML models often struggle to effectively capture the high-dimensional, sparse, and heterogeneous nature of dynamic behavioral data, thereby constraining the performance of ML-based detection methods~\cite{lecun2015dl}.

To capture these complex behaviors more effectively, DL models have emerged as a promising alternative~\cite{lecun2015dl, goodfellow2016dl}, given their ability to learn complex, high-dimensional, and sequential patterns~\cite{markus2019dl, goodfellow2016dl, vaswani2017dl_transformer}. Additionally, DL-based static methods such as Ea4mp~\cite{sun2024ea4mp} and Cerebro~\cite{zhang2025cerebro}, multimodal DL frameworks~\cite{iqbal2026clampd}, and LLM-based methods~\cite{guo2026pyguard} enable the extraction of deep semantic and contextual relationships from software artifacts. However, the effectiveness of DL models is fundamentally dependent on the quality and modality of input data~\cite{goodfellow2016dl}.

\begin{table*}[htbp]
\centering
\scriptsize
\renewcommand{\arraystretch}{1.25}
\caption{Comprehensive comparison of existing studies across software ecosystems, analysis types, and key methodological dimensions, including model, performance evaluation, feature representation, stability assessment, and explainability coverage.}
\label{tab:final_comparison}
\resizebox{\textwidth}{!}{
\begin{tabular}{|l|l|l|l|l|l|c|c|}
\hline
\tc{Study} 
& \tc{Ecosystem} 
& \tc{Type} 
& \tc{Model} 
& \tc{Performance} 
& \tc{Feature Analysis} 
& \tc{\makecell{\tc{Detection Model}\\\tc{Stability}}}
& \tc{\makecell{\tc{Explainability}\\\tc{(Per Package)}}} \\
\hline
Ladisa et al. (2023)\cite{ladisa2023crosslang} & PyPI/NPM & Static & ML (XGB) & PyPI: {$\mathbf{\mathcal{A}}$}=97.0\% & Statistical+domain expertise & Partial & $\times$ \\
\hline
PyRadar (2024)\cite{gao2024pyradar} & PyPI & Hybrid & ML (RF) & {$\mathbf{\mathcal{A}}$}=99.5\% & Statistical+feature importance & $\times$ & $\times$ \\
\hline
PyGuardEX (2024)\cite{gandhi2024pyguardex} & PyPI & Static & ML (RF) & {$\mathbf{\mathcal{A}}$}=97.0\% & Statistical+feature importance & $\times$ & $\times$ \\
\hline
MeMPtec (2024)\cite{halder2024memptec} & NPM & Metadata & ML (DRF) & {$\mathbf{\mathcal{A}}$}=99.61\% & ETM and DTM & Partial & $\times$ \\
\hline
OSCAR (2024)\cite{zheng2024oscar} & PyPI/NPM & Dynamic & Rule-based & PyPI: {$\mathbf{\mathcal{F}}$}1=91\% & Behavioral feature & $\times$ & $\times$ \\
\hline
Ea4mp (2024)\cite{sun2024ea4mp} & PyPI & Static & DL + Ensemble & {$\mathbf{\mathcal{F}}$}1=97.6\% & Code behavior sequences & Partial & $\times$ \\
\hline
DONAPI (2024)\cite{huang2024donapi} & NPM & Hybrid & ML (RF) & {$\mathbf{\mathcal{A}}$}=97.0\% & API-sequence + ML & $\times$ & $\times$ \\
\hline
Samaana et al. (2025)\cite{samaana2025mlpypi} & PyPI & Static & ML (Stacking) & {$\mathbf{\mathcal{F}}$}1=94.20\% & Statistical+feature importance & Partial & $\times$ \\
\hline
Cerebro (2025)\cite{zhang2025cerebro} & PyPI/NPM & Static & DL (RoBERTa) & PyPI: {$\mathbf{\mathcal{P}}$}=96.1\% & Sequence modeling & $\times$ & $\times$ \\
\hline
MalGuard (2025)\cite{gao2025malguard} & PyPI & Static & ML (RF) & {$\mathbf{\mathcal{F}}$}1=99.0\% & Graph + ML & $\times$ & Partial \\
\hline
PypiGuard (2025)\cite{iqbal2025pypiguard} & PyPI & Hybrid & Meta (Stacking) & {$\mathbf{\mathcal{A}}$}=98.43\% & Statistical + ML-based & $\checkmark$ & $\times$ \\
\hline
PyPIMalDet (2025)\cite{yan2026pypimaldet} & PyPI & Hybrid & ML (Stacking) & {$\mathbf{\mathcal{F}}$}1=97.91\% & DAE + stacking fusion & $\times$ & $\times$ \\
\hline
CLAMPD-Net (2026)\cite{iqbal2026clampd} & PyPI/NPM & Hybrid & DL (CNN-BiGRU) & AVG: {$\mathbf{\mathcal{A}}$}=97.23\% & Fusion + graph + semantic & Partial & Partial \\
\hline
PYGUARD (2026)\cite{guo2026pyguard} & PyPI/NPM & Hybrid & RAG-based LLM & PyPI: {$\mathbf{\mathcal{A}}$}=97.37\% & Semantic + taxonomy & $\times$ & $\times$ \\
\hline
DySec (2026)\cite{mehedi2025dysec} & PyPI & Dynamic & ML (RF) & {$\mathbf{\mathcal{A}}$}=95.99\% & Statistical + ML & $\times$ & $\times$ \\
\hline

\best{eDySec (Proposed)} & \best{PyPI} & \best{Dynamic} & \best{DL (MLP)} & \best{{$\mathbf{\mathcal{A}}$}=99.00\%} & \best{Statistical+ FLAML+ Probability} & \best{$\checkmark$} & \best{$\checkmark$} \\
\hline

\end{tabular}}
\end{table*}

\subsection{Supply Chain Security Datasets}

The effectiveness of detection models is heavily constrained by the quality and modality of available datasets\cite{goodfellow2016dl}. In this regard, existing datasets for malicious package detection can be broadly categorized into metadata-based, static, hybrid, and dynamic datasets. Metadata-based datasets, such as MalwareBench~\cite{zahan2024malwarebench}, PyRadar~\cite{gao2024pyradar}, BadSnakes~\cite{vu2023badsnakes}, and GuardDog~\cite{guarddog2023}, primarily rely on repository metadata, and heuristic indicators. While these datasets are large-scale, they lack behavioral depth and are susceptible to adversarial manipulation. Static and structural datasets, including Guo et al.~\cite{guo2023pypi}, PackageIntel~\cite{guo2024packageintel}, and Backstabber’s Knife~\cite{ohm2020backstabber}, incorporate AST features, API usage patterns, and structural representations, improving code-level understanding but still failing to capture next-gen attacks behavior. Hybrid datasets, such as PypiGuard~\cite{iqbal2025pypiguard}, DONAPI~\cite{huang2024donapi}, and OSS-MPAW~\cite{ossmpaw2024}, combine metadata, static, and limited behavioral traces. CLAMPD-1905~\cite{iqbal2026clampd} further extends this paradigm by integrating multimodal representations, including graph embeddings and semantic APIs, enabling richer feature spaces for DL models; however, these datasets still rely on abstracted representations rather than system-level behaviors.

Among these categories, dynamic datasets remain scarce but are essential for detecting next-gen real-world attacks. QUT-DV25~\cite{mehedi2025qutdv25} represents a significant advancement by capturing install-time and post-installation behaviors using eBPF-based monitoring. It includes 14,271 PyPI packages (7,127 malicious) and 36 system-level behavioral features, such as system calls, file operations, and network activity. Unlike datasets, including CLAMPD-1905~\cite{iqbal2026clampd}, PackageIntel~\cite{guo2024packageintel}, and DONAPI~\cite{huang2024donapi}, QUT-DV25 provides direct install and post-install time evidence, enabling the detection of multiphase malware execution, remote access activation, conditional triggers, and dynamic payload generation. This makes it particularly suitable for DL-based behavioral modeling.

\subsection{Stability, and Explainability Analysis}

Stability analysis is essential in rigorous DL evaluation frameworks for assessing model stability and consistency~\cite{demsar2006stability, benavoli2017stability, dror2018stability, garcia2010stability}. However, it remains largely overlooked and inconsistently evaluated in existing studies. Most studies, including PyRadar~\cite{gao2024pyradar}, PyGuardEX~\cite{gandhi2024pyguardex}, DONAPI~\cite{huang2024donapi}, Cerebro~\cite{zhang2025cerebro}, PyPIMalDet~\cite{yan2026pypimaldet}, and PYGUARD~\cite{guo2026pyguard}, focus primarily on predictive performance without incorporating explicit stability analysis. A subset of studies, such as Ladisa et al.~\cite{ladisa2023crosslang}, MeMPtec~\cite{halder2024memptec}, Ea4mp~\cite{sun2024ea4mp}, Samaana et al.~\cite{samaana2025mlpypi}, and CLAMPD-Net~\cite{iqbal2026clampd}, provide partial evidence of stability through repeated experiments and feature sensitivity analyses; however, these evaluations remain qualitative. They do not report standardized metrics such as standard deviation, average rank, or confidence intervals, which are commonly used for quantitative stability assessment~\cite{demsar2006stability, benavoli2017stability, dror2018stability, garcia2010stability}. Among all studies, only PypiGuard~\cite{iqbal2025pypiguard} explicitly evaluates detection model stability using quantitative metrics, demonstrating consistent performance across runs with low variance (e.g., $\pm$0.06\% accuracy). In contrast, MalGuard~\cite{gao2025malguard} focuses on interpretability and does not include stability analysis. Dynamic methods, including OSCAR~\cite{zheng2024oscar} and DySec~\cite{mehedi2025dysec}, provide implicit stability through real-world analysis, but lack statistical stability evaluation.

In addition to ensuring detection model stability, explainability is essential for understanding why detection models classify specific behaviors as malicious, thereby improving transparency and interpretability. However, explainability remains limited in the current literature. Most studies, including Ladisa et al.~\cite{ladisa2023crosslang}, PyRadar~\cite{gao2024pyradar}, PyGuardEX~\cite{gandhi2024pyguardex}, MeMPtec~\cite{halder2024memptec}, DONAPI~\cite{huang2024donapi}, Cerebro~\cite{zhang2025cerebro}, and PyPIMalDet~\cite{yan2026pypimaldet}, do not provide per-package explanations for model decisions, limiting transparency and interpretability. A few methods partially address this limitation: MalGuard~\cite{gao2025malguard} incorporates LIME-based explanations, while CLAMPD-Net~\cite{iqbal2026clampd} provides limited interpretability through multimodal feature representations. Dynamic methods, such as OSCAR~\cite{zheng2024oscar} and DySec~\cite{mehedi2025dysec}, enable analysts to inspect install-time and post-install activities; however, they do not offer any explainability techniques such as feature attribution or instance-level explanations. Although model-agnostic techniques such as LIME~\cite{ribeiro2016lime} and SHAP~\cite{lundberg2017shap} provide powerful frameworks for interpreting complex DL models~\cite{harikha2025xai, gaith2023xai, baniecki2024xai}, they are largely absent in existing software security studies.

A comprehensive comparison of existing studies across ecosystem, analysis type, model, performance, feature analysis, stability, and explainability coverage is presented in Table~\ref{tab:final_comparison}. These limitations collectively motivate the proposed eDySec framework, which advances the understanding of the strengths and limitations of dynamic datasets for developing an efficient, stable and explainable DL-based detectors against next-gen software supply chain attacks.

\section{eDySec Framework}
\label{label:framework}

\begin{figure}[htpb]
    \centering  \includegraphics[width=1\linewidth]{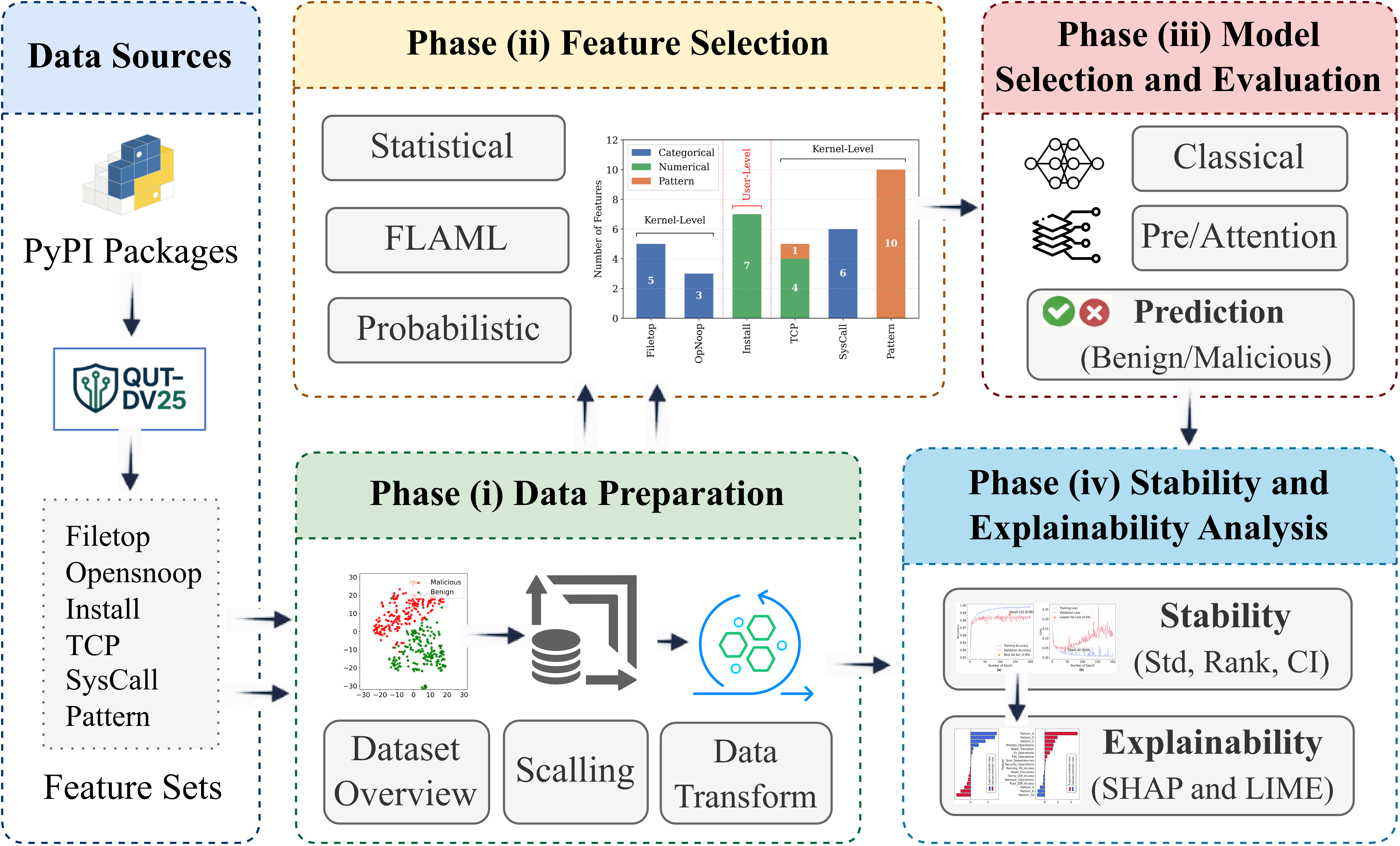}
    \caption{Proposed eDySec framework for detecting malicious PyPI packages.}
    \Description{}
    \label{fig:framework}
\end{figure}

We propose eDySec, an efficient, stable, and explainable DL-based dynamic analysis framework for detecting malicious Python packages. It is designed to address the high-dimensional, sparse, and heterogeneous nature of dynamic behavioral data. As illustrated in Figure~\ref{fig:framework}, eDySec consists of four main phases: (i) data preparation, (ii) feature selection, (iii) model selection and evaluation, and (iv) stability and explainability analysis. The notations used throughout the framework are summarized in \textit{Appendix Table~1}.

\textbf{Data preparation:} Let $\mathcal{D}=\{(\mathbf{x}_i,y_i)\}_{i=1}^{N}$ denote the dataset of $\textit{N}$ package, where $\mathbf{x}_i \in \mathbb{R}^{d}$ is the raw feature representation of the $i$-th package, $\mathbb{R}$ denotes the set of real numbers, and $y_i \in \{0,1\}$ is its class label, with $0$ and $1$ denoting benign and malicious packages, respectively. As shown in Figure~\ref{fig:framework}, the raw inputs are derived from trace-based feature sets, including Filetop, Opensnoop, Install, TCP, SysCall, and Pattern traces. These trace-derived features are organized into three feature types: numerical, categorical, and pattern-based features. The dataset is then partitioned into training, validation, and test sets, denoted by $\mathcal{D}_{tr}$, $\mathcal{D}_{vd}$, and $\mathcal{D}_{te}$, such that $\mathcal{D}_{tr}\cup\mathcal{D}_{vd}\cup\mathcal{D}_{te}=\mathcal{D}$, $\mathcal{D}_{tr}\cap\mathcal{D}_{vd}=\emptyset$, $\mathcal{D}_{tr}\cap\mathcal{D}_{te}=\emptyset$, and $\mathcal{D}_{vd}\cap\mathcal{D}_{te}=\emptyset$.To handle the heterogeneous nature of these features, a preprocessing function $\mathcal{P}(\cdot)$ is applied to obtain processed representations, $(\tilde{\mathcal{D}}_{tr},\tilde{\mathcal{D}}_{vd},\tilde{\mathcal{D}}_{te})=\mathcal{P}(\mathcal{D}_{tr},\mathcal{D}_{vd},\mathcal{D}_{te})$, where each processed set contains transformed samples of the form $(\tilde{\mathbf{x}}_i,y_i)$ or $(\mathbf{s}_i,y_i)$, depending on the model category. This phase includes data cleaning, feature scaling, and representation transformation.

More specifically, as the evaluated DL models span different categories, the selected features are prepared according to the input requirements of each model type rather than through a uniform preprocessing pipeline. For classical DL models, numerical features are standardized using z-score normalization, such that $\hat{x}_{i,j}=(x_{i,j}-\mu_j)/\sigma_j$, where $\mu_j$ and $\sigma_j$ denote the mean and standard deviation of the $j$-th feature estimated from the training set. Categorical and pattern-based features are independently transformed into TF-IDF representations learned from the corresponding training columns. If $\mathcal{T}_k(\cdot)$ denotes the TF-IDF mapping for the $k$-th categorical or pattern-based feature, the resulting text representation is $\mathbf{x}_i^{(\mathrm{text})}=[\mathcal{T}_1(x_{i,1}) \,\|\, \cdots \,\|\, \mathcal{T}_m(x_{i,m})]$, and the final processed sample is formed as $\tilde{\mathbf{x}}_i=[\hat{\mathbf{x}}_i^{(\mathrm{num})} \,\|\, \mathbf{x}_i^{(\mathrm{text})}] \in \mathbb{R}^{\tilde{d}}$. In contrast, for pre-attention and attention-based models, each sample is serialized into a textual sequence by concatenating feature value pairs across all selected columns. The resulting text is then tokenized, mapped into integer sequences using a vocabulary learned from the training set, and padded or truncated to a fixed length, yielding a sequential representation $\mathbf{s}_i \in \textit{N}^{L}$, where $L$ denotes the max sequence length.

\textbf{Feature selection:} Let the processed feature space be denoted by $\mathcal{F}=\{f_1,f_2,\dots,f_{\tilde{d}}\}$. As shown in Phase~(ii) of Figure~\ref{fig:framework}, feature selection is performed using multiple families of methods, including statistical, FLAML-based, and probabilistic selectors. Let $\mathcal{S}=\{S_j\}_{j=1}^{J}$ denote the set of candidate feature selection methods, where each method $S_j$ selects a subset $\mathcal{F}_j \subseteq \mathcal{F}$ according to its method-specific selection criterion, with cardinality $d_j=|\mathcal{F}_j|$. Each subset is evaluated using a baseline model $M^{base}$ on the reduced training and validation sets, producing a validation performance score $P_j$. The feature selection objective is defined as $J_{FS}(S_j)=\alpha P_j + (1-\alpha)\left(1-\frac{d_j}{\tilde{d}}\right)$, where $\alpha \in [0,1]$ controls the trade-off between predictive performance and feature reduction. The optimal feature selection method is then chosen as $S^{*}=\arg\max_{S_j \in \mathcal{S}} J_{FS}(S_j)$, yielding the best feature subset $\mathcal{F}^{*}$. The processed datasets are subsequently projected onto this subset as $\tilde{\mathcal{D}}_{tr}^{*}=\Pi_{\mathcal{F}^{*}}(\tilde{\mathcal{D}}_{tr})$, $\tilde{\mathcal{D}}_{vd}^{*}=\Pi_{\mathcal{F}^{*}}(\tilde{\mathcal{D}}_{vd})$, and $\tilde{\mathcal{D}}_{te}^{*}=\Pi_{\mathcal{F}^{*}}(\tilde{\mathcal{D}}_{te})$. The same feature subset is then used across all selected models, ensuring a common feature basis for fair comparison while preserving each model’s required input representation.

\textbf{Model selection and evaluation:} Using the reduced datasets $(\tilde{\mathcal{D}}_{tr}^{*},\tilde{\mathcal{D}}_{vd}^{*},\tilde{\mathcal{D}}_{te}^{*})$, a set of candidate DL models $\mathcal{M}=\{M_k\}_{k=1}^{K}$, including classical, pre-attention, and attention-based models is evaluated as shown in Phase~(iii) of Figure~\ref{fig:framework}. For each model $M_k$, the parameters $\Theta_k$ are optimized on the training set by minimizing the binary cross-entropy loss. Its validation performance score is denoted by $Q_k$ and computed as $Q_k=\Psi(M_k,\mathcal{D}_{tr}^{*},\mathcal{D}_{vd}^{*})$, where $\Psi(\cdot)$ denotes the evaluation function defined for the model-specific input representation. The optimal model is selected as $M^{*}=\arg\max_{M_k \in \mathcal{M}} Q_k$. Finally, the selected model $M^{*}$ is evaluated on the unseen test set, where predicted labels are obtained by thresholding output probabilities. The final performance is reported using standard classification metrics, including accuracy, precision, recall, and F1-score.

\begin{algorithm}[htpb]
\caption{Pseudo-code of the eDySec workflow}
\label{alg:framework}
\small
\KwIn{Dataset $\mathcal{D}=\{(\mathbf{x}_i,y_i)\}_{i=1}^{N}$, preprocessing function $\mathcal{P}(\cdot)$, feature selection methods $\mathcal{S}=\{S_j\}_{j=1}^{J}$, baseline model $M^{base}$, candidate DL models $\mathcal{M}=\{M_k\}_{k=1}^{K}$, explainability function $\mathcal{E}(\cdot)$}
\KwOut{Best feature subset $\mathcal{F}^{*}$, best model $M^{*}$, stability statistics $\mathcal{T}_{stab}$, explainability ranking $\mathcal{R}_{exp}$, overlap set $\mathcal{C}$}

\BlankLine
\tcp{\footnotesize Phase (i): Data Preparation}

Split $\mathcal{D}$ into $\mathcal{D}_{tr}$, $\mathcal{D}_{vd}$, and $\mathcal{D}_{te}$ \;
Apply preprocessing:
$(\tilde{\mathcal{D}}_{tr},\tilde{\mathcal{D}}_{vd},\tilde{\mathcal{D}}_{te}) \leftarrow \mathcal{P}(\mathcal{D}_{tr},\mathcal{D}_{vd},\mathcal{D}_{te})$ \;
Let $\mathcal{F}=\{f_1,f_2,\dots,f_{\tilde{d}}\}$ denote the processed feature space\;

\BlankLine
\tcp{\footnotesize Phase (ii): Feature Selection}

\ForEach{$S_j \in \mathcal{S}$}{
    Select subset $\mathcal{F}_j \subseteq \mathcal{F}$ according to its method-specific threshold, with $d_j=|\mathcal{F}_j|$\;
    Evaluate $\mathcal{F}_j$ using $M^{base}$ and obtain validation score $P_j$\;
    Compute $J_{FS}(S_j) \leftarrow \alpha P_j + (1-\alpha)\left(1-\frac{d_j}{\tilde{d}}\right)$ \;
}
Select $S^{*} \leftarrow \arg\max_{S_j \in \mathcal{S}} J_{FS}(S_j)$ and obtain $\mathcal{F}^{*}$ \;
Project datasets onto $\mathcal{F}^{*}$:
$\tilde{\mathcal{D}}_{tr}^{*} \leftarrow \Pi_{\mathcal{F}^{*}}(\tilde{\mathcal{D}}_{tr})$,
$\tilde{\mathcal{D}}_{vd}^{*} \leftarrow \Pi_{\mathcal{F}^{*}}(\tilde{\mathcal{D}}_{vd})$,
$\tilde{\mathcal{D}}_{te}^{*} \leftarrow \Pi_{\mathcal{F}^{*}}(\tilde{\mathcal{D}}_{te})$ \;
Construct sequence for pre/attention-based models \;

\BlankLine
\tcp{\footnotesize Phase (iii): Model Selection and Evaluation}

\ForEach{$M_k \in \mathcal{M}$}{
    Train $M_k$ on its corresponding selected training set \;
    Compute validation $Q_k \leftarrow \Psi(M_k,\mathcal{D}_{tr}^{*},\mathcal{D}_{vd}^{*})$ \;
}
Select $M^{*} \leftarrow \arg\max_{M_k \in \mathcal{M}} Q_k$ \;
Evaluate $M^{*}$ on the unseen test set \;

\BlankLine
\tcp{\footnotesize Phase (iv): Stability and Explainability}

Collect scores $\{q_r\}_{r=1}^{R}$ over runs and compute $\mathcal{T}_{stab}$ \;
\ForEach{$\mathbf{z}_i \in \mathcal{D}_{te}^{*}$}{
    Compute attribution vector $\mathbf{e}_i \leftarrow \mathcal{E}(M^{*},\mathbf{z}_i)$ \;
}
Aggregate attributions to obtain $\mathcal{R}_{exp}$ \;
Compute overlap $\mathcal{C} \leftarrow \mathcal{F}^{*}\cap \mathrm{Feat}(\mathcal{R}_{exp})$ \;

\BlankLine
\Return{$\mathcal{F}^{*}, M^{*}, \mathcal{T}_{stab}, \mathcal{R}_{exp}, \mathcal{C}$} \;
\end{algorithm}

\textbf{Stability and explainability analysis:} Beyond predictive performance, the selected model $M^{*}$ is further examined in terms of stability and interpretability. Let $\{q_r\}_{r=1}^{R}$ denote the performance scores obtained over $R$ repeated runs. The resulting stability summary is denoted by $\mathcal{T}_{stab}$ and includes the mean $\bar{q}$, standard deviation $\sigma_q$, confidence intervals, and rank-based comparisons across candidate models. To interpret the predictions of $M^{*}$, an explainability function $\mathcal{E}(\cdot)$ is applied to each test instance, producing a local attribution vector $\mathbf{e}_i=\mathcal{E}(M^{*},\mathbf{z}_i)$, where $\mathbf{z}_i$ denotes the model-specific input representation of the $i$-th test sample. These local attributions are aggregated across the test set to derive global feature importance scores $I(f)$ and the corresponding explainability ranking $\mathcal{R}_{exp}=\mathrm{Rank}(\{I(f): f \in \mathcal{F}^{*}\})$. Finally, the consistency between the selected feature subset and the model's learned behavior is assessed through $\mathcal{C}=\mathcal{F}^{*}\cap \mathrm{Feat}(\mathcal{R}_{exp})$, where a larger overlap indicates stronger agreement between the selected features and the features that most strongly influence model predictions. Algorithm~\ref{alg:framework} presents the pseudo-code for the proposed eDySec framework workflow.

\section{eDySec Design and Setup}
\label{label:design}

This section describes the experimental design of eDySec, including the computational environment, dataset overview, feature engineering and selection strategy, DL model comparison protocol, and the stability and explainability analyses implemented to support stable and interpretable evaluation.

\subsection{Experimental Setup}

The experiments were conducted using both dedicated hardware and a controlled software environment. The hardware platform comprised a 13th Gen Intel Core i9-13900K processor, 128~GB RAM, and an NVIDIA RTX A6000 GPU with 48~GB memory, running on a Ubuntu 22.04 LTS (64-bit) operating system. The software environment was configured with Python 3.10.20, Scikit-learn 1.2.2, and TensorFlow 2.11.0 for data preprocessing, feature selection, model development, training, and evaluation.

\subsection{Dataset Overview}

\begin{figure}[htpb]
    \centering
    \includegraphics[width=1\linewidth]{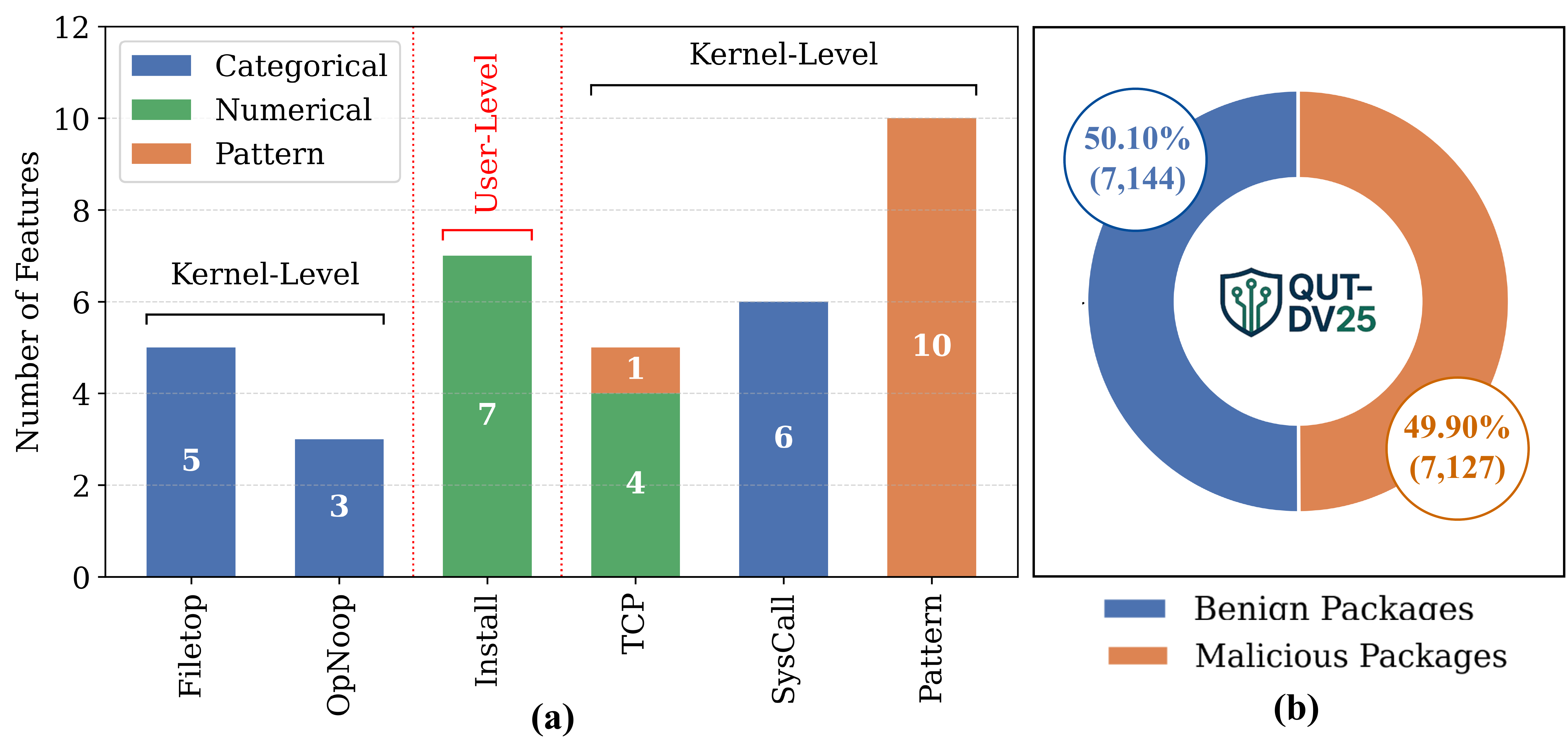}
    \caption{Overview of the QUT-DV25 dataset: (a) statistics of the feature sets across traces; (b) class distribution of benign and malicious packages.}
    \Description{}
    \label{fig:dataset_overview}
\end{figure}

The experiments were performed on the QUT-DV25 dataset, which captures the dynamic behaviors of PyPI packages during both install-time and post-installation execution. As illustrated in Figure~\ref{fig:dataset_overview}(a), the dataset contains 36 features organized into six dynamic trace categories: \textit{Filetop}, which captures file I/O activities that can reveal abnormal file access or missing critical files; \textit{Opensnoop}, which records file open attempts that can indicate access to sensitive or protected directories; \textit{Install}, which captures install-time events that can expose indirect or concealed dependency installation behavior; \textit{TCP}, which represents network activities that help identify communication with suspicious endpoints; \textit{SysCall}, which reflects low-level system interactions that can indicate sabotage, abuse, or privilege misuse; and \textit{Pattern}, which characterizes execution patterns that can reveal repeated loops or payload-triggering behaviors. Collectively, these traces span both user and kernel-level activities, enabling the dataset to capture diverse behaviors relevant to malicious package detection.

\begin{figure}[htpb]
    \centering
    \includegraphics[width=1\linewidth]{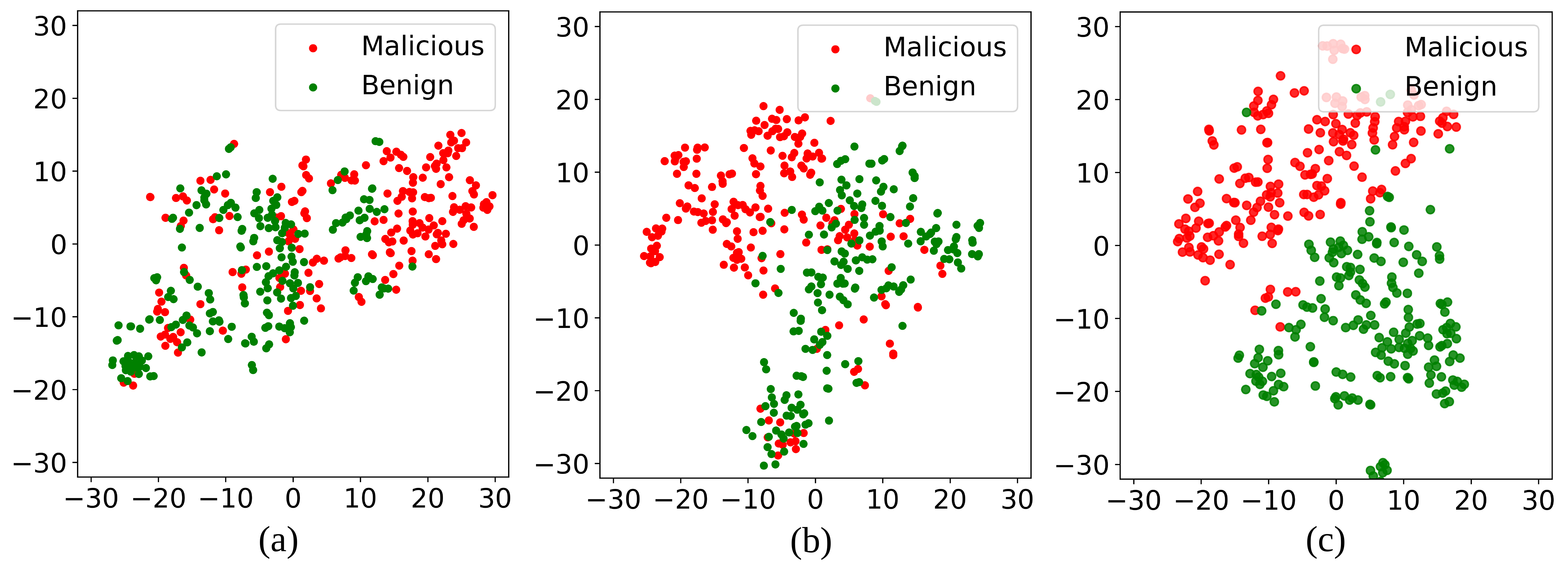}
    \caption{t-SNE visualization of 200 randomly selected samples (100 benign and 100 malicious) using (a) metadata, (b) static, and (c) dynamic features, where the QUT-DV25 dynamic features show clearer class separation.}
    \Description{}
    \label{fig:t-sne}
\end{figure}

As shown in Figure~\ref{fig:dataset_overview}(b), the dataset contains 14,271 samples, including 7,144 benign packages (50.10\%) and 7,127 malicious packages (49.90\%). This near-balanced class distribution supports reliable supervised learning and reduces the risk of class imbalance bias. In addition, the feature space includes numerical, categorical, and pattern-based attributes, resulting in a heterogeneous and behaviorally rich representation for comprehensive analysis. To further explore the discriminative structure of these representations, Figure~\ref{fig:t-sne} presents t-SNE visualizations of 200 randomly selected samples using metadata~\ref{fig:t-sne}(a), static~\ref{fig:t-sne}(b), and dynamic~\ref{fig:t-sne}(c) features from QUT-DV25 dataset. Compared with metadata and static-based representations, the dynamic feature space exhibits clearer separation between benign and malicious packages, suggesting that install and post-install behavioral signals provide stronger discriminatory information for malicious package detection.

\subsection{Feature Engineering and Selection}

The heterogeneous and high-dimensional nature of QUT-DV25 dataset necessitates feature engineering and selection to reduce redundancy while preserving discriminative information for model evaluation. To identify a robust and compact feature subset, we evaluated multiple complementary feature selection strategies on both the combined and individual feature sets. The evaluated methods include statistical, namely Analysis of Variance (ANOVA)~\cite{fisher1925anova_fs} and Correlation (CORR)~\cite{hall1999corr_fs}; an AutoML mechanism, Fast and Lightweight AutoML (FLAML)~\cite{wang2021flaml_fs}; and probabilistic optimization methods, namely Particle Swarm Optimization (PSO)~\cite{kennedy1995pso_fs} and Whale Optimization Algorithm (WOA)~\cite{mirjalili2016woa_fs}. ANOVA selects the most discriminative features based on F-score ranking, CORR preserves features strongly associated with the target while minimizing inter-feature redundancy, and FLAML retains features according to learned relative importance. By contrast, PSO and WOA perform probabilistic subset optimization using sigmoid-based update rules~\cite{kennedy1995pso_fs, mirjalili2016woa_fs}. To reduce heuristic bias and prevent data leakage, all method-specific thresholds and control settings were determined through extensive ablation studies, following the evaluation principle in~\cite{rahman2026evnexttrade}, using only the training and validation sets. These methods were selected because they represent complementary and widely adopted feature selection paradigms, enabling a balanced evaluation of predictive effectiveness, redundancy reduction, and subset compactness in high-dimensional learning settings~\cite{mehedi2025dysec, guyon2003fs, islam2023fs_1}.

\begin{figure}[htpb]
    \centering
    \includegraphics[width=0.95\linewidth]{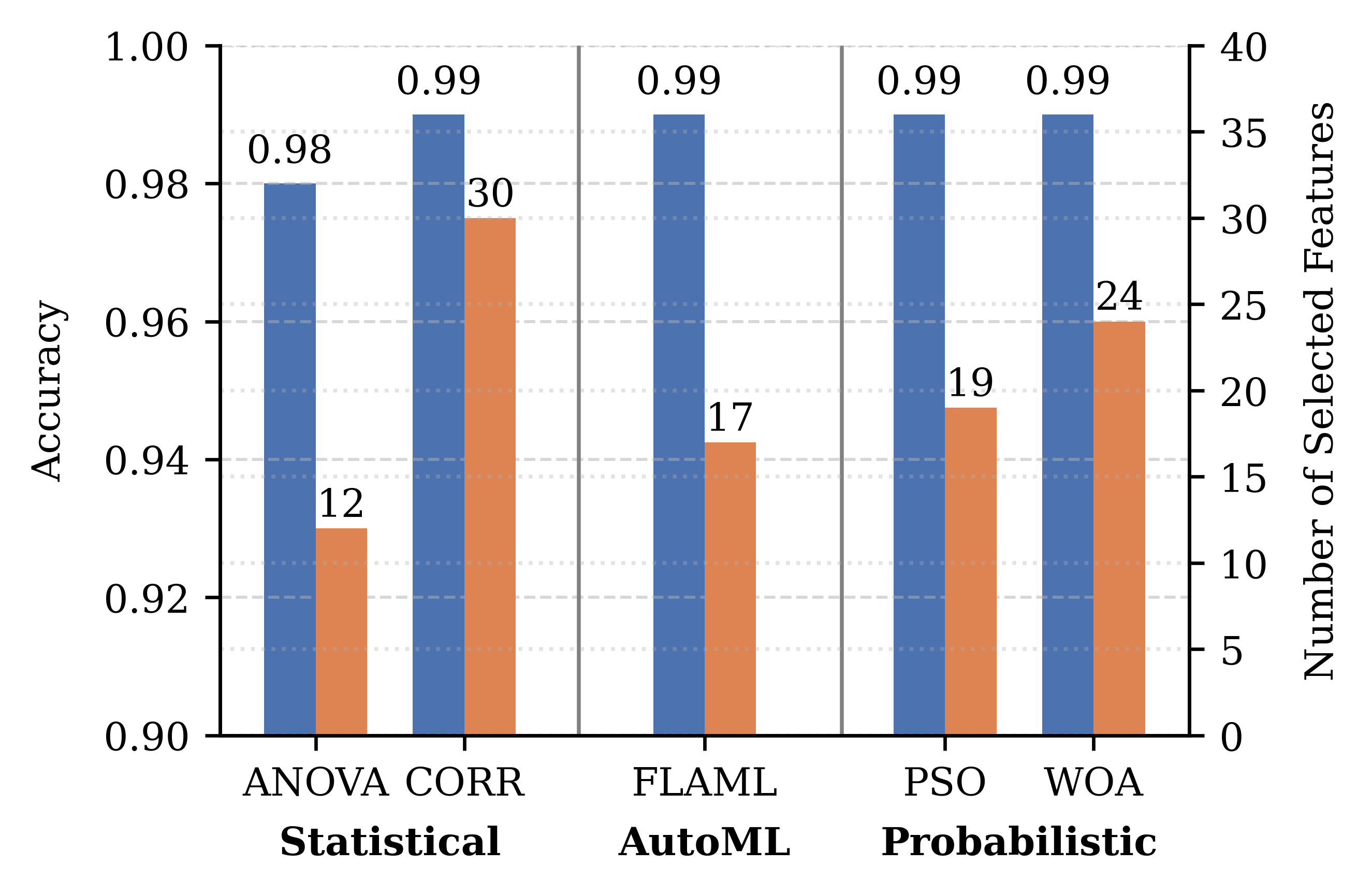}
    \caption{Performance comparison of feature selection methods on the QUT-DV25 dataset using the MLP model.}
    \label{fig:feature_selection}
    \Description{}
\end{figure}

Figure~\ref{fig:feature_selection} summarizes the feature selection results obtained using the MLP classifier on the combined dataset, where both classification accuracy and the number of retained features are considered; the corresponding results for individual trace categories are provided in \textit{Appendix Figure~1}. ANOVA achieves an accuracy of 0.98 using 12 selected features, whereas CORR, FLAML, PSO, and WOA each achieve 0.99 accuracy. However, these methods differ substantially in terms of retained dimensionality: CORR selects 30 features, WOA selects 24, PSO selects 19, and FLAML selects only 17. Among these methods, FLAML provides the most favorable trade-off between predictive performance and feature compactness, matching the highest observed accuracy while using the fewest features among the top-performing methods. Compared with DySec, this reduces feature dimensionality by 52.78\%. Therefore, FLAML is selected as the optimal feature selection method, and its 17-feature subset is applied in the subsequent model comparison phase across all DL models under their corresponding input representations.

\subsection{Learning Package Maliciousness}

After determining the optimal feature subset, we evaluated a diverse set of 10 candidate DL models grouped into three model categories: classical models, namely Convolutional Neural Network (CNN), Multilayer Perceptron (MLP), LeCun Network (LeNet), Multi-Dimensional CNN (MDCNN), and Neural Network (NN); pre-attention models, namely Long Short-Term Memory (LSTM) and Recurrent Neural Network (RNN); and attention-based models, namely Transformer, Bidirectional Encoder Representations from Transformers (BERT), and DistilGPT-2. This enables a structured comparison across progressively more expressive models, ranging from conventional feedforward and convolution-based models to sequential and attention-driven models.

The classical models serve as strong baselines for learning discriminative patterns from structured transformed features. Specifically, CNN and LeNet adopt convolution-based representation learning~\cite{lecun1998dl_cnn, tanzir2023dl}, whereas MLP and NN act as fully connected feedforward baselines trained via backpropagation~\cite{rumelhart1986dl_mlp}. In contrast, the pre-attention models, LSTM and RNN, are designed to capture sequential dependencies in serialized inputs~\cite{hochreiter1997dl_lstm, elman1990dl_rnn}. Extending this progression, the attention-based models, namely Transformer, BERT, and DistilGPT-2, implement self-attention and contextual representation learning to capture long-range dependencies within the input sequences~\cite{vaswani2017dl_transformer, devlin2019dl_bert, sanh2019dl_distilbert}. For each candidate model, the parameters were learned using the selected feature subset under the preprocessing configuration appropriate to the model family. The hyperparameter settings for all evaluated models are reported in \textit{Appendix Table~2}. The dataset was split into 70\% training, 15\% validation, and 15\% test sets. Model selection was based on validation performance, and the final selected model was then evaluated on the unseen test set.

\subsection{Stability and Explainability Integration}

eDySec further incorporates both stability and explainability analysis into the evaluation pipeline to strengthen the selected model. Stability analysis is conducted to assess the consistency of model performance across repeated runs. Specifically, stability is summarized using the mean performance, standard deviation, confidence intervals, and rank-based comparisons, which are widely adopted in rigorous DL evaluation frameworks to assess performance consistency~\cite{demsar2006stability, benavoli2017stability, dror2018stability, garcia2010stability}. In addition, multiple explainability techniques are considered to interpret the predictions of the selected model. Among them, SHapley Additive exPlanations (SHAP)~\cite{broeck2022shap} and Local Interpretable Model-agnostic Explanations (LIME)~\cite{damien2020lime} are employed as the primary explanation methods, as both provide effective and widely validated techniques for explaining complex DL models~\cite{harikha2025xai, gaith2023xai, baniecki2024xai}. SHAP provides both local and global interpretability by assigning contribution scores to features based on Shapley values from cooperative game theory, thereby quantifying how each feature influences model predictions across the dataset in a theoretically grounded and consistent manner. In contrast, LIME explains individual predictions by approximating the complex model locally with an interpretable surrogate model, thereby highlighting the most influential features for a specific sample. To provide multi-faceted insights, eDySec further integrates DL Important FeaTures (DeepLIFT), Integrated Gradients (IG), and Explain Like I’m 5 (ELI5)~\cite{alketbi2025xai}. However, SHAP and LIME remain the core of the eDySec due to their model-agnostic nature, strong theoretical foundations, broad applicability across architectures, and widespread adoption in explainable AI research.

\section{eDySec Evaluation}
\label{label:evaluation}

This section provides a comprehensive evaluation of eDySec by exploring the effectiveness of the selected feature sets and DL models, as well as their stability and explainability.

\subsection{Evaluation Metrics}

The evaluation of eDySec was performed using classification and stability measures. Classification performance was assessed using accuracy ($\mathbf{\mathcal{A}}$), precision ($\mathbf{\mathcal{P}}$), recall ($\mathbf{\mathcal{R}}$), and F1-score ($\mathbf{\mathcal{F}_1}$). $\mathbf{\mathcal{A}}$ captures the overall correctness of the detector. $\mathbf{\mathcal{P}}$ reflects its ability to minimize false positives (FP), where benign packages are incorrectly classified as malicious. $\mathbf{\mathcal{R}}$ assesses its ability to reduce false negatives (FN), where malicious packages are incorrectly classified as benign. The $\mathbf{\mathcal{F}_1}$ balances precision and recall, thereby providing a comprehensive measure of detection effectiveness and practical applicability. In addition, stability was assessed using mean performance, which summarizes average effectiveness across repeated runs; standard deviation, which measures variability; confidence intervals, which indicate the reliability of the estimated performance; and rank-based comparisons, which reflect the relative consistency of competing models. Based on these measures, the evaluation was guided by the following research questions (RQs):

\begin{itemize}
    \item RQ1: Which feature sets lead to the best performance?
    \item RQ2: How do DL models compare to ML models?
    \item RQ3: How do DL models ensure stability?
    \item RQ4: Can XAI provide meaningful interpretations? 
\end{itemize}

\subsection{Which feature sets lead to the best performance?}

\begin{figure}[htpb]
    \centering
    \includegraphics[width=1\linewidth]{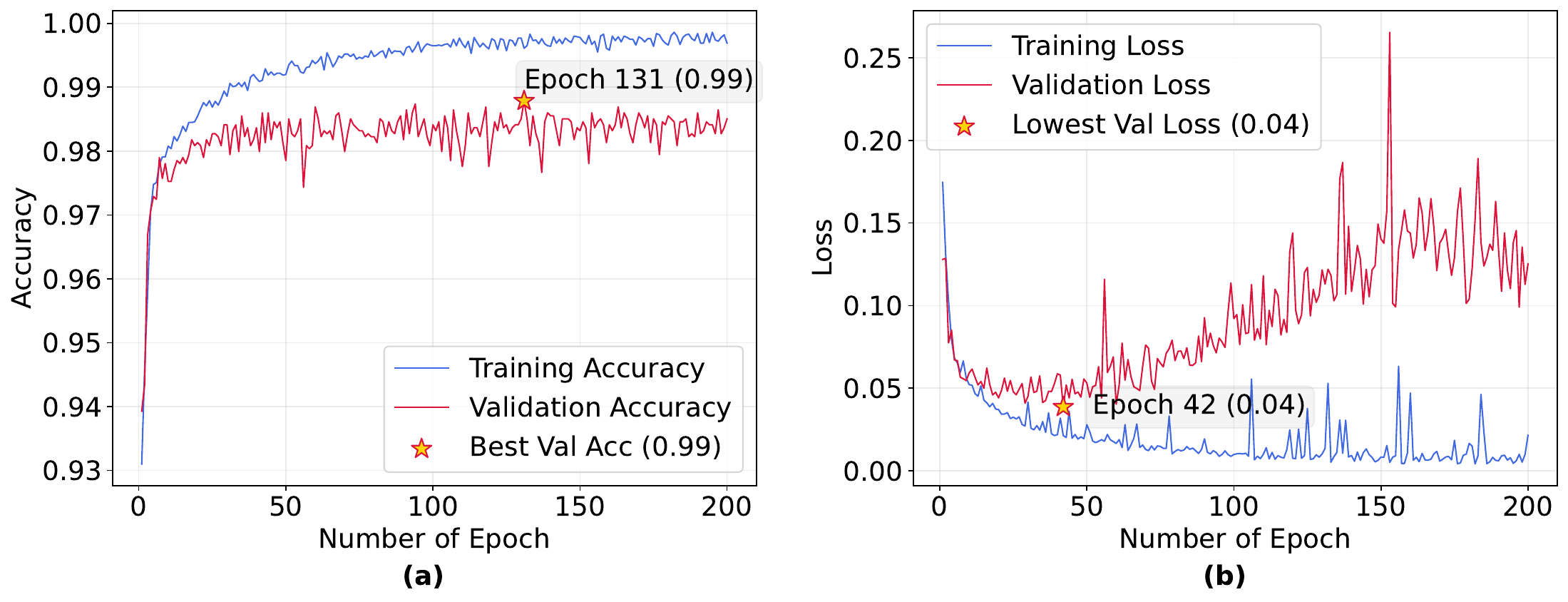}
    \caption{Performance of FLAML-based MLP model on the QUT-DV25 Combined traces dataset: (a) accuracy and (b) loss.}
    \Description{}
    \label{fig:combined_flaml_mlp}
\end{figure}

Identifying the most discriminative feature set is essential for understanding which behaviors contribute most effectively to malicious package detection. As shown in Table~\ref{tab:performance_combined_traces}, the Combined trace feature set delivers the strongest overall performance, achieving 0.99 accuracy, precision, recall, and F1-score under CORR, FLAML, PSO, and WOA, particularly when paired with lightweight classical models such as MLP, LeNet, MDCNN, and NN. Even under ANOVA, the Combined feature set remains highly effective, achieving peak scores of 0.98, which confirms that its superiority is not tied to a single trace, model, or experimental configuration. This advantage is further supported by the learning curves of the overall best-performing model, MLP, in Figure~\ref{fig:combined_flaml_mlp}(a) and Figure~\ref{fig:combined_flaml_mlp}(b), which demonstrate strong optimization behavior in terms of both accuracy and loss. This finding is additionally reinforced by the comparative results in \textit{Appendix Figure~2} and \textit{Appendix Figure~3}, where the FLAML-selected DL models on the Combined trace set exhibit strong convergence, clear class separation, and robust discriminative capability.

\begin{table}[htbp]
\centering
\scriptsize
\setlength{\tabcolsep}{5.5pt}
\renewcommand{\arraystretch}{1.10}
\caption{DL model performance across feature selection techniques on Combined Traces; bold green-shaded values indicate the best results.}
\label{tab:performance_combined_traces}
\begin{tabular}{|c| c| c| l | c c c c|}
\hline
\tc{$\mathbf{\mathcal{F_C}}$} &
\tc{$\mathbf{\mathcal{F}}$} &
\tc{$\mathbf{\mathcal{M_C}}$} &
\tc{$\mathbf{\mathcal{M}}$} &
\tc{$\mathbf{\mathcal{A}}$} &
\tc{$\mathbf{\mathcal{P}}$} &
\tc{$\mathbf{\mathcal{R}}$} &
\tc{$\mathbf{\mathcal{F}_1}$} \\
\hline
\multirow{18}{*}{\rotatebox{90}{Statistical}}
& \multirow{9}{*}{ANOVA}
& \multirow{5}{*}{Classical}
& CNN   & 0.97 & 0.98 & 0.97 & 0.97 \\
&  &  & MLP   & \best{0.98} & \best{0.98} & \best{0.98} & \best{0.98} \\
&  &  & LeNet & 0.97 & 0.98 & 0.98 & 0.97 \\
&  &  & MDCNN & \best{0.98} & \best{0.98} & \best{0.98} & \best{0.98} \\
&  &  & NN    & \best{0.98} & \best{0.98} & \best{0.98} & \best{0.98} \\
\cline{3-8}
&  & \multirow{2}{*}{Pre-Attention}
& LSTM  &  0.94 &  0.94 &  0.94 &  0.94 \\
&  &  & RNN & 0.95 & 0.95 & 0.95 & 0.95 \\
\cline{3-8}
&  & \multirow{3}{*}{Attention}
& Transformer & 0.95  & 0.93  & 0.96  & 0.95  \\
&  &  & BERT  & 0.97 & 0.97 & 0.96 & 0.97 \\
&  &  & DistilGPT2 & 0.97 & 0.98 & 0.96 & 0.97 \\
\cline{2-8}
& \multirow{9}{*}{CORR}
& \multirow{5}{*}{Classical}
& CNN   & 0.98 & 0.98 & 0.98 & 0.98 \\
&  &  & MLP   & \best{0.99} & \best{0.99} & \best{0.99} & \best{0.99} \\
&  &  & LeNet & 0.98 & 0.99 & 0.98 & 0.98 \\
&  &  & MDCNN & \best{0.99} & \best{0.99} & \best{0.99} & \best{0.99} \\
&  &  & NN    & \best{0.99} & \best{0.99} & \best{0.99} & \best{0.99} \\
\cline{3-8}
&  & \multirow{2}{*}{Pre-Attention}
& LSTM  & 0.98 & 0.98 & 0.98 & 0.98 \\
&  &  & RNN   & 0.98  & 0.98  & 0.98 &  0.98 \\
\cline{3-8}
&  & \multirow{3}{*}{Attention}
& Transformer & 0.97 & 0.98 & 0.97 & 0.97 \\
&  &  & BERT  &  0.95 &  0.96 &  0.95 &  0.95 \\
&  &  & DistilGPT2 & 0.96 & 0.96 & 0.96 & 0.96 \\
\hline
\multirow{9}{*}{\rotatebox{90}{AutoML}}
& \multirow{9}{*}{FLAML}
& \multirow{5}{*}{Classical}
& CNN   & 0.98 & 0.97 & 0.98 & 0.98 \\
&  &  & MLP   & \best{0.99} & \best{0.99} & \best{0.98} & \best{0.99} \\
&  &  & LeNet & \best{0.99} & \best{0.99} & \best{0.99} & \best{0.99} \\
&  &  & MDCNN & \best{0.99} & \best{0.98} & \best{0.99} & \best{0.99} \\
&  &  & NN    & \best{0.99} & \best{0.99} & \best{0.99} & \best{0.99} \\
\cline{3-8}
&  & \multirow{2}{*}{Pre-Attention}
& LSTM  & 0.98 & 0.97 & 0.99 & 0.98 \\
&  &  & RNN  & 0.97 & 0.97 & 0.97 & 0.97 \\
\cline{3-8}
&  & \multirow{3}{*}{Attention}
& Transformer & 0.97  & 0.97 & 0.98 & 0.97 \\
&  &  & BERT  &  0.97 &  0.96 &  0.98 &  0.97 \\
&  &  & DistilGPT2 & 0.98 & 0.98 & 0.98 & 0.98 \\
\hline
\multirow{18}{*}{\rotatebox{90}{Probabilistic}}
& \multirow{9}{*}{PSO}
& \multirow{5}{*}{Classical}
& CNN   & 0.98 & 0.98 & 0.98 & 0.98 \\
&  &  & MLP   & \best{0.99} & \best{0.99} & \best{0.99} & \best{0.99} \\
&  &  & LeNet & 0.98 & 0.99 & 0.99 & 0.98 \\
&  &  & MDCNN & \best{0.99} & \best{0.99} & \best{0.99} & \best{0.99} \\
&  &  & NN    & \best{0.99} & \best{0.99} & \best{0.99} & \best{0.99} \\
\cline{3-8}
&  & \multirow{2}{*}{Pre-Attention}
& LSTM  & 0.97 & 0.97 & 0.97 & 0.97 \\
&  &  & RNN   &   0.95 &  0.95  &  0.95 &  0.95 \\
\cline{3-8}
&  & \multirow{3}{*}{Attention}
& Transformer & 0.98 & 0.98 & 0.98 & 0.98 \\
&  &  & BERT  & 0.97 & 0.97 & 0.97 & 0.97 \\
&  &  & DistilGPT2 & 0.97 & 0.97 & 0.97 & 0.97 \\
\cline{2-8}
& \multirow{9}{*}{WOA}
& \multirow{5}{*}{Classical}
& CNN   & 0.98 & 0.98 & 0.98 & 0.98 \\
&  &  & MLP   & \best{0.99} & \best{0.99} & \best{0.99} & \best{0.99} \\
&  &  & LeNet & 0.98 & 0.98 & 0.98 & 0.98 \\
&  &  & MDCNN & \best{0.99} & \best{0.99} & \best{0.99} & \best{0.99} \\
&  &  & NN    & \best{0.99} & \best{0.99} & \best{0.99} & \best{0.99} \\
\cline{3-8}
&  & \multirow{2}{*}{Pre-Attention}
& LSTM  & 0.98  & 0.98 & 0.98 & 0.98 \\
&  &  & RNN   &  0.98 & 0.97  &  0.98 &  0.98 \\
\cline{3-8}
&  & \multirow{3}{*}{Attention}
& Transformer & 0.98 & 0.98 & 0.98 & 0.98  \\
&  &  & BERT  & 0.98 & 0.98 & 0.98 & 0.98 \\
&  &  & DistilGPT2 &  0.98 & 0.98 & 0.98 & 0.98 \\
\hline
\end{tabular}
\end{table}

\begin{table*}[htpb]
\centering
\scriptsize
\caption{Performance comparison between DySec models and the proposed eDySec models on the QUT-DV25 dataset; bold and green-shaded values indicate the best performance.}
\renewcommand{\arraystretch}{1.0}
\label{tab:performance_comparison}
\resizebox{\textwidth}{!}{
\begin{tabular}{|c|c|c|c|c|c|c|c|c|c|c|c|c|}
\hline
\tc{\multirow{2}{*}{Framework}} & 
\tc{\multirow{2}{*}{Methods}} &
\tc{\multirow{2}{*}{Category}} & 
\tc{\multirow{2}{*}{Model}} & 
\tc{\multirow{2}{*}{Features}} & 
\tc{\multirow{2}{*}{Test AUC}} & 
\tc{\multirow{2}{*}{F1 Score}} & 
\tc{\multirow{2}{*}{FPR(\%)}} & 
\tc{\multirow{2}{*}{FNR(\%)}} & 
\multicolumn{4}{c|}{\tc{Confusion Matrix}} \\ 
\cline{10-13}
 &  &  &  &  &  &  &  &  & 
\tc{TP} ↑ & \tc{TN} ↑ & \tc{FP} ↓ & \tc{FN} ↓ \\ 
\hline
\multirow{4}{*}{\makecell{DySec~\cite{mehedi2025dysec}}}
 & \multirow{4}{*}{\makecell{ML\\Models}}
 & \multirow{2}{*}{Tree} & RF & \multirow{2}{*}{36} & 0.96 & 0.96 & 3.23 & 4.77 & 1038 & 1017 & 34 & 52\\
 &  &  & DT &  & 0.94 & 0.94 & 1.74 & 9.52 & 1055 & 958 & 17 & 111  \\
\cline{3-13}
 &  & Kernel & SVM & 36 & 0.95 & 0.95 & 5.76 & 3.63 & 1009 & 1031 & 63 & 38 \\
\cline{3-13}
 &  & Ensemble & GB & 36 & 0.94 & 0.94 & 1.84 & 9.29 & 1054 & 961 & 18 & 108 \\ 
\hline
\multirow{10}{*}{\makecell{eDySec \\ {[Proposed]}}}
 & \multirow{10}{*}{\makecell{DL\\Models}}
 & \multirow{5}{*}{Classic} & CNN & \multirow{5}{*}{17} & 0.98 & 0.98 &  2.41 & 1.32 & 1046 & 1055 & 26 & 14 \\
 &  &  & \best{MLP} &  & \best{0.99} & \best{0.99} & \best{0.56} & 1.02 & \best{1066} & 1058 & \best{6} & 11 \\
 &  &  & \best{LeNet} &  & \best{0.99} & \best{0.99} & 0.93 & \best{0.75} & 1062 & \best{1061} & 10 & \best{8} \\
 &  &  & \best{MDCNN} &  & \best{0.99} & \best{0.99} & 1.12 & 0.93 & 1060 & 1059 & 12 & 10\\
 &  &  & \best{NN} &  & \best{0.99} & \best{0.99} & \best{0.56} & 0.93 & \best{1066} & 1059 & \best{6} & 10\\
\cline{3-13}
 &  & \multirow{2}{*}{Pre-Attention} & LSTM & \multirow{2}{*}{17} & 0.98 & 0.98 & 2.48 & 0.85 & 1045 & 1060 & 27 & 9 \\
 &  &  & RNN & & 0.97 & 0.97 & 2.52 & 2.42 & 1045 & 1043 & 27 & 26 \\
\cline{3-13}
 &  & \multirow{3}{*}{Attention} & Transformer & \multirow{3}{*}{17} & 0.97 & 0.97 & 2.78 & 1.79 & 1042 & 1050 & 30 & 19 \\
 &  &  & BERT &  & 0.97 & 0.97 & 3.22 & 1.61 & 1037 & 1052 & 35 &  17\\
 &  &  & DistilGPT2 &  & 0.98 & 0.98 & 1.86 & 1.41 & 1052 & 1054 & 20 & 15 \\
\hline
\end{tabular}}
\end{table*}

\begin{figure}[htpb]
    \centering
    \includegraphics[width=1\linewidth]{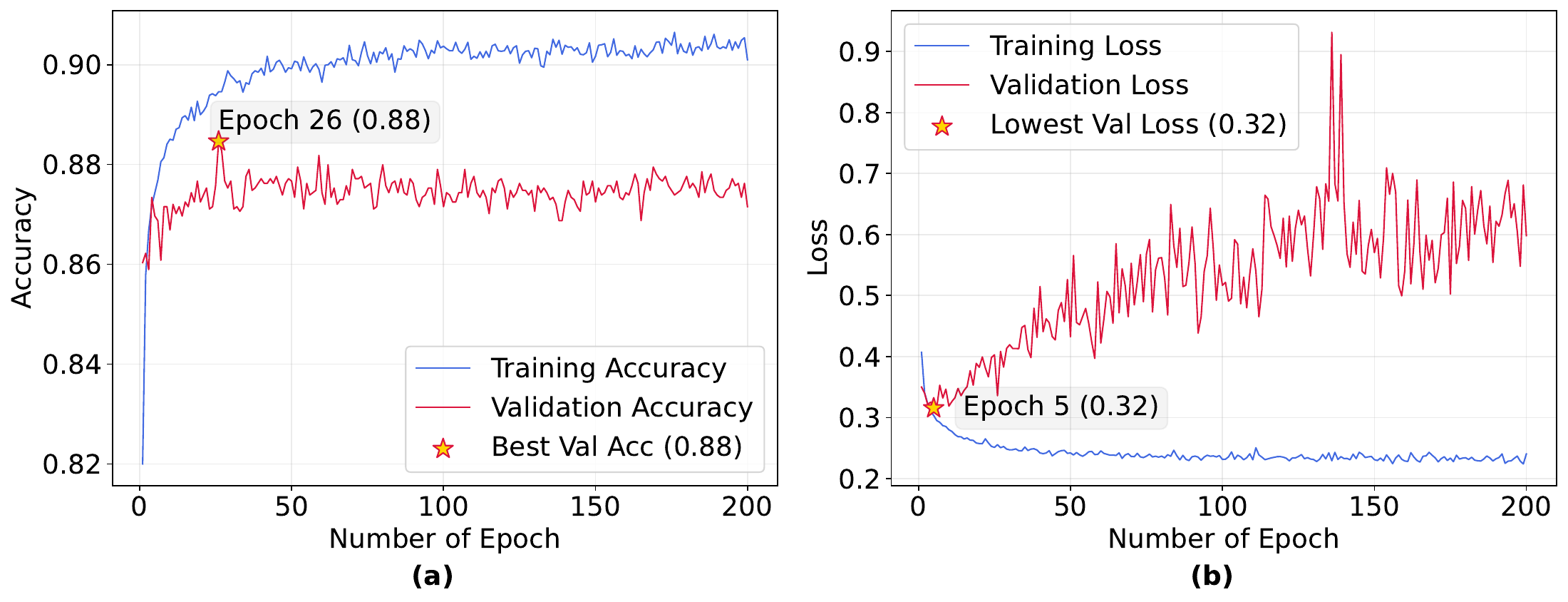}
    \caption{Performance of FLAML-based MLP model on the QUT-DV25 Pattern traces dataset: (a) accuracy and (b) loss.}
    \Description{}
    \label{fig:pattern_flaml_mlp}
\end{figure}

A comparison with the individual trace sets further reinforces this conclusion. As summarized in \textit{Appendix Tables~3--9}, no single trace matches the performance of the Combined representation. Among the individual traces, Filetop and SysCall are the strongest, each reaching approximately 0.94 F1-score at their best while also maintaining comparatively low false-positive and false-negative rates, as reported in \textit{Appendix Table~9}. This indicates that low-level system activity and fine-grained file interaction behaviors provide highly informative signals for malicious package detection. Opensnoop forms a second tier, with peak performance around 0.91 F1-score, whereas Pattern and TCP achieve best results of roughly 0.88 F1-score, suggesting that they provide useful but less discriminative evidence when used in isolation. The training behavior of a single-trace setting is further illustrated by Figure~\ref{fig:pattern_flaml_mlp}(a) and Figure~\ref{fig:pattern_flaml_mlp}(b), which present the corresponding accuracy and loss curves for the Pattern trace under the FLAML-MLP configuration. By contrast, Install is clearly the weakest feature set, with performance generally capped around 0.80 F1-score despite occasionally achieving very high recall, indicating that only install trace behaviors are insufficiently distinctive and tend to produce less balanced detection outcomes. This observation is further supported by \textit{Appendix Figure~4} and \textit{Appendix Figure~5}, which show that even the FLAML-selected DL models on a single-trace settings remain inferior to the Combined representation. Overall, these results show that the best individual feature sets are Filetop and SysCall, whereas the Combined trace with 17 features remains decisively superior, because it integrates complementary behavioral evidence across multiple views and therefore provides the most discriminative and stable basis for malicious package detection.

\subsection{How do DL models compare to ML models?}

Table~\ref{tab:performance_comparison} presents a quantitative comparison between the DySec ML baselines and the proposed eDySec DL models on the same dynamic behavior dataset. The results show that eDySec consistently surpasses the DySec models while operating with a substantially smaller feature set. Specifically, DySec relies on 36 features, whereas eDySec achieves superior results using only 17 selected features, corresponding to a 52.78\% reduction in dimensionality. Among the DySec baselines, RF delivers the strongest performance, reaching an accuracy and F1-score of 0.96, but still producing 34 false positives and 52 false negatives. In contrast, the best eDySec model, MLP, achieves an accuracy and F1-score of 0.99, while reducing false positives and false negatives by 82.35\% and 78.85\%, respectively. These results indicate that eDySec provides not only higher predictive accuracy but also stronger error minimization.

This advantage is not confined to a single model, but is consistently observed across the classical DL family. MLP, LeNet, MDCNN, and NN all achieve an F1-score of 0.99, while also leading substantially lower FPR and FNR than the DySec baselines. Relative to RF, these models reduce false positives from 34 to as low as 6 and false negatives from 52 to as low as 11, demonstrating a substantially better trade-off between attack sensitivity and false-alarm control. This pattern is particularly important in malicious package detection, where false negatives allow threats to evade detection and false positives undermine the practical usability of the detector.

Beyond quantitative performance, these findings suggest that the advantage of eDySec arises not only from reduced feature dimensionality, but also from its ability to capture complex interactions in high-dimensional, sparse dynamic data. Specifically, the DL-based formulation appears to extract richer predictive structure from fewer features, indicating that dynamic behavioral signals are more effectively captured through learned nonlinear representations than through conventional ML decision boundaries. The results further suggest that increased architectural complexity does not necessarily translate into superior detection performance. Although pre-attention and attention-based models remain competitive, the most consistent gains are achieved by the classical DL models, implying that the selected feature representation is already sufficiently informative for feedforward and convolution-based models.

\subsection{How do DL models ensure stability?}

We evaluated model consistency and overall predictive performance across repeated runs. As summarized in Table~\ref{tab:mean_std_rank}, the classical DL models exhibit most consistent behavior. To be precise, MLP, MDCNN, and NN achieve the highest mean F1-score of 0.988, the lowest standard deviation of 0.004, the best average rank of 2.1, and a stability score of 0.996, which matches the highest observed value. Their 95\% confidence intervals further suggest that this predictive performance is consistently maintained across repeated evaluations rather than arising from a small number of favorable runs.

\begin{table}[htbp]
\centering
\scriptsize
\renewcommand{\arraystretch}{1.05}
\caption{Performance of models across all feature selection methods, bold green-shaded values indicate the best results.}
\label{tab:mean_std_rank}
\resizebox{3.3in}{!}{
\begin{tabular}{|l|c|c|c|c|c|}
\hline
\tc{Model} & \tc{Mean F1} & \tc{Std} & \tc{Avg Rank} & \tc{95\% CI} & \tc{Stability} \\ \hline
MLP   & \best{0.988} & \best{0.004} & \best{2.1} & \best{[0.984, 0.990]} & \best{0.996} \\ \hline
MDCNN & \best{0.988} & \best{0.004} & \best{2.1} & \best{[0.984, 0.990]} & \best{0.996} \\ \hline
NN    & \best{0.988} & \best{0.004} & \best{2.1} & \best{[0.984, 0.990]} & \best{0.996} \\ \hline
LeNet & 0.980 & 0.007 & 5.1 & [0.974, 0.986] & 0.993 \\ \hline
CNN   & 0.978 & 0.004 & 5.8 & [0.974, 0.980] & 0.996 \\ \hline
DistilGPT2  & 0.972 & 0.008 & 7.1 & [0.966, 0.978] & 0.992 \\ \hline
LSTM        & 0.970 & 0.017 & 7.3 & [0.954, 0.980] & 0.983 \\ \hline
Transformer & 0.970 & 0.012 & 7.5 & [0.960, 0.978] & 0.988 \\ \hline
BERT        & 0.968 & 0.011 & 7.9 & [0.958, 0.976] & 0.989 \\ \hline
RNN & 0.966 & 0.015 & 8.0 & [0.954, 0.978] & 0.985 \\
\hline
\end{tabular}}
\end{table}

\begin{figure}[htpb]
    \centering
    \includegraphics[width=1\linewidth]{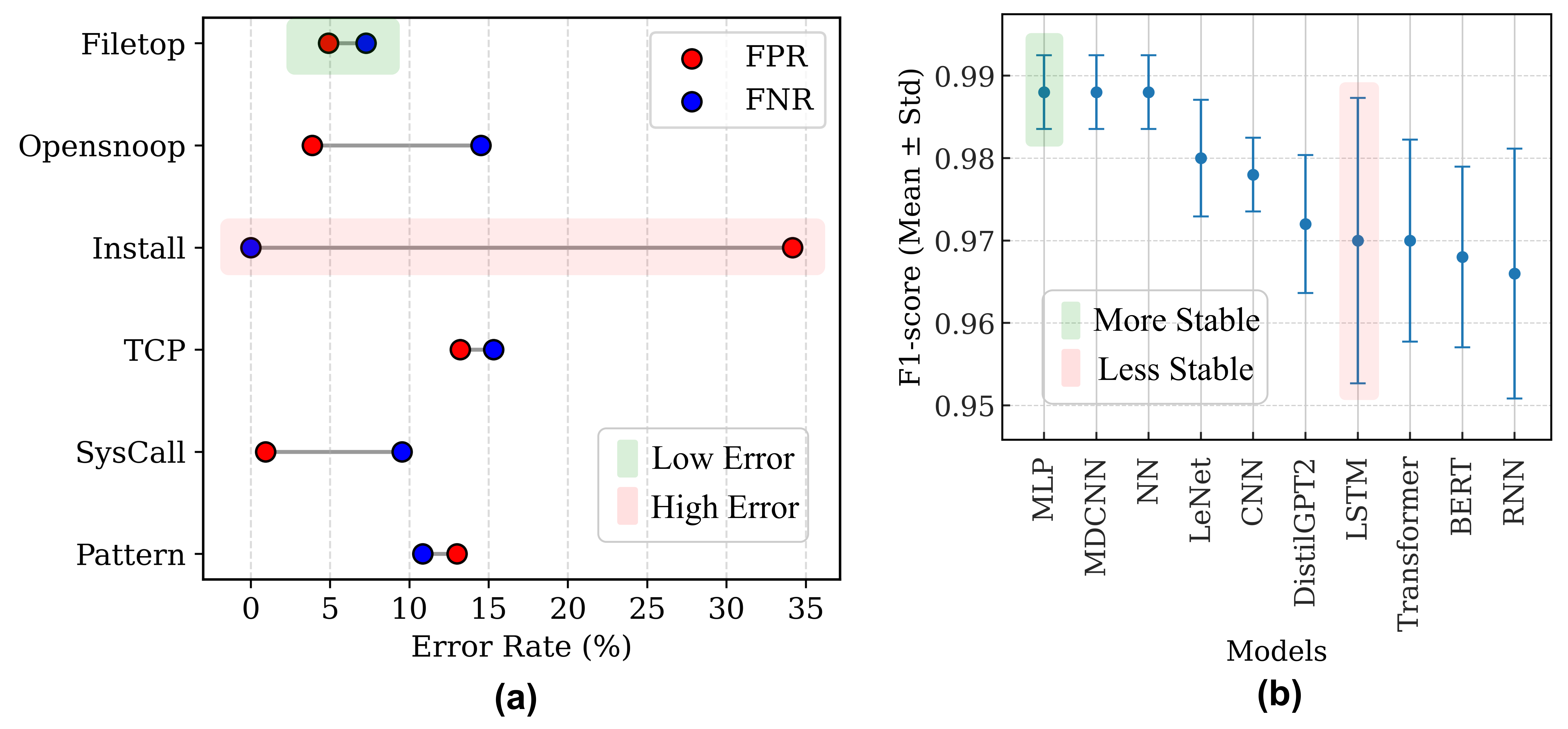}
    \caption{Comparison of (a) FPR and FNR across trace types using the MLP model, and (b) model stability across all trace types measured by F1-score.}
    \Description{}
    \label{fig:fpr_fnr}
\end{figure}

Figure~\ref{fig:fpr_fnr}(b) supports this finding by showing that MLP, MDCNN, and NN combine high mean F1-scores with relatively small uncertainty bands, reflecting a favorable trade-off between effectiveness and consistency. CNN is similarly stable, with a stability score of 0.996 and low dispersion, although its mean F1-score is slightly lower (0.978). By comparison, the pre-attention and attention-based models exhibit greater variability and weaker average performance. LSTM records the largest standard deviation (0.017), indicating the highest sensitivity to training variation, while BERT and RNN produce the lowest mean F1-scores, with RNN ranking last as a whole. These findings suggest that, for the selected dynamic feature representation, classical DL models provide a better balance between predictive accuracy and stability.

We further analyzed error-rate behavior across trace types using the most stable MLP model. As shown in Figure~\ref{fig:fpr_fnr}(a), the combined dataset consists of multiple trace types, each with a different false positive and false negative errors. Filetop achieves the most favorable trade-off between FPR and FNR, while Pattern and SysCall also exhibit comparatively low error rates. By contrast, Install produces the highest FPR despite a relatively low FNR, indicating a stronger tendency to over-flag benign samples when considered in isolation. TCP and Opensnoop demonstrate intermediate behavior. These findings provide a feature-level explanation for the observed stability trends, suggesting that the most stable model benefits from trace families with more balanced error behavior.

Computational efficiency further strengthens this conclusion. Table~\ref{tab:model_structure} shows that eDySec achieves a balance between predictive performance and computational cost compared with the DySec baselines. Here, ER denotes the error rate, that is, the percentage of incorrectly classified samples, where lower values indicate better performance. Within DySec, RF is the strongest baseline, achieving the lowest ER of 4.00\%, with a training time of 523.59s and a testing time of 0.41s. In contrast, the best-performing eDySec model, MLP, reduces the ER substantially to 0.65\%, while requiring 318.59s for training and only 0.17s for testing. This corresponds to a 58.54\% reduction in testing time relative to RF, despite achieving markedly lower classification error. Although some larger models, such as BERT and DistilGPT2, contain 109,580,801 and 82,167,425 parameters, respectively, they do not deliver corresponding gains in performance. Notably, BERT is the least efficient model, exhibiting both the longest training time (17,450.98s) and the slowest testing time (4.61s). In summary, these results indicate that the improved stability of eDySec is achieved more effectively through classical DL models than through increasing architectural complexity alone.

\begin{table}[htpb]
\centering
\scriptsize
\caption{Comparison of parameters, error rate, training, and testing time on the Combined traces; bold green-shaded values indicate the strongest baseline.}
\label{tab:model_structure}
\renewcommand{\arraystretch}{1.05}
\resizebox{3.3in}{!}{
\begin{tabular}{|c|l|r|c|r|c|}
\hline
\tc{$\mathbf{\mathcal{M_C}}$} &
\tc{$\mathbf{\mathcal{M}}$} &
\tc{$\mathbf{Params}$} &
\tc{$\mathbf{ER}$ (\%)} &
\tc{$\mathbf{\mathcal{T}_{train}}$ (s)} &
\tc{$\mathbf{\mathcal{T}_{test}}$ (s)} \\ \hline
\multirow{4}{*}{\makecell{DySec~\cite{mehedi2025dysec}}}
&  \best{RF}  &  \best{--} &  \best{4.00} &  \best{523.59} &  \best{0.41} \\
\cline{2-6}
& DT  & -- & 5.63 & 66.30 & 0.05 \\
\cline{2-6}
& SVM & -- & 4.70 & 16181.15 & 4.41 \\
\cline{2-6}
& GB  & -- & 5.57 & 46.96 & 0.03 \\ 
\hline
\multirow{10}{*}{\makecell{eDySec \\ {[Proposed]}}}
& CNN         & 33,281      & 1.54 & 383.88  & 0.18 \\
\cline{2-6}
& \best{MLP}         & \best{1,273,501}   & \best{0.65} & \best{318.59}  & \best{0.17} \\ 
\cline{2-6}
& LeNet       & 1,921,551   & 0.93 & 446.04  & 0.20 \\ 
\cline{2-6}
& MDCNN       & 2,238,833   & 0.80 & 354.37  & 0.25 \\
\cline{2-6}
& NN          & 109,753     & 0.65 & 330.81  & 0.22 \\ 
\cline{2-6}
& LSTM        & 547,137     & 1.67 & 936.58   & 0.28 \\ 
\cline{2-6}
& RNN         & 169,217     & 2.47 & 7173.79  & 0.38 \\
\cline{2-6}
& Transformer & 452,353     & 2.29 & 1725.78  & 0.40 \\ 
\cline{2-6}
& BERT        & 109,580,801 & 2.42 & 17450.98 & 4.61 \\
\cline{2-6}
& DistilGPT2  & 82,167,425  & 1.64 & 4794.64  & 1.08 \\
\hline
\end{tabular}}
\end{table}

\subsection{Can XAI provide meaningful interpretation?}

\begin{figure}[htpb]
    \centering
    \includegraphics[width=0.90\linewidth]{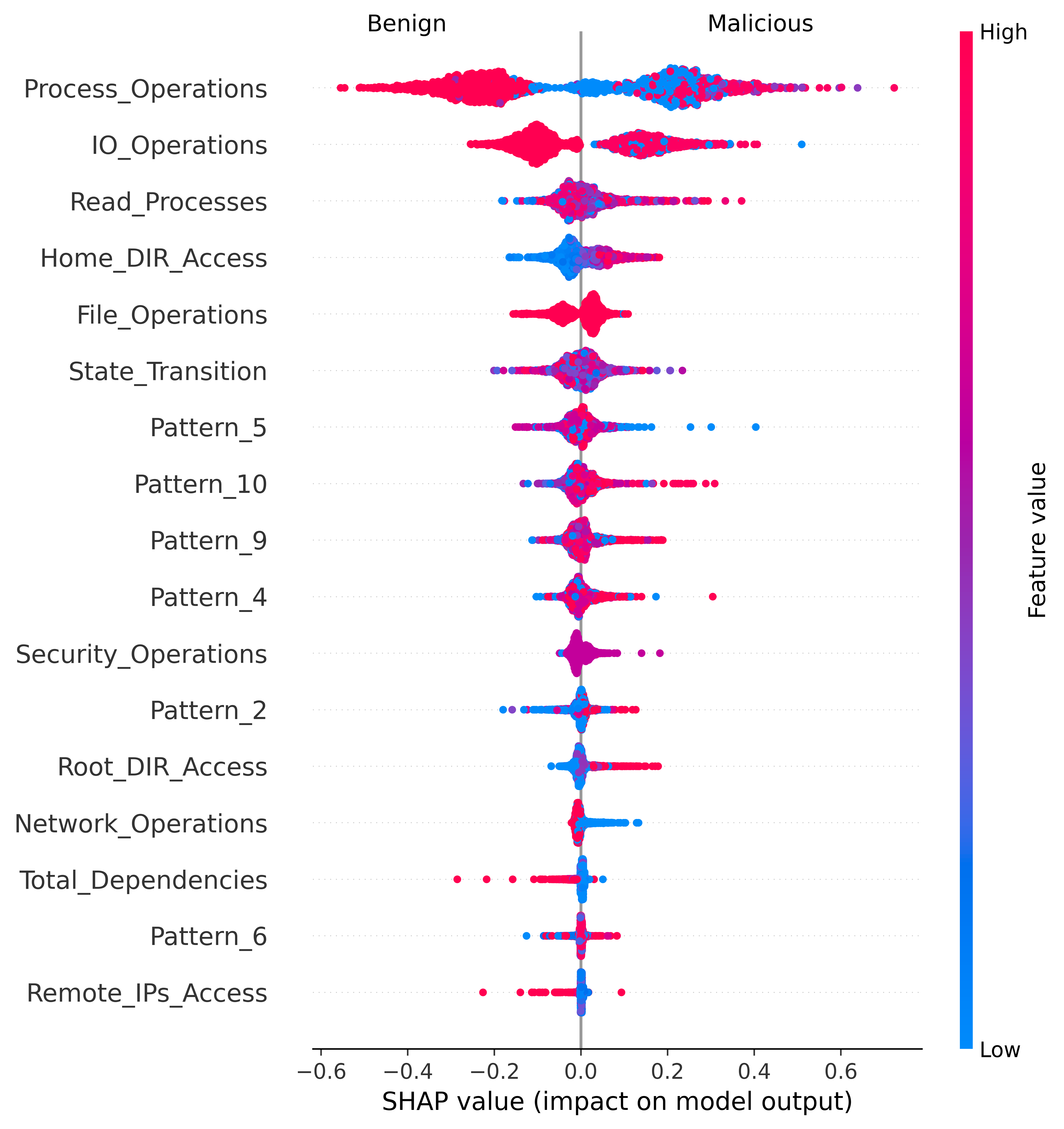}
    \caption{Global SHAP summary of the most influential features for malicious package detection. The distribution of SHAP values shows how each feature contributes to benign (left) or malicious (right) predictions across all samples. Color indicates feature magnitude (blue: low, red: high).}
    \Description{}
    \label{fig:shap_summary_1}
\end{figure}

To assess whether the proposed framework is not only accurate but also interpretable, we further analyze the predictions of the overall best-performing MLP model using explainability techniques. We first apply SHAP to identify globally important features, then apply local SHAP to explain representative instance-level predictions, and finally employ LIME to explore the consistency of these local interpretations. This progression enables us to evaluate whether the detector relies on meaningful behavioral evidence at both the global and sample-specific levels. Figure~\ref{fig:shap_summary_1} presents the global SHAP summary of the most influential features, showing both their overall importance and directional effect on the model output. Among these, \textit{Process\_Operations} has the highest mean absolute SHAP value, indicating that process-level activity during install-time and post-installation is the most influential factor in the model's global decision behavior. This is consistent with the fact that malicious packages often exhibit suspicious behavior, such as process spawning, child-process creation, and multi-stage execution. In contrast, \textit{Remote\_IPs\_Access} has relatively low importance, suggesting that it is a less stable indicator and more dependent on execution context. As a whole, the prominence of features related to I/O activity, process inspection, filesystem access, and state transitions suggests that the detector relies on install-time and post-installation behavioral evidence. The high ranking of several learned pattern features further indicates that the model captures higher-order behavioral regularities in the execution traces.

The SHAP distribution further clarifies how these features influence class decisions. Negative SHAP values push predictions toward the benign class, whereas positive SHAP values push them toward the malicious class. Color encodes the original feature magnitude, with blue denoting low values and red denoting high values. This allows us to assess not only whether a feature is influential, but also whether high or low values tend to support a particular class. Notably, \textit{Process\_Operations} and \textit{IO\_Operations} exhibit broad distributions on both sides of the decision boundary, indicating that their contribution is context dependent rather than uniformly directional. A similar pattern is observed for \textit{Read\_Processes}, \textit{Home\_DIR\_Access}, and \textit{File\_Operations}, reinforcing the importance of process and filesystem-related behaviors. However, \textit{Home\_DIR\_Access} appears to be a more distinct marker, as lower feature values are consistently associated with benign samples, whereas higher values are more frequently associated with malicious ones. In contrast, lower-ranked features such as \textit{Remote\_IPs\_Access}, \textit{Pattern\_6}, and \textit{Total\_Dependencies} exhibit more compact SHAP distributions, suggesting weaker but still complementary contributions.

\begin{figure}[htpb]
    \centering
    \includegraphics[width=1\linewidth]{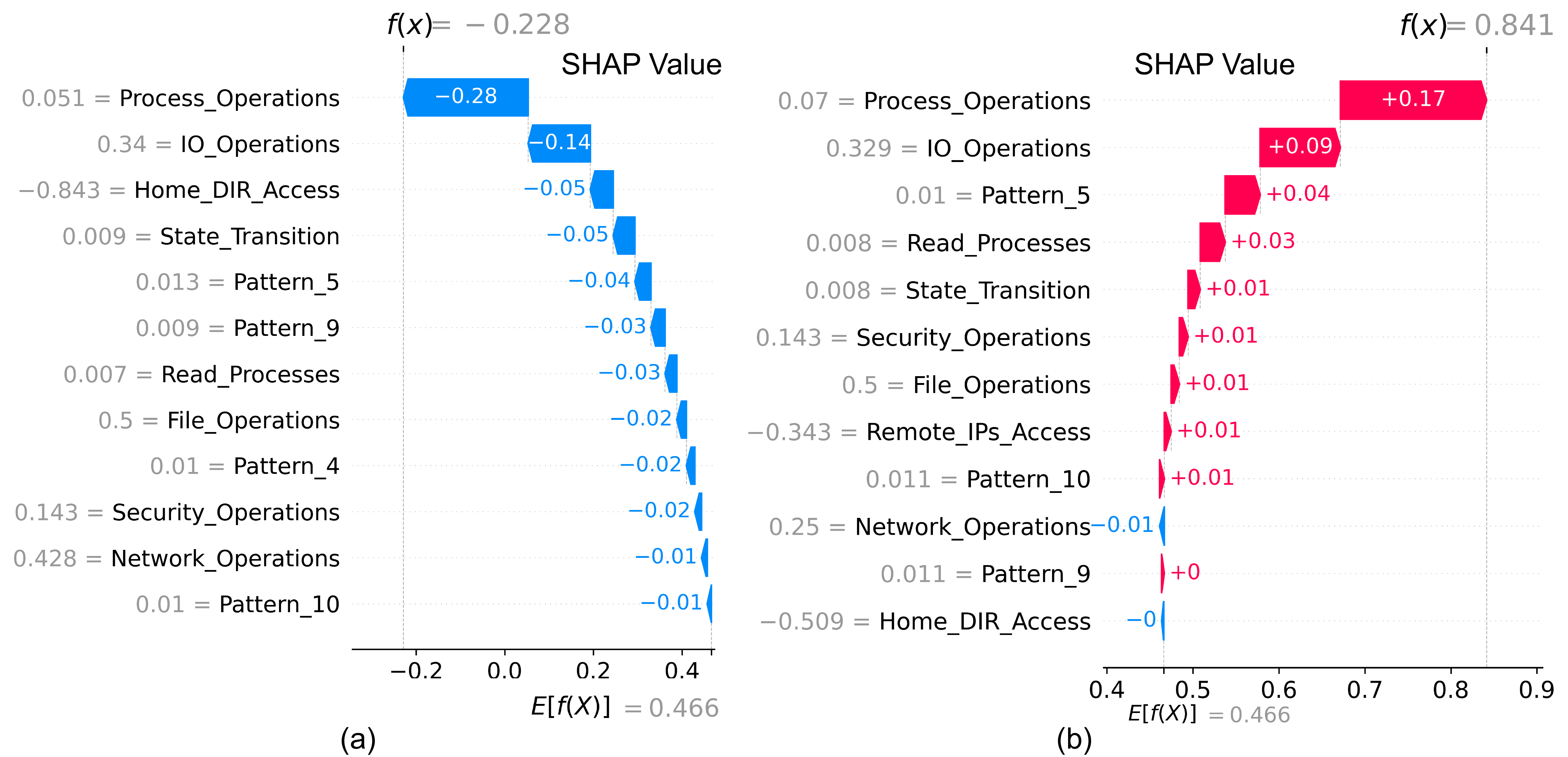}
    \caption{SHAP waterfall explanations for representative samples. (a) Benign sample; (b) malicious sample, showing how individual features contribute to the final prediction.}
    \Description{}
    \label{fig:shap_summary_2}
\end{figure}

While the global SHAP summary identifies influential features overall, it does not explain how these features combine within individual predictions. To address this, Figure~\ref{fig:shap_summary_2} presents local SHAP waterfall explanations for representative benign and malicious samples. In a SHAP waterfall plot, the prediction for an individual sample is expressed as \(f(x) = E[f(X)] + \sum_{i=1}^{M} \phi_i\), where \(E[f(X)]\) denotes the baseline model output over the background dataset, \(\phi_i\) is the contribution of feature \(i\), and \(f(x)\) is the final prediction for the sample. In both cases, the prediction starts from the baseline \(E[f(X)] = 0.466\), and each feature shifts the output upward or downward according to its SHAP value. For the benign instance in Figure~\ref{fig:shap_summary_2}(a), the cumulative contributions move the output downward to \(f(x) = -0.228\), driven mainly by negative effects from \textit{Process\_Operations}, \textit{IO\_Operations}, \textit{Home\_DIR\_Access}, \textit{State\_Transition}, \textit{Pattern\_5}, and \textit{Pattern\_9}. By contrast, for the malicious instance in Figure~\ref{fig:shap_summary_2}(b), the contributions shift the output upward to \(f(x) = 0.841\), primarily due to positive effects from \textit{Process\_Operations}, \textit{IO\_Operations}, \textit{Pattern\_5}, \textit{Read\_Processes}, \textit{State\_Transition}, \textit{Security\_Operations}, and \textit{File\_Operations}. Importantly, the same broad feature families remain influential across both samples, but their direction and magnitude vary according to the observed behavioral profile. This suggests that the detector learns class-discriminative behavioral relationships rather than relying solely on rigid heuristics.

\begin{figure}[htpb]
    \centering
    \includegraphics[width=1\linewidth]{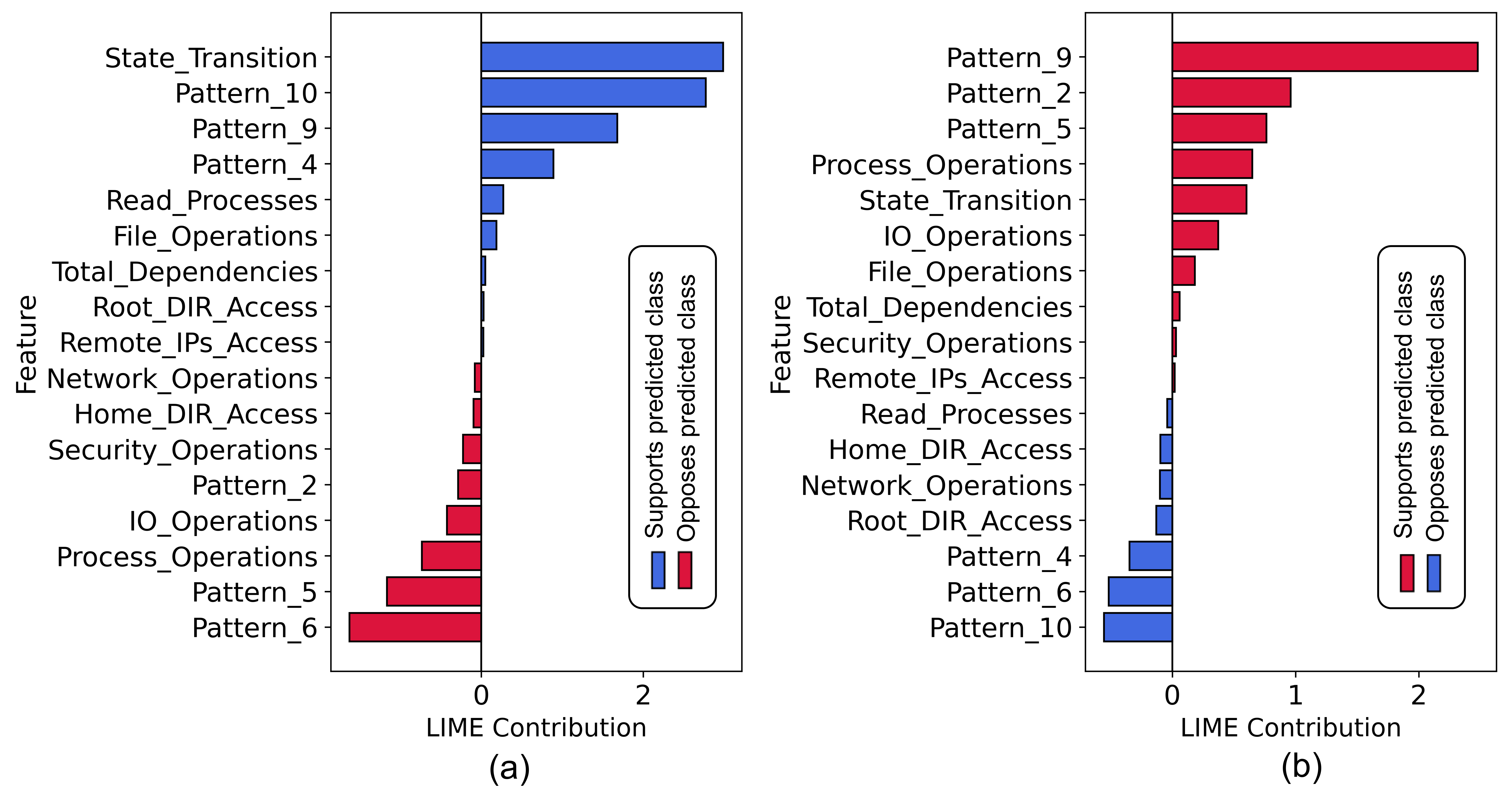}
    \caption{Local explanations for representative samples. (a) Benign sample and (b) malicious sample explanations.}
    \Description{}
    \label{fig:lime_summary}
\end{figure}

To further explore the consistency of local explanations, Figure~\ref{fig:lime_summary} presents LIME interpretations for representative benign and malicious samples. For the benign sample in Figure~\ref{fig:lime_summary}(a), the prediction is mainly supported by \textit{State\_Transition}, \textit{Pattern\_10}, \textit{Pattern\_9}, \textit{Read\_Processes}, and \textit{File\_Operations}, whereas \textit{Pattern\_6}, \textit{Pattern\_5}, \textit{Process\_Operations}, and \textit{IO\_Operations} act against the predicted class. For the malicious sample in Figure~\ref{fig:lime_summary}(b), the strongest supporting factors are \textit{Pattern\_9}, \textit{Pattern\_2}, \textit{Pattern\_5}, \textit{Process\_Operations}, \textit{State\_Transition}, and \textit{IO\_Operations}, while \textit{Pattern\_10}, \textit{Pattern\_6}, and several access-related features contribute in the opposite direction. Although SHAP and LIME are based on different explanation techniques, both highlight a consistent set of influential behavioral factors centered on process activity, I/O behavior, state transitions, and execution patterns. This consistency strengthens confidence that the proposed framework relies on meaningful behavioral evidence rather than incidental correlations. To support transparency and reproducibility, the eDySec repository~\footnote{eDySec GitHub: \url{https://github.com/tanzirmehedi/eDySec}} publicly provides all local explanations for the 2,141 test samples, along with fine-grained evidence covering process activity, I/O behavior, state transitions, and execution patterns.

\subsection{Findings and Implications}

\textbf{Implications for security developers:} The study highlights the importance of monitoring install-time and post-installation behaviors as part of software supply chain defense, since these stages can expose malicious actions that remain hidden during metadata and static code inspection. The results further show that strong detection performance can be achieved using classical DL models with low inference latency, making dynamic behavior-based detection more practical for integration into real-world software supply chain security pipelines. In addition, stable and explainable outputs can improve analyst trust and make automated decisions more transparent in high-volume package review settings.

\textbf{Implications for repository platforms:} These results also have important implications for repository-level software supply chain protection. In particular, metadata and static-based screening should be complemented with dynamic behavioral analysis to better identify malicious packages before they propagate through the dependency ecosystem. A layered defense pipeline, in which suspicious packages are escalated to sandboxed install-time and post-installation analysis, would strengthen the ability of repository platforms to detect sophisticated supply chain attacks prior to distribution. The results further suggest that process activity, I/O behavior, state transitions, and execution patterns should be treated as high-value supply chain risk signals in future package vetting systems. Moreover, stable and explainable detection can support more transparent and defensible moderation decisions at scale.

\textbf{Implications for researchers:} This study reinforces the importance of dynamic behavioral analysis in advancing software supply chain security research. The results suggest that, for malicious package detection, feature quality and behavioral representation may matter more than model complexity alone. It also indicates that XAI can help verify whether detection models are learning supply chain-relevant behavioral patterns rather than spurious correlations. Accordingly, future research should focus on additional dynamic datasets, cross-ecosystem evaluation, attack scenarios, and kernel-level AI agent security monitoring frameworks that combine metadata, static, and dynamic evidence.

\section{Threats to Validity and Limitations}
\label{label:limitation}

\textbf{Internal validity:} The reported performance depends on the correctness of dataset labeling, trace collection, preprocessing, and experimental configuration. Noise in labeling, incomplete behavioral capture, or implementation errors in feature extraction and model training could affect the observed results. Moreover, optimization-based feature selection methods such as PSO and WOA may yield minor variations across runs, and post-hoc explainability methods such as SHAP and LIME may produce slightly different interpretations depending on sampling strategy, and hardware conditions. More capable hardware can also provide more stable and efficient execution, which may positively influence experimental outcomes. To reduce these risks, all compared models were evaluated under the same dataset, preprocessing pipeline, feature selection procedure, explainability setting, and experimental environment.

\textbf{External validity:} This study is limited to PyPI packages and to behaviors observed during install-time and post-installation phase. While this setting remains highly relevant to software supply chain attacks in general, especially because the dataset captures both user and kernel-level behavioral signals, the findings may not generalize directly to other ecosystems. Extending the framework to npm, Maven, or RubyGems would likely require retraining, feature adaptation, and domain-specific calibration to reflect differences in package structure, and attack patterns.

\textbf{Construct validity:} eDySec relies on monitored dynamic execution and therefore inherits the constraints of the analysis environment. It may miss behaviors triggered by user interaction, command-line arguments, or time-delayed execution, where the malicious payload is activated only after a predefined waiting period. The current setup may also fail to expose environment-sensitive attacks that activate only under specific operating systems or hardware profiles. Moreover, because eBPF support is currently Linux-centered, the proposed framework is most representative of Linux-based environments. This study is further limited by malicious packages that spawn child processes capable of partially evading eBPF-based monitoring. Extending the framework to other platforms would require platform-specific sources, such as ETW providers combined with Security Audit events on Windows to capture process creation, registry modifications, privilege usage, and scheduled tasks, or kqueue, FSEvents, and network activity monitoring on macOS to support developer-oriented environments.

\section{Conclusion and Future Works}
\label{label:conclusion}

This study presents eDySec, an efficient, stable, and explainable DL-based framework for malicious package detection using the dynamic QUT-DV25 dataset. The results show that dynamic analysis is highly effective for software supply chain attack detection. In particular, eDySec significantly outperforms the state-of-the-art framework by reducing feature dimensionality by 52.78\%, lowering false positives and false negatives by 82\% and 79\%, respectively, improving accuracy by 3\%, and maintaining an inference latency of 170ms per package. Among the evaluated models, lightweight classical DL models, more specifically, MLP, deliver the best overall performance. Beyond predictive accuracy, the findings further demonstrate that effective feature selection is essential for fully exploiting the potential of dynamic, as certain feature combinations can degrade detection capability. By integrating detection model stability analysis and XAI techniques, eDySec also improves the stability and transparency of detection decisions. Ultimately, this study establishes a foundation for efficient, stable, and explainable DL-based detection of next-gen software supply chain attacks. Future work will extend this evaluation to additional dynamic datasets, ecosystems, attacks, and kernel-level AI agent security monitoring to further assess the generalizability of the proposed framework.

\newpage

\bibliographystyle{ACM-Reference-Format}
\bibliography{references}

\newpage

\appendix

\setcounter{figure}{0}
\setcounter{table}{0}
\setcounter{footnote}{0}

\section*{Appendix}

This appendix provides details on the open science artifacts, ethical considerations, and a summary of all analyses conducted in the eDySec framework.

\section{Open Science}

This paper follows open science principles by providing access to all artifacts necessary to evaluate and reproduce the results of eDySec.

\textbf{Artifacts availability:} The complete implementation of the proposed eDySec framework is made publicly available through a GitHub repository at the time of submission\footnote{eDySec GitHub: \url{https://github.com/tanzirmehedi/eDySec}}. The repository includes the full project structure covering all four phases of the pipeline: (i) data preparation, (ii) feature selection, (iii) deep learning model selection and evaluation, and (iv) stability and explainability analysis. It provides Jupyter notebooks for each experimental component, an automated execution utility (\texttt{edysec\_runner.py}) to reproduce the workflow, and a \texttt{requirements.txt} file for dependency management. Detailed documentation, including setup instructions for both local environments and GitHub Dev/Codespaces, is included to ensure ease of use.

\textbf{Dataset access:} The experimental evaluation is conducted using the QUT-DV25 dataset, a large-scale dynamic behavioral dataset comprising 14,271 Python packages (including 7,127 malicious samples) from the PyPI ecosystem. The dataset includes multiple execution trace categories, namely Filetop, Opensnoop, Install, TCP, SysCall, and Pattern traces, along with a combined feature representation. The dataset is publicly available via DOI\footnote{ QUT-DV25 Dataset: \url{https://doi.org/10.7910/DVN/LBMXJY}}. Due to security and ethical considerations associated with redistributing potentially malicious artifacts, the repository provides processed and structured data representations (e.g., extracted traces and feature sets) rather than raw malicious packages. The expected dataset directory structure and integration steps are explicitly documented to ensure correct usage within the pipeline.

\textbf{Reproducibility details:} To facilitate full reproducibility of the experimental results, the repository provides:
\begin{itemize}
    \item Source code with version control covering all phases
    \item Structured Jupyter notebooks for each step of the pipeline
    \item An automated runner script to execute notebooks in the correct order while preserving relative paths
    \item Environment configuration details, including Python version (3.10.20), system setup, and required dependencies
    \item Explicit hardware and software specifications used in the original experiments
    \item Execution workflows (phase-wise and end-to-end)
    \item Generated outputs, including evaluation metrics (e.g., confusion matrices, ROC curves, learning curves), feature selection results, and training logs
    \item Explainability artifacts, including SHAP and LIME outputs for both global and local interpretations
    \item Stability analysis outputs, including statistical comparisons and performance summaries
\end{itemize}

Additionally, step-by-step instructions are provided for reproducing results either manually (via phase-wise notebook execution) or automatically using the provided runner script.

\textbf{Limitations and constraints:} While the repository ensures a high degree of transparency and reproducibility, certain raw artifacts (e.g., original malicious package binaries) are not redistributed due to legal, ethical, and security constraints. However, all necessary intermediate representations, extracted features, and execution traces are provided to enable independent verification of the results.

\section{Ethical Considerations}

This study uses the published QUT-DV25 dataset, which contains traces derived from malicious and benign Python packages. Since the dataset includes malware-related behavioral artifacts, all stages of the eDySec pipeline, including data handling, preprocessing, feature extraction, model training, and evaluation, were performed in isolated and controlled environments to reduce operational risk and ensure safe experimentation. The dataset was used exclusively for research and defensive purposes to advance malicious package detection in software supply chains. No personal, private, or user-generated data were processed in this study. The proposed framework, derived features, and outputs are intended only for cybersecurity research and defensive applications.

\onecolumn

\section{Detailed Result Analysis}

Table~\ref{tab:notation} summarizes the notations used throughout the proposed framework and its mathematical formulation with description.

\begin{table}[htbp]
\centering
\small
\renewcommand{\arraystretch}{1.3}
\caption{Notations used in the proposed framework.}
\label{tab:notation}
\begin{tabular}{p{2.2cm} p{12cm}}
\toprule
\textbf{Notation} & \textbf{Description} \\
\midrule
$\mathcal{D}$ & Full dataset. \\
$N$ & Total number of packages/samples. \\
$\mathbf{x}_i$ & Raw feature representation of the $i$-th package. \\
$y_i$ & Class label of the $i$-th package. \\
$d$ & Dimension of the raw feature space. \\
$\mathcal{D}_{tr}, \mathcal{D}_{vd}, \mathcal{D}_{te}$ & Training, validation, and test sets. \\
$\mathcal{P}(\cdot)$ & Preprocessing function. \\
$\tilde{\mathbf{x}}_i$ & Processed feature vector for classical DL models. \\
$\mathbf{s}_i$ & Sequential input representation for pre-attention and attention-based models. \\
$\mathcal{F}$ & Processed feature space. \\
$\mathcal{F}^{*}$ & Optimal selected feature subset. \\
$\mathcal{S}$ & Set of candidate feature selection methods. \\
$S^{*}$ & Optimal feature selection method. \\
$\mathcal{M}$ & Set of candidate DL models. \\
$M_k$ & The $k$-th candidate DL model. \\
$M^{*}$ & Selected optimal DL model. \\
$\Theta_k$ & Learnable parameters of model $M_k$. \\
$J_{FS}(S_j)$ & Feature selection objective function. \\
$\alpha$ & Trade-off parameter between performance and feature reduction. \\
$Q_k$ & Validation performance of model $M_k$. \\
$\{q_r\}_{r=1}^{R}$ & Performance scores obtained over repeated runs. \\
$R$ & Total number of repeated runs. \\
$\bar{q}$ & Mean performance score across repeated runs. \\
$\sigma_q$ & Standard deviation of performance scores across repeated runs. \\
$\mathcal{T}_{stab}$ & Stability statistics summarizing repeated-run performance. \\
$\mathcal{E}(\cdot)$ & Explainability function. \\
$\mathbf{e}_i$ & Local attribution vector for the $i$-th sample. \\
$I(f)$ & Global importance score of feature $f$. \\
$\mathcal{C}$ & Consistency between selected and explanation-important features. \\
\bottomrule
\end{tabular}
\end{table}

\newpage

Table~\ref{tab:hyperparameters_dl_models} summarizes the final hyper-parameter settings of the selected DL models, including architectural, training, regularization, and attention-specific configurations, chosen through extensive hyperparameter exploration based on the best validation performance.

\begin{table*}[htbp]
\centering
\scriptsize
\renewcommand{\arraystretch}{1.6}
\caption{Summary of the final hyper-parameter settings of the selected DL models.}
\label{tab:hyperparameters_dl_models}

\resizebox{\textwidth}{!}{
\begin{tabular}{|c|l|lllll|ll|lll|}
\hline

\multirow{2}{*}{\textbf{C}} & \multirow{2}{*}{\textbf{Parameters}}
& \multicolumn{5}{c|}{\textbf{Classical}}
& \multicolumn{2}{c|}{\textbf{Pre-Attention}}
& \multicolumn{3}{c|}{\textbf{Attention}} \\

\cline{3-12}

&
& \textbf{CNN}
& \textbf{MLP}
& \textbf{LeNet}
& \textbf{MDCNN}
& \textbf{NN}
& \textbf{LSTM}
& \textbf{RNN}
& \textbf{Transformer}
& \textbf{BERT}
& \textbf{DistilGPT2} \\

\hline

\multirow{6}{*}{\rotatebox{90}{Architecture}}
& Num of Hidden Layers & 3 & 3 & 3 & 3 & 2 & 2 & 2 & 4 & 12+1 & 6+1 \\
\cline{2-12}
& Units per Hidden Layer & 64,128,64 & 500,500,500 & 5,20,500 & 8,8,64 & 68,68 & 64,32 & 128,64 & 256,128,64 & 768,128 & 768,128 \\
\cline{2-12}
& Embedding Dimension & - & - & - & - & - & 64 & 128 & 128 & 768 & 768 \\
\cline{2-12}
& Kernel Size & 3 & - & 5 & 5 & - & - & - & - & - & - \\
\cline{2-12}
& Stride & 1 & - & 1 & 1 & - & - & - & - & - & - \\
\cline{2-12}
& Pooling Type & GlobalMax & - & Max & Max & - & - & - & GlobalAvg & CLS & LastToken/Avg \\

\hline

\multirow{7}{*}{\rotatebox{90}{Training}}
& Batch Size & 16 & 16 & 16 & 16 & 16 & 16 & 16 & 16 & 16 & 16 \\
\cline{2-12}
& Number of Epochs & 200 & 200 & 200 & 200 & 200 & 200 & 200 & 200 & 200 & 200 \\
\cline{2-12}
& Learning Rate & $1\times10^{-3}$ & $1\times10^{-3}$ & $1\times10^{-3}$ & $1\times10^{-3}$ & $1\times10^{-3}$ & $1\times10^{-3}$ & $1\times10^{-4}$ & $3\times10^{-4}$ & $2\times10^{-4}$ & $2\times10^{-4}$ \\
\cline{2-12}
& Optimizer & Adam & Adam & Adam & Adam & Adam & Adam & Adam & Adam & Adam & AdamW \\
\cline{2-12}
& Loss Function & Binary & Binary & Binary & Binary & Binary & Binary & Binary & Binary & Binary & Binary \\
\cline{2-12}
& Hidden Layer Activation & ReLU & ReLU & ReLU & ReLU & ReLU & tanh/ReLU & tanh/ReLU & ReLU & ReLU & GELU \\
\cline{2-12}
& Output Activation Fun & Sigmoid & Sigmoid & Sigmoid & Sigmoid & Sigmoid & Sigmoid & Sigmoid & Sigmoid & Sigmoid & Sigmoid \\

\hline

\multirow{3}{*}{\rotatebox{90}{Regular.}}
& Dropout Rate & - & 0.1,0.2,0.3 & - & - & - & 0.3 & 0.3,0.3 & 0.2,0.3,0.2 & 0.3 & 0.1 \\
\cline{2-12}
& Weight Initialization & Default & Default & Default & Default & Default & Default & Default & Default & Pretrained & Pretrained \\
\cline{2-12}
& L2 Regularization & - & - & - & - & - & - & - & $10^{-5}$ & - & $10^{-4}$ \\

\hline

\multirow{3}{*}{\rotatebox{90}{Attention}}
& Attention Type & - & - & - & - & - & - & - & Self-Attention & Self-Attention & Self-Attention \\
\cline{2-12}
& Attention Heads & - & - & - & - & - & - & - & 4 & 12 & 12 \\
\cline{2-12}
& Attention Dimension & - & - & - & - & - & - & - & 32 & 64 & 64 \\

\hline

\end{tabular}}
\end{table*}

\newpage

\begin{figure*}[htpb]
    \centering
    \includegraphics[width=0.8\linewidth]{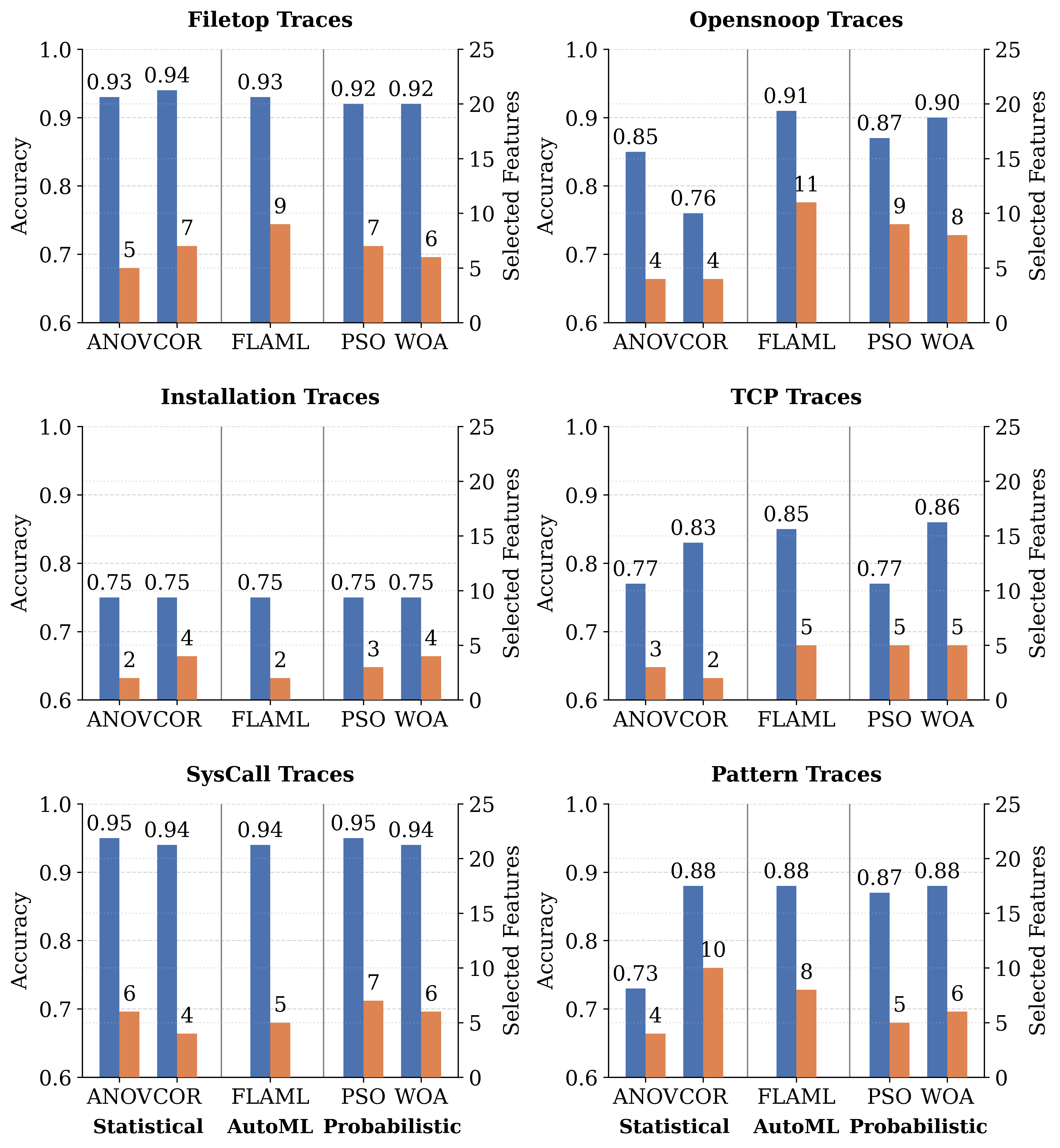}
    \captionsetup{justification=centering}
    \vspace{10pt}
    \caption{Performance comparison of different feature selection methods on the QUT-DV25 dataset using MLP model.}
    \Description{}
    \label{fig:six_feature_selection}
\end{figure*}

Figure~\ref{fig:six_feature_selection} presents a comparative evaluation of different feature selection methods on the QUT-DV25 dataset using the MLP model. It illustrates how each method influences detection performance, providing a clear view of the relative performance of the selected feature subsets across the evaluated settings.

\newpage

\begin{figure*}[htp]
    \centering
    \includegraphics[width=0.88\linewidth]{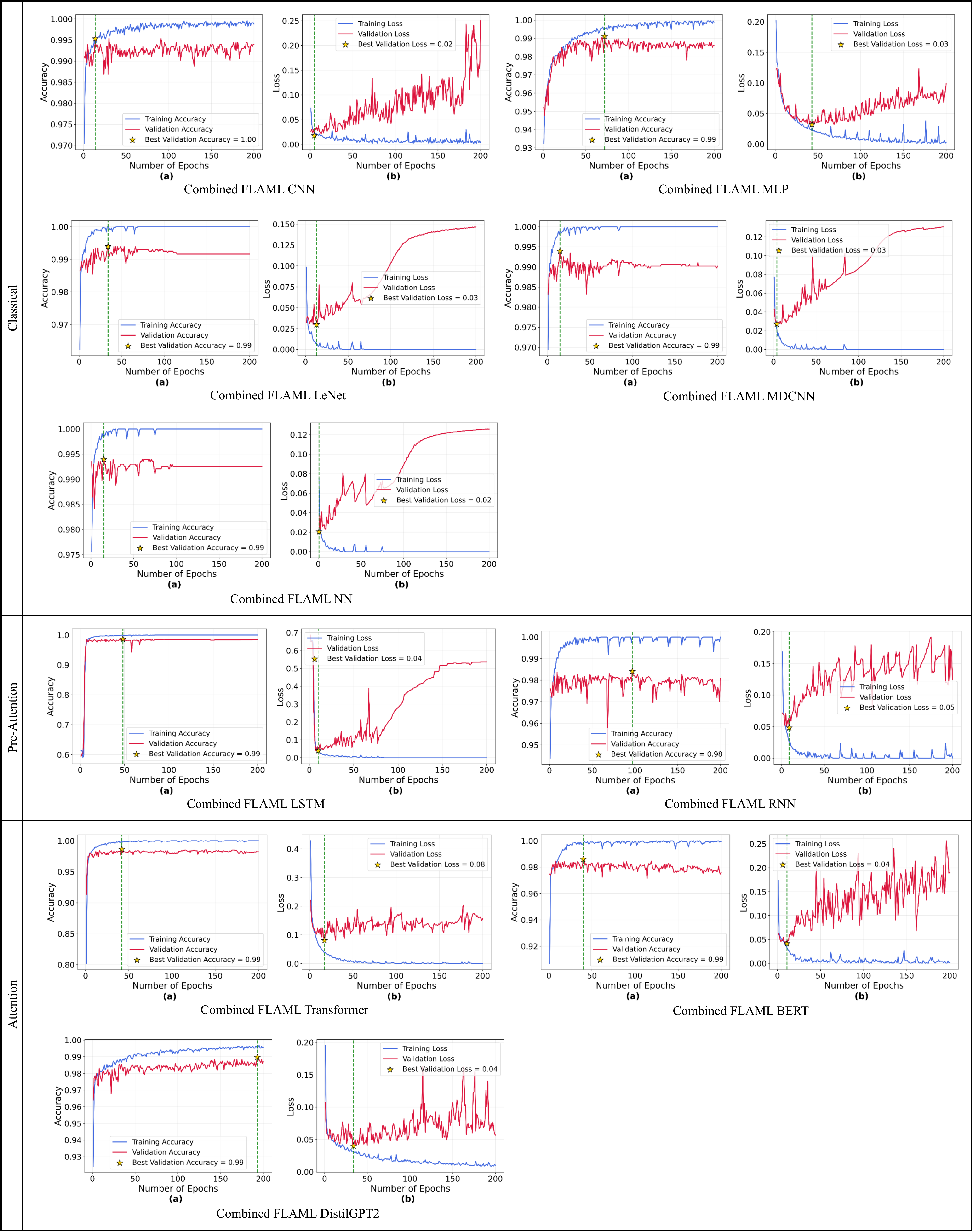}
    \caption{Performance of the FLAML-DL models on the QUT-DV25 Combined dataset: (a) accuracy and (b) loss.}
    \Description{}
    \label{fig:combined_flaml_results}
\end{figure*}

Figure~\ref{fig:combined_flaml_results} presents the performance of the FLAML-selected DL models on the QUT-DV25 Combined dataset. It shows both the accuracy and loss curves, providing a clear view of the training behavior and comparative convergence characteristics.

\newpage

\begin{figure*}[htpb]
    \centering
    \includegraphics[width=0.86\linewidth]{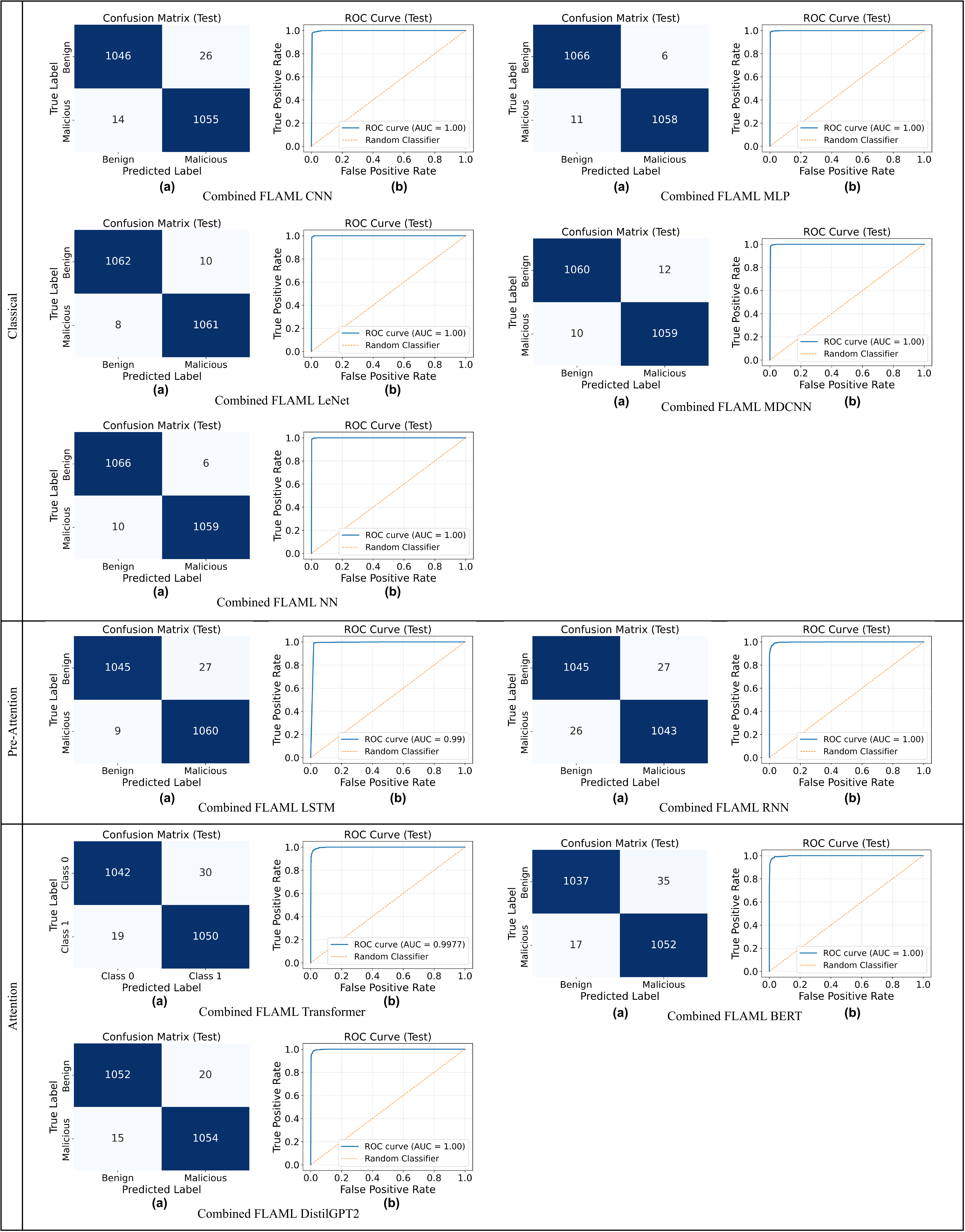}
    \caption{Performance of the FLAML-DL models on the QUT-DV25 Combined dataset: (a) Confusion matrix and (b) ROC curve.}
    \Description{}
    \label{fig:combined_flaml_roc_results}
\end{figure*}

Figure~\ref{fig:combined_flaml_roc_results} presents the classification performance of the FLAML-selected DL models on the QUT-DV25 Combined dataset. It shows the confusion matrices and ROC curves, providing a clear view of the predictive discrimination and class-wise detection behavior of the evaluated models.

\newpage

\begin{figure*}[htpb]
    \centering
    \includegraphics[width=0.88\linewidth]{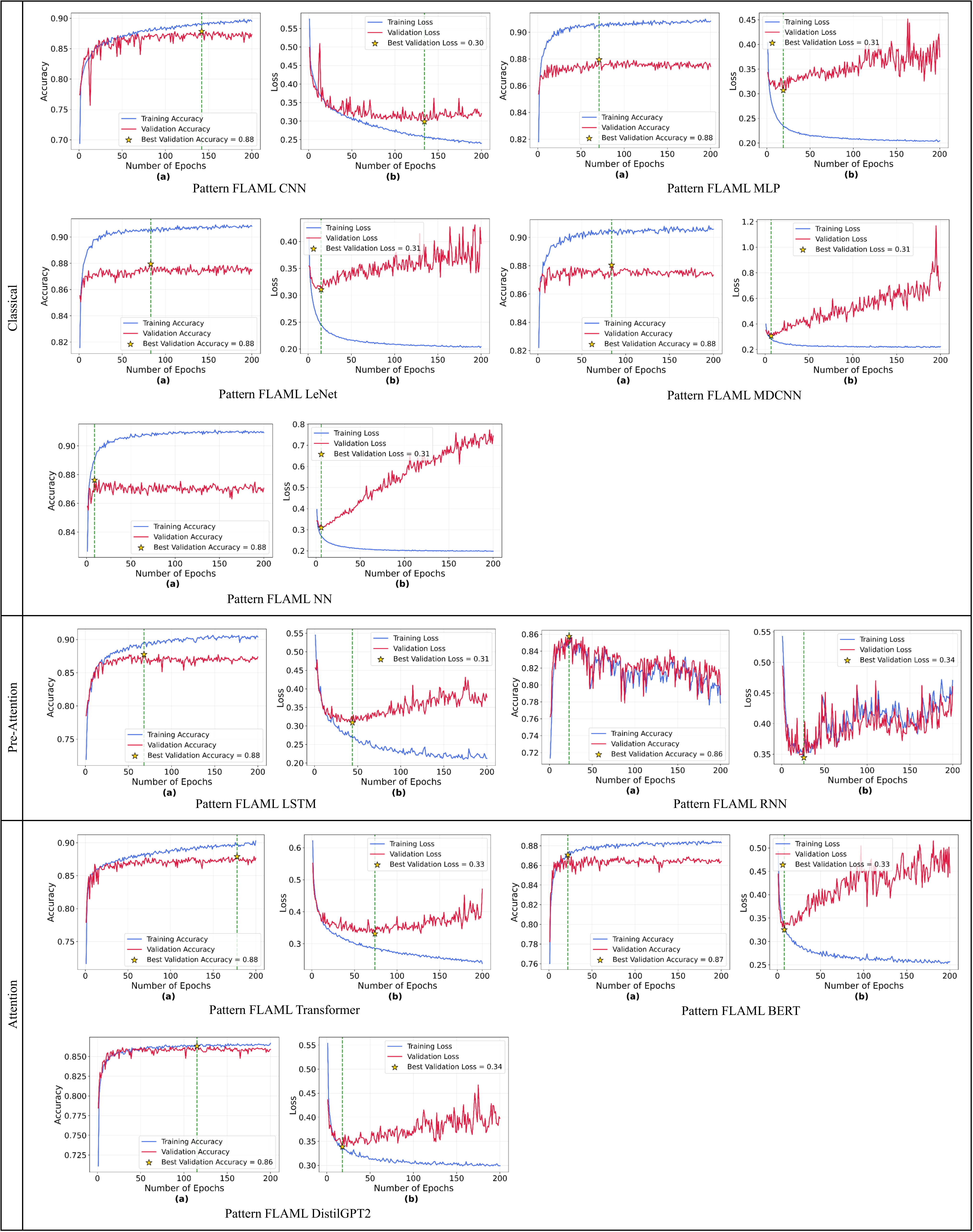}
    \caption{Performance of the FLAML-DL models on the Pattern trace dataset: (a) accuracy and (b) loss.}
    \Description{}
    \label{fig:pattern_flaml_results}
\end{figure*}

Figure~\ref{fig:pattern_flaml_results} presents the performance of the FLAML-selected DL models on the Pattern trace dataset. It shows both the accuracy and loss curves, providing a clear view of the training behavior and comparative convergence characteristics.

\newpage

\begin{figure*}[htpb]
    \centering
    \includegraphics[width=0.88\linewidth]{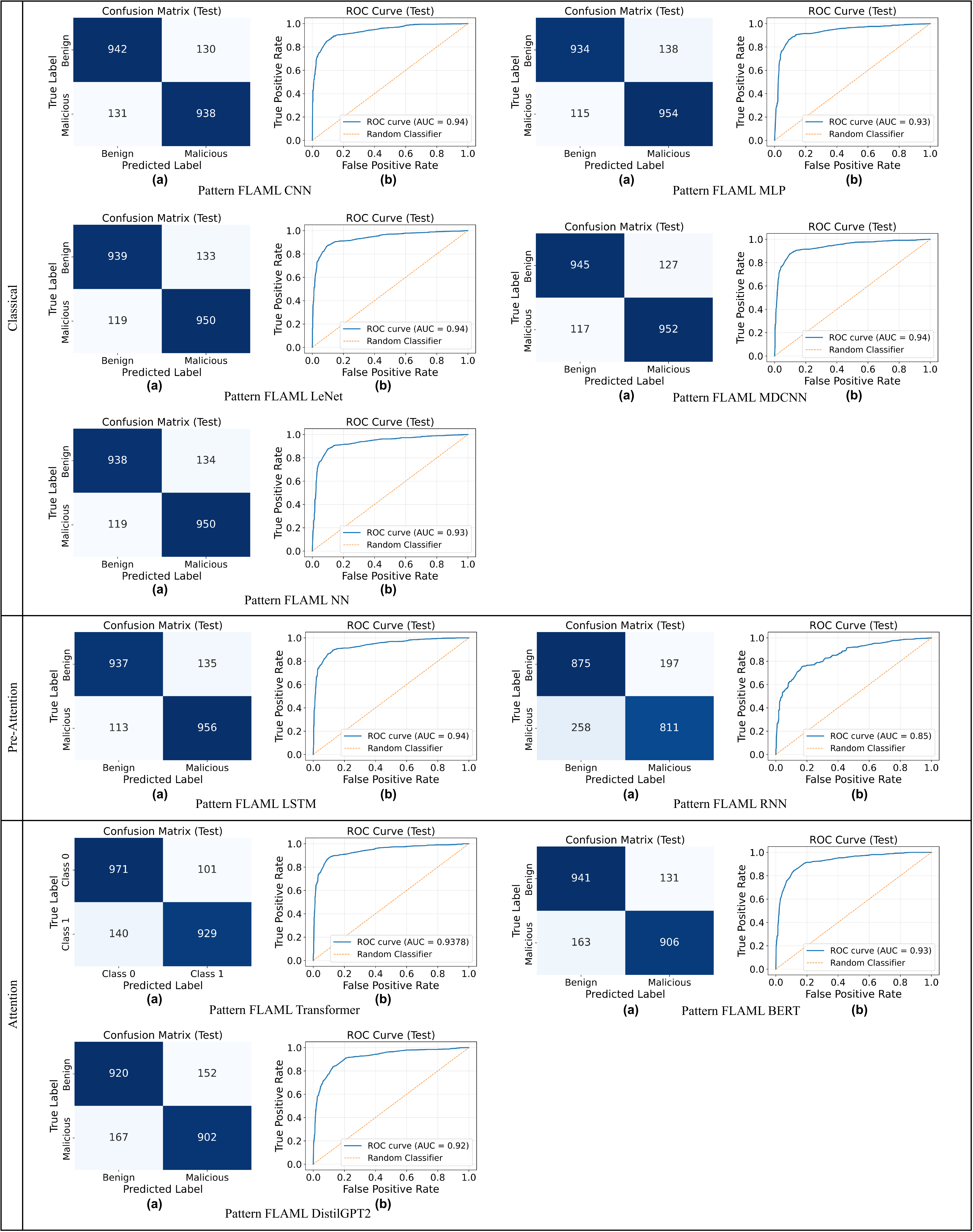}
    \caption{Performance of the FLAML-DL models on the Pattern trace dataset: (a) Confusion matrix and (b) ROC curve.}
    \Description{}
    \label{fig:pattern_flaml_roc_results}
\end{figure*}

Figure~\ref{fig:pattern_flaml_roc_results} presents the classification performance of the FLAML-selected DL models on the Pattern trace dataset. It shows the confusion matrices and ROC curves, providing a clear view of the predictive discrimination and class-wise detection behavior of the evaluated models.

\twocolumn

\begin{table}[htbp]
\centering
\scriptsize
\setlength{\tabcolsep}{5.5pt}
\renewcommand{\arraystretch}{1.5}
\caption{DL model performance across feature selection techniques on Filetop Traces, bold and green-shaded values indicate the best performance.}
\label{tab:filetop}
\begin{tabular}{|c| c| c| l | c c c c|}
\hline
$\mathbf{\mathcal{F_C}}$ &
$\mathbf{\mathcal{F}}$ &
$\mathbf{\mathcal{M_C}}$ &
$\mathbf{\mathcal{M}}$ &
$\mathbf{\mathcal{A}}$ &
$\mathbf{\mathcal{P}}$ &
$\mathbf{\mathcal{R}}$ &
$\mathbf{\mathcal{F}_1}$ \\
\hline

\multirow{18}{*}{\rotatebox{90}{Statistical}}
& \multirow{9}{*}{ANOVA}
& \multirow{5}{*}{Classical}
& CNN & 0.85 & 0.85 & 0.85 & 0.85 \\
&  &  & \best{MLP}   & \best{0.93} & \best{0.93} & \best{0.93} & \best{0.93} \\
&  &  & LeNet & 0.92 & 0.92 & 0.92 & 0.92 \\
&  &  & MDCNN & 0.92 & 0.92 & 0.92 & 0.92 \\
&  &  & NN    & 0.92 & 0.92 & 0.92 & 0.92 \\
\cline{3-8}
&  & \multirow{2}{*}{Pre-Attention}
& LSTM  & 0.91 & 0.90 & 0.91 & 0.91 \\
&  &  & RNN  & 0.91 & 0.90 & 0.91 & 0.91 \\
\cline{3-8}
&  & \multirow{3}{*}{Attention}
& Transformer & 0.90 & 0.89  & 0.93  & 0.91  \\
&  &  & BERT  & 0.89 & 0.87 & 0.91 & 0.89 \\
&  &  & DistilGPT2  & 0.88 & 0.86 & 0.91 & 0.88 \\
\cline{2-8}

& \multirow{9}{*}{CORR}
& \multirow{5}{*}{Classical}
& CNN   & 0.89 & 0.89 & 0.89 & 0.89  \\
&  &  & \best{MLP}   & \best{0.94} & \best{0.94} & \best{0.94} & \best{0.94} \\
&  &  & LeNet & 0.93 & 0.93 & 0.93 & 0.93 \\
&  &  & MDCNN & 0.93 & 0.93 & 0.93 & 0.93 \\
&  &  & NN    & 0.93 & 0.93 & 0.93 & 0.93 \\
\cline{3-8}
&  & \multirow{2}{*}{Pre-Attention}
& LSTM  & 0.90 & 0.90 & 0.90 & 0.90 \\
&  &  & RNN  & 0.86 & 0.86 & 0.86 & 0.86 \\
\cline{3-8}
&  & \multirow{3}{*}{Attention}
& Transformer & 0.87 & 0.87 & 0.87 & 0.87 \\
&  &  & BERT  & 0.83 & 0.83 & 0.83 & 0.83 \\
&  &  & DistilGPT2  & 0.92 & 0.92 & 0.92 & 0.92 \\
\hline

\multirow{9}{*}{\rotatebox{90}{AutoML}}
& \multirow{9}{*}{FLAML}
& \multirow{5}{*}{Classical}
& CNN   & 0.91 & 0.91 & 0.91 & 0.91 \\
&  &  & \best{MLP}   & \best{0.93} & \best{0.93} & \best{0.93} & \best{0.93} \\
&  &  & \best{LeNet} & \best{0.93} & \best{0.93} & \best{0.93} & \best{0.93} \\
&  &  & \best{MDCNN} & \best{0.93} & \best{0.93} & \best{0.93} & \best{0.93} \\
&  &  & \best{NN}    & \best{0.93} & \best{0.93} & \best{0.93} & \best{0.93} \\
\cline{3-8}
&  & \multirow{2}{*}{Pre-Attention}
& LSTM  & 0.91 & 0.90 & 0.92 & 0.91 \\
&  &  & RNN & 0.85 & 0.80 & 0.93 & 0.86 \\
\cline{3-8}
&  & \multirow{3}{*}{Attention}
& Transformer & 0.89 & 0.87 & 0.92 & 0.89 \\
&  &  & BERT  & 0.85 & 0.83 & 0.89 & 0.86 \\
&  &  & DistilGPT2  & 0.92 & 0.92 & 0.93 & 0.92 \\
\hline

\multirow{18}{*}{\rotatebox{90}{Probabilistic}}
& \multirow{9}{*}{PSO}
& \multirow{5}{*}{Classical}
& CNN   & 0.91 & 0.91 & 0.91 & 0.91 \\
&  &  & \best{MLP}   & \best{0.92} & \best{0.92} & \best{0.92} & \best{0.92} \\
&  &  & \best{LeNet} & \best{0.92} & \best{0.92} & \best{0.92} & \best{0.92} \\
&  &  & \best{MDCNN} & \best{0.92} & \best{0.92} & \best{0.92} & \best{0.92} \\
&  &  & \best{NN}    & \best{0.92} & \best{0.92} & \best{0.92} & \best{0.92} \\
\cline{3-8}
&  & \multirow{2}{*}{Pre-Attention}
& LSTM  & 0.67 & 0.79 & 0.67 & 0.63 \\
&  &  & RNN  & 0.89 & 0.89 & 0.89 & 0.89 \\
\cline{3-8}
&  & \multirow{3}{*}{Attention}
& Transformer & 0.87 & 0.87  & 0.87  & 0.87  \\
&  &  & BERT  & 0.87  & 0.87 & 0.87 & 0.87 \\
&  &  & DistilGPT2  & 0.91 & 0.91 & 0.91 & 0.91 \\
\cline{2-8}

& \multirow{9}{*}{WOA}
& \multirow{5}{*}{Classical}
& CNN   & 0.88 & 0.88 & 0.88 & 0.88 \\
&  &  & \best{MLP}   & \best{0.92} & \best{0.92} & \best{0.92} & \best{0.92} \\
&  &  & LeNet & 0.91 & 0.91 & 0.91 & 0.91 \\
&  &  & MDCNN & 0.91 & 0.91 & 0.91 & 0.91 \\
&  &  & NN    & 0.92 & 0.92 & 0.92 & 0.92 \\
\cline{3-8}
&  & \multirow{2}{*}{Pre-Attention}
& LSTM  & 0.90 & 0.90 & 0.90 & 0.90 \\
&  &  & RNN  & 0.89 & 0.89 & 0.89 & 0.89 \\
\cline{3-8}
&  & \multirow{3}{*}{Attention}
& Transformer & 0.88 & 0.84 & 0.92 & 0.88 \\
&  &  & BERT  & 0.85 & 0.85 & 0.85 & 0.85 \\
&  &  & DistilGPT2  & 0.87 & 0.87 & 0.87 & 0.87 \\
\hline
\end{tabular}
\end{table}

\begin{table}[htbp]
\centering
\scriptsize
\setlength{\tabcolsep}{5.5pt}
\renewcommand{\arraystretch}{1.5}
\caption{DL model performance across feature selection techniques on Opensnoop Traces, bold and green-shaded values indicate the best performance.}
\label{tab:opensnoop}
\begin{tabular}{|c| c| c| l | c c c c|}
\hline
$\mathbf{\mathcal{F_C}}$ &
$\mathbf{\mathcal{F}}$ &
$\mathbf{\mathcal{M_C}}$ &
$\mathbf{\mathcal{M}}$ &
$\mathbf{\mathcal{A}}$ &
$\mathbf{\mathcal{P}}$ &
$\mathbf{\mathcal{R}}$ &
$\mathbf{\mathcal{F}_1}$ \\
\hline

\multirow{18}{*}{\rotatebox{90}{Statistical}}
& \multirow{9}{*}{ANOVA}
& \multirow{5}{*}{Classical}
& CNN & 0.84 & 0.84 & 0.84 & 0.84 \\
&  &  & \best{MLP}   & \best{0.85} & \best{0.85} & \best{0.85} & \best{0.85} \\
&  &  & LeNet & 0.84 & 0.85 & 0.84 & 0.84 \\
&  &  & MDCNN & 0.84 & 0.84 & 0.84 & 0.84 \\
&  &  & \best{NN}   & \best{0.85} & \best{0.85} & \best{0.85} & \best{0.85}\\
\cline{3-8}
&  & \multirow{2}{*}{Pre-Attention}
& LSTM  & 0.70 & 0.70 & 0.71 & 0.71 \\
&  &  & RNN  & 0.70 & 0.69 & 0.72 & 0.71 \\
\cline{3-8}
&  & \multirow{3}{*}{Attention}
& Transformer & 0.73 & 0.72  & 0.74  &  0.73 \\
&  &  & BERT  & 0.81 & 0.82 & 0.79 & 0.80 \\
&  &  & DistilGPT2  & 0.84 & 0.89 & 0.78 & 0.83 \\
\cline{2-8}

& \multirow{9}{*}{CORR}
& \multirow{5}{*}{Classical}
& CNN   & 0.79 & 0.79 & 0.79 & 0.79 \\
&  &  & MLP   & 0.76 & 0.76 & 0.76 & 0.76 \\
&  &  & LeNet & 0.72 & 0.72 & 0.72& 0.72 \\
&  &  & MDCNN & 0.68 & 0.68 & 0.68 & 0.68 \\
&  &  & NN    & 0.73 & 0.73 & 0.73 & 0.73 \\
\cline{3-8}
&  & \multirow{2}{*}{Pre-Attention}
& LSTM  & 0.60 & 0.63 & 0.60 & 0.58 \\
&  &  & RNN  & 0.60 & 0.62 & 0.60 & 0.58  \\
\cline{3-8}
&  & \multirow{3}{*}{Attention}
& Transformer & 0.62 & 0.61 & 0.65 & 0.63 \\
&  &  & \best{BERT}  & \best{0.80} & \best{0.80} & \best{0.80} & \best{0.80} \\
&  &  & DistilGPT2  & 0.79 & 0.79 & 0.79 & 0.79 \\
\hline

\multirow{9}{*}{\rotatebox{90}{AutoML}}
& \multirow{9}{*}{FLAML}
& \multirow{5}{*}{Classical}
& CNN   & 0.89 & 0.89 & 0.89 & 0.89 \\
&  &  & \best{MLP}   & \best{0.91} & \best{0.90} & \best{0.90} & \best{0.91} \\
&  &  & LeNet & 0.88 & 0.88 & 0.88 & 0.88 \\
&  &  & MDCNN & 0.88 & 0.89 & 0.88 & 0.88 \\
&  &  & NN    & 0.90 & 0.90 & 0.90 & 0.90 \\
\cline{3-8}
&  & \multirow{2}{*}{Pre-Attention}
& LSTM  & 0.76 & 0.73 & 0.83 & 0.78 \\
&  &  & RNN   & 0.75 & 0.77 & 0.71 & 0.74 \\
\cline{3-8}
&  & \multirow{3}{*}{Attention}
& Transformer & 0.73 & 0.70 & 0.79 & 0.74 \\
&  &  & BERT  & 0.90 & 0.94 & 0.88 & 0.90 \\
&  &  & DistilGPT2  & 0.89 & 0.92 & 0.86 & 0.89 \\
\hline

\multirow{18}{*}{\rotatebox{90}{Probabilistic}}
& \multirow{9}{*}{PSO}
& \multirow{5}{*}{Classical}
& CNN   & 0.87 & 0.87 & 0.87 & 0.87 \\
&  &  & MLP   & 0.87 & 0.87 & 0.875 & 0.87 \\
&  &  & LeNet & 0.88 & 0.88 & 0.88 & 0.88 \\
&  &  & MDCNN & 0.87 & 0.87 & 0.87 & 0.87 \\
&  &  & \best{NN} & \best{0.90} & \best{0.90} & \best{0.90} & \best{0.90} \\
\cline{3-8}
&  & \multirow{2}{*}{Pre-Attention}
& LSTM  & 0.72 & 0.72 & 0.72 & 0.72 \\
&  &  & RNN  & 0.72 & 0.72 & 0.72 & 0.72 \\
\cline{3-8}
&  & \multirow{3}{*}{Attention}
& Transformer & 0.66 & 0.61  & 0.84  & 0.71  \\
&  &  & BERT  & 0.88 & 0.91 & 0.84 & 0.87 \\
&  &  & DistilGPT2  & 0.88 & 0.88 & 0.88 & 0.88 \\
\cline{2-8}

& \multirow{9}{*}{WOA}
& \multirow{5}{*}{Classical}
& CNN   & 0.88 & 0.88 & 0.88 & 0.88 \\
&  &  & MLP & 0.82 & 0.82 & 0.82 & 0.82 \\
&  &  & LeNet & 0.87 & 0.87 & 0.87 & 0.87 \\
&  &  & MDCNN & 0.88 & 0.88 & 0.88 & 0.88 \\
&  &  & \best{NN} & \best{0.90} & \best{0.90} & \best{0.90} & \best{0.90} \\
\cline{3-8}
&  & \multirow{2}{*}{Pre-Attention}
& LSTM  & 0.74 & 0.74 & 0.74 & 0.74 \\
&  &  & RNN  & 0.75 & 0.75 & 0.75 & 0.75 \\
\cline{3-8}
&  & \multirow{3}{*}{Attention}
& Transformer & 0.75 & 0.75  & 0.74  & 0.75  \\
&  &  & BERT  & 0.80 & 0.80 & 0.80 & 0.80 \\
&  &  & DistilGPT2  & 0.88  & 0.88 & 0.88 & 0.88 \\
\hline
\end{tabular}
\end{table}

\begin{table}[htbp]
\centering
\scriptsize
\setlength{\tabcolsep}{5.5pt}
\renewcommand{\arraystretch}{1.5}
\caption{DL model performance across feature selection techniques on Install Traces, bold and green-shaded values indicate the best performance.}
\label{tab:install}
\begin{tabular}{|c| c| c| l | c c c c|}
\hline
$\mathbf{\mathcal{F_C}}$ &
$\mathbf{\mathcal{F}}$ &
$\mathbf{\mathcal{M_C}}$ &
$\mathbf{\mathcal{M}}$ &
$\mathbf{\mathcal{A}}$ &
$\mathbf{\mathcal{P}}$ &
$\mathbf{\mathcal{R}}$ &
$\mathbf{\mathcal{F}_1}$ \\
\hline

\multirow{18}{*}{\rotatebox{90}{Statistical}}
& \multirow{9}{*}{ANOVA}
& \multirow{5}{*}{Classical}
& \best{CNN} & \best{0.75} & \best{0.83} & \best{0.75} & \best{0.74} \\
&  &  & \best{MLP}   & \best{0.75} & \best{0.83} & \best{0.74} & \best{0.74} \\
&  &  & \best{LeNet} & \best{0.75} & \best{0.83} & \best{0.75} & \best{0.74} \\
&  &  & \best{MDCNN} & \best{0.75} & \best{0.83} & \best{0.75} & \best{0.74} \\
&  &  & \best{NN}    & \best{0.75} & \best{0.83} & \best{0.75} & \best{0.74} \\
\cline{3-8}
&  & \multirow{2}{*}{Pre-Attention}
& \best{LSTM}  & \best{0.75} & \best{0.67} & \best{1.00} & \best{0.80} \\
&  &  & \best{RNN}  & \best{0.75} & \best{0.67} & \best{1.00} & \best{0.80} \\
\cline{3-8}
&  & \multirow{3}{*}{Attention}
& Transformer & 0.74 & 0.67  & 1.00  & 0.79  \\
&  &  & \best{BERT}  & \best{0.75} & \best{0.67} & \best{1.00} & \best{0.80} \\
&  &  & \best{DistilGPT2}  & \best{0.75} & \best{0.67} & \best{1.00} & \best{0.80} \\
\cline{2-8}

& \multirow{9}{*}{CORR}
& \multirow{5}{*}{Classical}
& CNN   & 0.73 & 0.82 & 0.73 & 0.71 \\
&  &  & MLP   & 0.73 & 0.82 & 0.74 & 0.72 \\
&  &  & \best{LeNet} & \best{0.75} & \best{0.83} & \best{0.75} & \best{0.74} \\
&  &  & \best{MDCNN} & \best{0.75} & \best{0.83} & \best{0.75} & \best{0.74} \\
&  &  & \best{NN}    & \best{0.75} & \best{0.83} & \best{0.75} & \best{0.74} \\
\cline{3-8}
&  & \multirow{2}{*}{Pre-Attention}
& \best{LSTM}  & \best{0.75} & \best{0.83} & \best{0.75} & \best{0.74} \\
&  &  & \best{RNN}  & \best{0.75} & \best{0.83} & \best{0.75} & \best{0.74} \\
\cline{3-8}
&  & \multirow{3}{*}{Attention}
& \best{Transformer} & \best{0.75} & \best{0.66} & \best{0.99} & \best{0.80} \\
&  &  & \best{BERT}  & \best{0.75} & \best{0.83} & \best{0.75} & \best{0.74} \\
&  &  & \best{DistilGPT2}  & \best{0.75} & \best{0.83} & \best{0.75} & \best{0.73} \\
\hline

\multirow{9}{*}{\rotatebox{90}{AutoML}}
& \multirow{9}{*}{FLAML}
& \multirow{5}{*}{Classical}
& CNN   & 0.72 & 0.81 & 0.72 & 0.70 \\
&  &  & \best{MLP}   & \best{0.75} & \best{0.82} & \best{0.74} & \best{0.72} \\
&  &  & \best{LeNet} & \best{0.75} & \best{0.83} & \best{0.75} & \best{0.74} \\
&  &  & \best{MDCNN} & \best{0.75} & \best{0.83} & \best{0.75} & \best{0.74} \\
&  &  & \best{NN}    & \best{0.75} & \best{0.83} & \best{0.75} & \best{0.74} \\
\cline{3-8}
&  & \multirow{2}{*}{Pre-Attention}
& \best{LSTM}  & \best{0.75} & \best{0.67} & \best{1.00} & \best{0.80} \\
&  &  & RNN   & 0.72 & 0.73 & 0.99 & 0.69 \\
\cline{3-8}
&  & \multirow{3}{*}{Attention}
& \best{Transformer} & \best{0.75} & \best{0.67} & \best{1.00} & \best{0.80} \\
&  &  & BERT  & 0.69 & 0.69 & 1.00 & 0.66 \\
&  &  & \best{DistilGPT2}  & \best{0.75} & \best{0.67} & \best{1.00} & \best{0.80} \\
\hline

\multirow{18}{*}{\rotatebox{90}{Probabilistic}}
& \multirow{9}{*}{PSO}
& \multirow{5}{*}{Classical}
& \best{CNN}   & \best{0.75} & \best{0.83} & \best{0.75} & \best{0.74} \\
&  &  & \best{MLP}   & \best{0.75} & \best{0.83} & \best{0.75} & \best{0.73} \\
&  &  & \best{LeNet} & \best{0.75} & \best{0.83} & \best{0.75} & \best{0.74} \\
&  &  & \best{MDCNN} & \best{0.75} & \best{0.83} & \best{0.75} & \best{0.74} \\
&  &  & \best{NN}    & \best{0.75} & \best{0.83} & \best{0.75} & \best{0.73} \\
\cline{3-8}
&  & \multirow{2}{*}{Pre-Attention}
& \best{LSTM}  & \best{0.75} & \best{0.83} & \best{0.75} & \best{0.74} \\
&  &  & RNN  & 0.74 & 0.82 & 0.74 & 0.72 \\
\cline{3-8}
&  & \multirow{3}{*}{Attention}
& Transformer & 0.74 & 0.66 & 0.99 & 0.79 \\
&  &  & BERT  & 0.69 & 0.74 & 0.70 & 0.77 \\
&  &  & \best{DistilGPT2}  & \best{0.75} & \best{0.83} & \best{0.75} & \best{0.74} \\
\cline{2-8}

& \multirow{9}{*}{WOA}
& \multirow{5}{*}{Classical}
& \best{CNN}   & \best{0.75} & \best{0.83} & \best{0.75} & \best{0.74} \\
&  &  & \best{MLP}   & \best{0.75} & \best{0.83} & \best{0.75} & \best{0.74} \\
&  &  & \best{LeNet} & \best{0.75} & \best{0.83} & \best{0.75} & \best{0.74} \\
&  &  & \best{MDCNN} & \best{0.75} & \best{0.83} & \best{0.75} & \best{0.74} \\
&  &  & \best{NN}    & \best{0.75} & \best{0.83} & \best{0.75} & \best{0.74} \\
\cline{3-8}
&  & \multirow{2}{*}{Pre-Attention}
& \best{LSTM}  & \best{0.75} & \best{0.83} & \best{0.75} & \best{0.74} \\
&  &  & RNN  & 0.72 & 0.76 & 0.72 & 0.74 \\
\cline{3-8}
&  & \multirow{3}{*}{Attention}
& \best{Transformer} & \best{0.75} & \best{0.67} & \best{1.00} & \best{0.80} \\
&  &  & BERT  & 0.69 & 0.80 & 0.69 & 0.65 \\
&  &  & \best{DistilGPT2}  & \best{0.75} & \best{0.83} & \best{0.75} & \best{0.74} \\
\hline
\end{tabular}
\end{table}

\begin{table}[htbp]
\centering
\scriptsize
\setlength{\tabcolsep}{5.5pt}
\renewcommand{\arraystretch}{1.5}
\caption{DL model performance across feature selection techniques on TCP Traces, bold and green-shaded values indicate the best performance.}
\label{tab:tcp}
\begin{tabular}{|c| c| c| l | c c c c|}
\hline
$\mathbf{\mathcal{F_C}}$ &
$\mathbf{\mathcal{F}}$ &
$\mathbf{\mathcal{M_C}}$ &
$\mathbf{\mathcal{M}}$ &
$\mathbf{\mathcal{A}}$ &
$\mathbf{\mathcal{P}}$ &
$\mathbf{\mathcal{R}}$ &
$\mathbf{\mathcal{F}_1}$ \\
\hline

\multirow{18}{*}{\rotatebox{90}{Statistical}}
& \multirow{9}{*}{ANOVA}
& \multirow{5}{*}{Classical}
& CNN & 0.76 & 0.76 & 0.76 & 0.76 \\
&  &  & MLP & 0.72 & 0.72 & 0.72 & 0.72 \\ 
&  &  & LeNet & 0.76 & 0.76 & 0.76 & 0.76 \\
&  &  & MDCNN & 0.76 & 0.76 & 0.76 & 0.76 \\
&  &  & \best{NN} &\best{0.77} & \best{0.77} & \best{0.77} & \best{0.77} \\
\cline{3-8}
&  & \multirow{2}{*}{Pre-Attention}
& LSTM  & 0.72 & 0.73 & 0.71 & 0.72 \\
&  &  & RNN  & 0.68 & 0.69 & 0.65 & 0.67 \\
\cline{3-8}
&  & \multirow{3}{*}{Attention}
& Transformer & 0.68 & 0.68 & 0.68 & 0.68 \\
&  &  & BERT  & 0.76 & 0.76 & 0.74 & 0.75 \\
&  &  & DistilGPT2  & 0.76 & 0.76 & 0.77 & 0.76 \\
\cline{2-8}

& \multirow{9}{*}{CORR}
& \multirow{5}{*}{Classical}
& CNN   & 0.79 & 0.79 & 0.79 & 0.79 \\
&  &  & MLP   & 0.83 & 0.83 & 0.83 & 0.83 \\
&  &  & LeNet & 0.83 & 0.83 & 0.83 & 0.83 \\
&  &  & MDCNN & 0.82 & 0.82 & 0.82 & 0.82 \\
&  &  & NN    & 0.81 & 0.81 & 0.81 & 0.81 \\
\cline{3-8}
&  & \multirow{2}{*}{Pre-Attention}
& \best{LSTM}  & \best{0.84} & \best{0.84} & \best{0.84} & \best{0.84} \\
&  &  & RNN  & 0.80 & 0.81 & 0.80 & 0.80 \\
\cline{3-8}
&  & \multirow{3}{*}{Attention}
& Transformer & 0.83 & 0.84 & 0.82 & 0.83 \\
&  &  & \best{BERT}  & \best{0.84} & \best{0.84} & \best{0.84} & \best{0.84} \\
&  &  & DistilGPT2  & 0.77 & 0.77 & 0.77 & 0.77 \\
\hline

\multirow{9}{*}{\rotatebox{90}{AutoML}}
& \multirow{9}{*}{FLAML}
& \multirow{5}{*}{Classical}
& CNN   & 0.83 & 0.84 & 0.83 & 0.83 \\
&  &  & \best{MLP} & \best{0.85} & \best{0.86} & \best{0.86} & \best{0.85} \\
&  &  & \best{LeNet} & \best{0.85} & \best{0.85} & \best{0.85} & \best{0.85} \\
&  &  & \best{MDCNN} & \best{0.85} & \best{0.85} & \best{0.85} & \best{0.85} \\
&  &  & NN & 0.84 & 0.84 & 0.84 & 0.84\\
\cline{3-8}
&  & \multirow{2}{*}{Pre-Attention}
& \best{LSTM}  & \best{0.85} & \best{0.85} & \best{0.84} & \best{0.85} \\
&  &  & RNN   & 0.80  & 0.80 & 0.80 & 0.80 \\
\cline{3-8}
&  & \multirow{3}{*}{Attention}
& Transformer & 0.83 & 0.84 & 0.83 & 0.83 \\
&  &  & \best{BERT}  & \best{0.85} & \best{0.85} & \best{0.84} & \best{0.85} \\
&  &  & DistilGPT2  & 0.79 & 0.81 & 0.76 & 0.78 \\
\hline

\multirow{18}{*}{\rotatebox{90}{Probabilistic}}
& \multirow{9}{*}{PSO}
& \multirow{5}{*}{Classical}
& \best{CNN}   & \best{0.85} & \best{0.86} & \best{0.85} & \best{0.85} \\
&  &  & MLP   & 0.77 & 0.81 & 0.77 & 0.77 \\
&  &  & LeNet & 0.83 & 0.83 & 0.83 & 0.83 \\
&  &  & MDCNN & 0.84 & 0.84 & 0.84 & 0.84 \\
&  &  & \best{NN}    & \best{0.85} & \best{0.85} & \best{0.85} & \best{0.85} \\
\cline{3-8}
&  & \multirow{2}{*}{Pre-Attention}
& LSTM  & 0.69 & 0.69 & 0.69 & 0.69 \\
&  &  & RNN  & 0.71 & 0.71 & 0.71 & 0.71 \\
\cline{3-8}
&  & \multirow{3}{*}{Attention}
& Transformer & 0.71 & 0.73 & 0.68 & 0.70 \\
&  &  & BERT  & 0.76 & 0.76 & 0.76 & 0.76 \\
&  &  & DistilGPT2  & 0.82 & 0.82 & 0.82 & 0.82 \\
\cline{2-8}

& \multirow{9}{*}{WOA}
& \multirow{5}{*}{Classical}
& CNN & 0.87 & 0.88 & 0.87 & 0.87 \\
&  &  & MLP & 0.83 & 0.84 & 0.83 & 0.82 \\
&  &  & LeNet & 0.85 & 0.85 & 0.85 & 0.85 \\
&  &  & MDCNN & 0.85 & 0.85 & 0.85 & 0.85 \\
&  &  & \best{NN} & \best{0.88} & \best{0.88} & \best{0.88} & \best{0.88} \\   
\cline{3-8}
&  & \multirow{2}{*}{Pre-Attention}
& LSTM  & 0.73 & 0.73 & 0.73 & 0.73 \\
&  &  & RNN  & 0.73 & 0.73 & 0.73 & 0.73 \\
\cline{3-8}
&  & \multirow{3}{*}{Attention}
& Transformer & 0.74 & 0.76 & 0.71 & 0.74 \\
&  &  & BERT  & 0.84 & 0.84 & 0.84 & 0.84 \\
&  &  & DistilGPT2  & 0.84 & 0.85 & 0.84 & 0.84 \\
\hline
\end{tabular}
\end{table}

\begin{table}[htbp]
\centering
\scriptsize
\setlength{\tabcolsep}{5.5pt}
\renewcommand{\arraystretch}{1.5}
\caption{DL model performance across feature selection techniques on SysCall Traces, bold and green-shaded values indicate the best performance.}
\label{tab:syscall}
\begin{tabular}{|c| c| c| l | c c c c|}
\hline
$\mathbf{\mathcal{F_C}}$ &
$\mathbf{\mathcal{F}}$ &
$\mathbf{\mathcal{M_C}}$ &
$\mathbf{\mathcal{M}}$ &
$\mathbf{\mathcal{A}}$ &
$\mathbf{\mathcal{P}}$ &
$\mathbf{\mathcal{R}}$ &
$\mathbf{\mathcal{F}_1}$ \\
\hline

\multirow{18}{*}{\rotatebox{90}{Statistical}}
& \multirow{9}{*}{ANOVA}
& \multirow{5}{*}{Classical}
& \best{CNN}  & \best{0.95} & \best{0.92} & \best{0.99} & \best{0.95} \\
&  &  & \best{MLP}  & \best{0.95} & \best{0.91} & \best{0.99} & \best{0.95} \\
&  &  & \best{LeNet} & \best{0.95} & \best{0.92} & \best{0.99} & \best{0.95} \\
&  &  & \best{MDCNN} & \best{0.95} & \best{0.92} & \best{0.99} & \best{0.95} \\
&  &  & \best{NN}  &  \best{0.95} & \best{0.92} & \best{0.99} & \best{0.95} \\
\cline{3-8}
&  & \multirow{2}{*}{Pre-Attention}
& LSTM  & 0.94 & 0.90 & 0.99 & 0.94 \\
&  &  & RNN  & 0.94 & 0.90 & 0.99 & 0.94 \\
\cline{3-8}
&  & \multirow{3}{*}{Attention}
& \best{Transformer} & \best{0.95} & \best{0.92}  & \best{0.98}  & \best{0.95}  \\
&  &  & BERT  & 0.94 & 0.94 & 0.98 & 0.94 \\
&  &  & DistilGPT2  & 0.94 & 0.99 & 0.90 & 0.94 \\
\cline{2-8}

& \multirow{9}{*}{CORR}
& \multirow{5}{*}{Classical}
& \best{CNN}  & \best{0.94} & \best{0.98} & \best{0.89} & \best{0.93} \\
&  &  & \best{MLP} & \best{0.94} & \best{0.98} & \best{0.89} & \best{0.93} \\
&  &  & \best{LeNet} & \best{0.94} & \best{0.99} & \best{0.90} & \best{0.93} \\
&  &  & \best{MDCNN} & \best{0.94} & \best{0.98} & \best{0.89} & \best{0.93} \\
&  &  & \best{NN} & \best{0.94} & \best{0.99} & \best{0.89} & \best{0.94} \\
\cline{3-8}
&  & \multirow{2}{*}{Pre-Attention}
& \best{LSTM}  & \best{0.94} & \best{0.98} & \best{0.89} & \best{0.93} \\
&  &  & \best{RNN}  & \best{0.94} & \best{0.98} & \best{0.89} & \best{0.94} \\
\cline{3-8}
&  & \multirow{3}{*}{Attention}
& \best{Transformer} & \best{0.94} & \best{0.98} & \best{0.89} & \best{0.93} \\
&  &  & BERT  & 0.93 & 0.93 & 0.93 & 0.93 \\
&  &  & \best{DistilGPT2} & \best{0.94} & \best{0.98} & \best{0.89} & \best{0.93} \\
\hline

\multirow{9}{*}{\rotatebox{90}{AutoML}}
& \multirow{9}{*}{FLAML}
& \multirow{5}{*}{Classical}
& \best{CNN}  & \best{0.94} & \best{0.90} & \best{0.98} & \best{0.94} \\
&  &  & \best{MLP}   & \best{0.94} & \best{0.99} & \best{0.89} & \best{0.94} \\
&  &  & \best{LeNet} & \best{0.94} & \best{0.90} & \best{0.98} & \best{0.94} \\
&  &  & \best{MDCNN} & \best{0.94} & \best{0.90} & \best{0.98} & \best{0.94} \\
&  &  & \best{NN}    & \best{0.94} & \best{0.90} & \best{0.98} & \best{0.94} \\
\cline{3-8}
&  & \multirow{2}{*}{Pre-Attention}
& \best{LSTM}  & \best{0.94} & \best{0.99} & \best{0.89} & \best{0.93} \\
&  &  & RNN  & 0.93 & 0.99 & 0.88 & 0.93 \\
\cline{3-8}
&  & \multirow{3}{*}{Attention}
& \best{Transformer} & \best{0.94} & \best{0.99}  & \best{0.89} & \best{0.94}  \\
&  &  & \best{BERT}  & \best{0.94} & \best{0.98} & \best{0.89} & \best{0.93} \\
&  &  & \best{DistilGPT2}  & \best{0.94} & \best{0.90} & \best{0.98} & \best{0.94} \\
\hline

\multirow{18}{*}{\rotatebox{90}{Probabilistic}}
& \multirow{9}{*}{PSO}
& \multirow{5}{*}{Classical}
& \best{CNN}  & \best{0.95} & \best{0.92} & \best{0.99} & \best{0.95} \\
&  &  & \best{MLP}  & \best{0.95} & \best{0.91} & \best{0.99} & \best{0.95} \\
&  &  & \best{LeNet} & \best{0.95} & \best{0.92} & \best{0.99} & \best{0.95} \\
&  &  & \best{MDCNN} & \best{0.95} & \best{0.92} & \best{0.99} & \best{0.95} \\
&  &  & \best{NN} & \best{0.95} & \best{0.92} & \best{0.99} & \best{0.95} \\
\cline{3-8}
&  & \multirow{2}{*}{Pre-Attention}
& LSTM  & 0.94 & 0.90 & 0.99 & 0.94 \\
&  &  & RNN  & 0.94 & 0.90 & 0.99 & 0.94 \\
\cline{3-8}
&  & \multirow{3}{*}{Attention}
& Transformer & 0.94 & 0.89 & 0.99 & 0.94 \\
&  &  & BERT  & 0.92 & 0.87 & 1.00 & 0.93 \\
&  &  & DistilGPT2  & 0.94 & 0.91 & 0.99 & 0.95 \\
\cline{2-8}

& \multirow{9}{*}{WOA}
& \multirow{5}{*}{Classical}
& \best{CNN}  & \best{0.94} & \best{0.91} & \best{0.98} & \best{0.94} \\
&  &  & \best{MLP} & \best{0.94} & \best{0.99} & \best{0.88} & \best{0.93} \\
&  &  & \best{LeNet} & \best{0.94} & \best{0.98} & \best{0.90} & \best{0.94} \\
&  &  & \best{MDCNN} & \best{0.94} & \best{0.91} & \best{0.98} & \best{0.94} \\
&  &  & \best{NN} & \best{0.94} & \best{0.91} & \best{0.98} & \best{0.94} \\
\cline{3-8}
&  & \multirow{2}{*}{Pre-Attention}
& \best{LSTM}  & \best{0.94} & \best{0.99} & \best{0.89} & \best{0.94} \\
&  &  & \best{RNN}  & \best{0.94} & \best{0.90} & \best{0.98} & \best{0.94} \\
\cline{3-8}
&  & \multirow{3}{*}{Attention}
& \best{Transformer} & \best{0.94} & \best{0.90} & \best{0.98} & \best{0.94}  \\
&  &  & BERT  & 0.93 & 0.99 & 0.88 & 0.93  \\
&  &  & \best{DistilGPT2}  & \best{0.94} & \best{0.90} & \best{0.98} & \best{0.94} \\
\hline
\end{tabular}
\end{table}

\begin{table}[htbp]
\centering
\scriptsize
\setlength{\tabcolsep}{5.5pt}
\renewcommand{\arraystretch}{1.5}
\caption{DL model performance across feature selection techniques on Pattern Traces, bold and green-shaded values indicate the best performance.}
\label{tab:pattern}
\begin{tabular}{|c| c| c| l | c c c c|}
\hline
$\mathbf{\mathcal{F_C}}$ &
$\mathbf{\mathcal{F}}$ &
$\mathbf{\mathcal{M_C}}$ &
$\mathbf{\mathcal{M}}$ &
$\mathbf{\mathcal{A}}$ &
$\mathbf{\mathcal{P}}$ &
$\mathbf{\mathcal{R}}$ &
$\mathbf{\mathcal{F}_1}$ \\
\hline

\multirow{18}{*}{\rotatebox{90}{Statistical}}
& \multirow{9}{*}{ANOVA}
& \multirow{5}{*}{Classical}
& \best{CNN} & \best{0.76} & \best{0.77} & \best{0.76} & \best{0.75} \\
&  &  & MLP   & 0.71 & 0.76 & 0.71 & 0.70 \\
&  &  & LeNet & 0.73 & 0.76 & 0.72 & 0.72 \\
&  &  & MDCNN & 0.75 & 0.76 & 0.75 & 0.75 \\
&  &  & NN    & 0.72 & 0.74 & 0.72 & 0.72 \\
\cline{3-8}
&  & \multirow{2}{*}{Pre-Attention}
& LSTM  & 0.71 & 0.70 & 0.72 & 0.71 \\
&  &  & RNN  & 0.71 & 0.70 & 0.72 & 0.71 \\
\cline{3-8}
&  & \multirow{3}{*}{Attention}
& Transformer & 0.70 & 0.66  & 0.83  & 0.73  \\
&  &  & \best{BERT}  & \best{0.76} & \best{0.71} & \best{0.88} & \best{0.79} \\
&  &  & DistilGPT2  & 0.71 & 0.70 & 0.75 & 0.72 \\
\cline{2-8}

& \multirow{9}{*}{CORR}
& \multirow{5}{*}{Classical}
& CNN   & 0.87 & 0.87 & 0.87 & 0.87 \\
&  &  & \best{MLP}   & \best{0.88} & \best{0.88} & \best{0.88} & \best{0.88} \\
&  &  & \best{LeNet} & \best{0.88} & \best{0.88} & \best{0.88} & \best{0.77} \\
&  &  & \best{MDCNN} & \best{0.88} & \best{0.88} & \best{0.88} & \best{0.88} \\
&  &  & \best{NN}    & \best{0.88} & \best{0.88} & \best{0.88} & \best{0.88} \\
\cline{3-8}
&  & \multirow{2}{*}{Pre-Attention}
& \best{LSTM}  & \best{0.88} & \best{0.88} & \best{0.88} & \best{0.88} \\
&  &  & \best{RNN}  & \best{0.88} & \best{0.88} & \best{0.88} & \best{0.88} \\
\cline{3-8}
&  & \multirow{3}{*}{Attention}
& Transformer & 0.86 & 0.87 & 0.84 & 0.85 \\
&  &  & BERT  & 0.84 & 0.84 & 0.84 & 0.84 \\
&  &  & DistilGPT2  & 0.83 & 0.83 & 0.83 & 0.83 \\
\hline

\multirow{9}{*}{\rotatebox{90}{AutoML}}
& \multirow{9}{*}{FLAML}
& \multirow{5}{*}{Classical}
& CNN   & 0.87 & 0.87 & 0.87 & 0.87 \\
&  &  & \best{MLP}   & \best{0.88} & \best{0.87} & \best{0.89} & \best{0.88} \\
&  &  & \best{LeNet} & \best{0.88} & \best{0.87} & \best{0.88} & \best{0.88} \\
&  &  & \best{MDCNN} & \best{0.88} & \best{0.88} & \best{0.89} & \best{0.88} \\
&  &  & \best{NN}    & \best{0.88} & \best{0.87} & \best{0.88} & \best{0.88} \\
\cline{3-8}
&  & \multirow{2}{*}{Pre-Attention}
& \best{LSTM}  & \best{0.88} & \best{0.87} & \best{0.89} & \best{0.88} \\
&  &  & RNN  & 0.78 & 0.80 & 0.75 & 0.78 \\
\cline{3-8}
&  & \multirow{3}{*}{Attention}
& \best{Transformer} & \best{0.88} & \best{0.90}  & \best{0.86}  & \best{0.88}  \\
&  &  & BERT  & 0.86 & 0.87 & 0.84 & 0.86 \\
&  &  & DistilGPT2  & 0.85 & 0.85 & 0.84 & 0.84 \\
\hline

\multirow{18}{*}{\rotatebox{90}{Probabilistic}}
& \multirow{9}{*}{PSO}
& \multirow{5}{*}{Classical}
& CNN   & 0.86 & 0.86 & 0.86 & 0.86 \\
&  &  & \best{MLP}   & \best{0.87} & \best{0.87} & \best{0.875} & \best{0.87} \\
&  &  & \best{LeNet} & \best{0.87} & \best{0.87} & \best{0.87} & \best{0.87} \\
&  &  & \best{MDCNN} & \best{0.87} & \best{0.87} & \best{0.87} & \best{0.87} \\
&  &  & \best{NN}    & \best{0.87} & \best{0.87} & \best{0.87} & \best{0.87} \\
\cline{3-8}
&  & \multirow{2}{*}{Pre-Attention}
& LSTM  & 0.86 & 0.87 & 0.86 & 0.86 \\
&  &  & RNN  & 0.81 & 0.81 & 0.81 & 0.81 \\
\cline{3-8}
&  & \multirow{3}{*}{Attention}
& Transformer & 0.84 & 0.83 & 0.85 & 0.84 \\
&  &  & BERT  & 0.83 & 0.83 & 0.83 & 0.83 \\
&  &  & DistilGPT2  & 0.84 & 0.84 & 0.84 & 0.84 \\
\cline{2-8}

& \multirow{9}{*}{WOA}
& \multirow{5}{*}{Classical}
& CNN   & 0.87 & 0.87 & 0.87 & 0.87 \\
&  &  & \best{MLP}   & \best{0.88} & \best{0.88} & \best{0.88} & \best{0.88} \\
&  &  & \best{LeNet} & \best{0.88} & \best{0.88} & \best{0.88} & \best{0.88} \\
&  &  & \best{MDCNN} & \best{0.88} & \best{0.88} & \best{0.88} & \best{0.88} \\
&  &  & NN    & 0.87 & 0.87 & 0.87 & 0.87 \\
\cline{3-8}
&  & \multirow{2}{*}{Pre-Attention}
& \best{LSTM}  & \best{0.88} & \best{0.88} & \best{0.88} & \best{0.88} \\
&  &  & RNN  & 0.78 & 0.80 & 0.78 & 0.78 \\
\cline{3-8}
&  & \multirow{3}{*}{Attention}
& Transformer & 0.87 & 0.87  & 0.87  &  0.87 \\
&  &  & BERT  & 0.83 & 0.83 & 0.83 & 0.83 \\
&  &  & DistilGPT2  & 0.84 & 0.84 & 0.84 & 0.84 \\
\hline
\end{tabular}
\end{table}

\onecolumn

\begin{table*}[htpb]
\centering
\scriptsize
\caption{Performance of the proposed eDySec using FLAML-selected features across all individual DL models; bold green-shaded values indicate the best results.}
\renewcommand{\arraystretch}{1.25}
\label{tab:performance_comparison_appendix}
\resizebox{5.6in}{!}{
\begin{tabular}{|c|c|c|c|c|c|c|c|c|c|c|c|}
\hline
\multirow{2}{*}{\textbf{Traces}} & 
\multirow{2}{*}{\textbf{Category}} & 
\multirow{2}{*}{\textbf{Model}} & 
\multirow{2}{*}{\textbf{Features}} & 
\multirow{2}{*}{\textbf{Test AUC}} & 
\multirow{2}{*}{\textbf{F1 Score}} & 
\multirow{2}{*}{\textbf{FPR(\%)}} & 
\multirow{2}{*}{\textbf{FNR(\%)}} & 
\multicolumn{4}{c|}{\textbf{Confusion Matrix}} \\ 
\cline{9-12}
 &  &  &  &  &  &  &  & 
\textbf{TP} ↑ & \textbf{TN} ↑ & \textbf{FP} ↓ & \textbf{FN} ↓ \\ 
\hline

\multirow{10}{*}{\makecell{Filetop \\Traces}}
 & \multirow{5}{*}{Classic} & CNN & \multirow{5}{*}{9} & 0.91 & 0.91 & 7.27 & 10.10 & 997 & 957 & 75 & 112 \\
 &  & \best{MLP} &  & \best{0.93} & \best{0.93} & \best{4.90} & 7.27 & 1021 & 989 & 51 & 80 \\
 &  & \best{LeNet} &  & \best{0.93} & \best{0.93} & 6.38 & 6.69 & 1004 & 997 & 68 & 72 \\
 &  & \best{MDCNN} &  & \best{0.93} & \best{0.93} & 5.24 & 6.78 & 1017 & 995 & 55 & 74 \\
 &  & \best{NN} &  & \best{0.93} & \best{0.93} & 6.60 & \best{6.10} & 1001 & \best{1004} & 71 & \best{65} \\
\cline{2-12}
 & \multirow{2}{*}{Pre-Attention} & LSTM & \multirow{2}{*}{9} & 0.91 & 0.91 & 9.67 & 7.56 & 966 & 990 & 106 & 79 \\
 &  & RNN &  & 0.85 & 0.86 & 19.03 & 7.21 & 836 & 1004 & 236 & 65 \\
\cline{2-12}
 & \multirow{3}{*}{Attention} & Transformer & \multirow{3}{*}{9} & 0.89 & 0.89 & 12.61 & 8.37 & 930 & 984 & 142 & 85 \\
 &  & BERT & & 0.85 & 0.86 & 16.34 & 16.65 & 886 & 952 & 186 & 177  \\
 &  & DistilGPT2 & & 0.92 & 0.92 & 7.90 & 6.36 & 986 & 1002 & 86 & 67 \\
\hline

\multirow{10}{*}{\makecell{Opensnoop \\Traces}}
 & \multirow{5}{*}{Classic} & CNN & \multirow{5}{*}{11} & 0.89 & 0.89 & 8.84 & 11.78 & 981 &  938 & 91 & 131 \\
 &  & \best{MLP} &  & \best{0.91} & \best{0.91} & \best{3.88} & 14.52 & \best{1036} & 893 & \best{36} & 176 \\
 &  & LeNet &  & 0.88 & 0.88 & 9.82 & 12.68 & 971 & 928 & 101 & 141 \\
 &  & MDCNN &  & 0.88 & 0.88 & 8.06 & 13.59 & 992 & 913 & 80 & 156 \\
 &  & NN &  & 0.90 & 0.90 & 9.70 & \best{9.45} & 968 & \best{968} & 104 & \best{101} \\
\cline{2-12}
 & \multirow{2}{*}{Pre-Attention} & LSTM & \multirow{2}{*}{11} & 0.76 & 0.78 & 26.67 & 18.25 & 746 & 896 & 326 & 173 \\
 &  & RNN &  & 0.75 & 0.74 & 22.79 & 26.60 & 847 & 762 & 225 & 307 \\
\cline{2-12}
 & \multirow{3}{*}{Attention} & Transformer & \multirow{3}{*}{11} & 0.73 & 0.74 & 29.55 & 23.36 & 715 & 851 & 357 & 218 \\ 
 &  & \best{BERT} &  & \best{0.91} & \best{0.91} & 5.68 & 10.80 & 1015 & 946 & 57 & 123 \\
 &  & DistilGPT2 &  & 0.89 & 0.89 & 7.66 & 12.41 & 995 & 928 & 77 & 141 \\
\hline

\multirow{10}{*}{\makecell{Install \\Traces}}
 & \multirow{5}{*}{Classic} & CNN & \multirow{5}{*}{2} & 0.72 & 0.70 & 35.32 & 1.40 & 492 & 1062 & 580 & 7 \\
 &  & \best{MLP} &  & \best{0.75} & \best{0.74} & \best{32.72} & \best{0.00} & 517 & \best{1069} & 555 & \best{0} \\
 &  & LeNet &  & 0.75 & 0.74 & \best{32.72} & \best{0.00} & \best{552} & \best{1069} & \best{520} & \best{0} \\
 &  & MDCNN &  & 0.75 & 0.74 & 32.77 & \best{0.00} & 551 & \best{1069} & 521 & \best{0} \\
 &  & NN &  & 0.75 & 0.74 & \best{32.72} & \best{0.00} & \best{552} & \best{1069} & \best{520} & \best{0} \\
\cline{2-12}
 & \multirow{2}{*}{Pre-Attention} & \best{LSTM} & \multirow{2}{*}{2} & \best{0.75} & \best{0.80} & \best{32.72} & \best{0.00} & \best{552} & \best{1069} & \best{520} & \best{0} \\
 &  & RNN &  & 0.56 & 0.69 & 46.58 & 4.58 & 146 & 1062 & 926 & 7 \\
\cline{2-12}
 & \multirow{3}{*}{Attention} & \best{Transformer} & \multirow{3}{*}{2} & \best{0.75} & \best{0.80} & 32.81 & \best{0.00} & 550 & \best{1069} & 522 & \best{0} \\
 &  & BERT &  & 0.59 & 0.66 & 43.31 & \best{0.00} & 255 & \best{1069} & 817 & \best{0} \\
 &  & \best{DistilGPT2} &  & \best{0.75} & \best{0.80} & \best{32.72} & \best{0.00} & \best{552} & \best{1069} & \best{520} & \best{0} \\
\hline

\multirow{10}{*}{\makecell{TCP \\Traces}}
 & \multirow{5}{*}{Classic} & CNN & \multirow{5}{*}{5} & 0.83 & 0.83 & \best{11.34} & 19.90 & \best{966} & 829 & \best{106} & 240 \\
 &  & MLP &  & 0.85 & 0.85 & 13.21 & 15.31 & 935 & 900 & 137 & 169 \\
 &  & LeNet &  & 0.85 & 0.85 & 12.64 & 16.31 & 944 & 885 & 128 & 184 \\
 &  & MDCNN &  & 0.86 & 0.86 & 12.15 & 14.31 & 946 & 911 & 126 & 158 \\
 &  & NN &  & 0.84 & 0.84 & 15.61 & 16.33 & 907 & 892 & 165 & 177 \\
\cline{2-12}
 & \multirow{2}{*}{Pre-Attention} & LSTM & \multirow{2}{*}{5} & 0.85 & 0.85 & 14.05 & 15.52 & 925 & 899 & 147 & 170 \\
 &  & RNN &  & 0.80 & 0.80 & 19.36 & 19.31 & 865 & 862 & 207 & 207 \\
\cline{2-12}
 & \multirow{3}{*}{Attention} & Transformer & \multirow{3}{*}{5} & 0.83 & 0.83 & 15.86 & 16.45 & 904 & 891 & 168 & 178 \\
 &  & \best{BERT} &  & \best{0.88} & \best{0.88} & 12.83 & \best{10.65} & 931 & \best{958} & 141 & \best{111} \\
 &  & DistilGPT2 &  & 0.79 & 0.78 & 18.29 & 22.33 & 890 & 813 & 182 & 256 \\
\hline

\multirow{10}{*}{\makecell{SysCall \\Traces}}
 & \multirow{5}{*}{Classic} & \best{CNN} & \multirow{5}{*}{5}  & \best{0.94} & \best{0.94} & 9.05 & \best{1.42} & 967 & \best{1055} & 105 & \best{14} \\
 &  & \best{MLP} & & \best{0.94} & \best{0.94} & 0.93 & 9.53 & 1063 & 957 & 9 & 112 \\
 &  & \best{LeNet} & & \best{0.94} & \best{0.94} & 9.13 & \best{1.42} & 966 & \best{1055} & 106 & \best{14}  \\
 &  & \best{MDCNN} &  & \best{0.94} & \best{0.94} & 9.05 & \best{1.42} & 967 & \best{1055} & 105 & \best{14} \\
 &  & \best{NN} &  & \best{0.94} & \best{0.94} & 9.13 & \best{1.42} & 966 & \best{1055} & 106 & \best{14}  \\
\cline{2-12}
 & \multirow{2}{*}{Pre-Attention} & LSTM & \multirow{2}{*}{5} & 0.94 & 0.93 & 0.83 & 9.90 & \best{1064} & 952 & 8 & 117 \\
 &  & RNN &  & 0.93 & 0.93 & \best{0.73} & 10.42 & \best{1065} & 945 & \best{7} & 124 \\
\cline{2-12}
 & \multirow{3}{*}{Attention} & \best{Transformer} & \multirow{3}{*}{5} & \best{0.94} & \best{0.94} & 0.83 & 9.60 & \best{1064} & 956 & 8 & 113 \\
 &  & \best{BERT} &  & \best{0.94} & \best{0.94} & 1.23 & 9.47 & 1060 & 958 & 12 & 111 \\
 &  & \best{DistilGPT2} &  & \best{0.94} & \best{0.94} & 9.13 & \best{1.42} & 966 & \best{1055} & 106 & \best{14} \\
\hline

\multirow{10}{*}{\makecell{Pattern \\Traces}}
 & \multirow{5}{*}{Classic} & CNN & \multirow{5}{*}{8} & 0.87 & 0.87 & 12.17 & 12.21 & 942 & 938 & 130 & 131 \\
 &  & \best{MLP} &  & \best{0.88} & \best{0.88} & 12.64 & 10.96 & 934 & 954 & 138 & 115 \\
 &  & \best{LeNet} &  & \best{0.88} & \best{0.88} & 12.28 & 11.25 & 939 & 950 & 133 & 119 \\
 &  & \best{MDCNN} &  & \best{0.88} & \best{0.88} & 11.77 & 11.02 & 945 & 952 & 127 & 117 \\
 &  & NN &  & 0.88 & 0.88 & 12.36 & 11.26 & 938 & 950 & 134 & 119 \\
\cline{2-12}
 & \multirow{2}{*}{Pre-Attention} & \best{LSTM} & \multirow{2}{*}{8} & \best{0.88} & \best{0.88} & 12.37 & \best{10.76} & 937 & \best{956} & 135 & \best{113} \\
 &  & RNN &  & 0.78 & 0.78 & 19.54 & 22.77 & 875 & 811 & 197 & 258 \\
\cline{2-12}
 & \multirow{3}{*}{Attention} & \best{Transformer} & \multirow{3}{*}{8} & \best{0.88} & \best{0.88} & \best{9.80} & 12.60 & \best{971} & 929 & \best{101} & 140 \\
 &  & BERT &  & 0.86 & 0.86 & 12.63 & 14.76  & 941 & 906 & 131 & 163 \\
 &  & DistilGPT2 &  & 0.85 & 0.84 & 14.42 & 15.36 & 920 & 902 & 152 & 167 \\
\hline

\end{tabular}}
\end{table*}

\end{document}